\documentclass[3p]{elsarticle}

\usepackage{amssymb,bm}
\usepackage{mathrsfs,amsmath}
\usepackage{amsfonts}
\usepackage{subfigure}
\usepackage{color}
\usepackage{graphicx}
\usepackage{tabu}
\usepackage{lineno,hyperref}
\usepackage{multirow}
\usepackage{enumitem}
\usepackage{lipsum}
\usepackage{etoolbox}
\usepackage{diagbox}
\usepackage[]{algorithm2e}

\newcommand{\captionabove}[2][]{%
	\vskip-\abovecaptionskip
	\vskip+\belowcaptionskip
	\ifx\@nnil#1\@nnil
	\caption{#2}%
	
	\else
	\caption[#1]{#2}%
	\fi
	\vskip+\abovecaptionskip
	\vskip-\belowcaptionskip
}

\usepackage{amsmath}
\DeclareMathOperator*{\argmax}{arg\,max}

\newtheorem{theorem}{Theorem}

\newtheorem{remark}[theorem]{Remark}

\numberwithin{equation}{section}

\makeatletter
\def\@mkboth#1#2{}
\newlength\appendixwidth
\preto\appendix{\addtocontents{toc}{\protect\patchl@section}}
\newcommand{\patchl@section}{%
	\settowidth{\appendixwidth}{\textbf{Appendix }}%
	\addtolength{\appendixwidth}{1.5em}%
	\patchcmd{\l@section}{1.5em}{\appendixwidth}{}{\ddt}%
}
\makeatother

\journal{}

\begin{document}
	
	\begin{frontmatter}
		\title{Cell division in deep material networks applied to multiscale strain localization modeling}
		\address[Zeliangaddress]{Ansys Inc., Livermore, CA, USA}
		\cortext[mycorrespondingauthor]{Corresponding author.}
		\author[Zeliangaddress]{Zeliang Liu\corref{mycorrespondingauthor}}
		\ead{zeliang.academic@gmail.com}
		%% Absract
		\begin{abstract}
			
			Despite the increasing importance of strain localization modeling (e.g., failure analysis) in computer-aided engineering,  there is a lack of effective approaches to capturing relevant material behaviors consistently across multiple length scales. We aim to address this gap within the framework of deep material networks (DMN) -- a machine learning model with embedded mechanics in the building blocks. A new cell-division scheme is proposed to track the scale transition through the network, and its consistency is ensured by the physics of fitting parameters. Essentially, each microscale node in the bottom layer is described by an ellipsoidal cell with its dimensions back-propagated from the macroscale material point. New crack surfaces in the cell are modeled by enriching cohesive layers, and failure algorithms are developed for crack initiation and evolution in the implicit DMN analysis. Besides studies on a single material point, we apply the multiscale model to concurrent multiscale simulations for the dynamic crush of a particle-reinforced composite tube and various tests on carbon fiber reinforced polymer composites. For the latter, experimental validations on an off-axis tensile test specimen are also provided.
		\end{abstract}
		%% Keywords
	\end{frontmatter}
	
	%% Main content
	\tableofcontents
	%\linenumbers
	\section{Introduction}
	\subsection{Motivation and background}
	Over the past decades, multiscale modeling has become a key enabling technology for computer simulation and materials design in the aerospace, automotive, and consumer electronics industries. When dealing with materials with expressive microstructures, engineers often find a single-scale phenomenological model fails to capture the anisotropic nonlinear behaviors due to complex physical interactions across scales. In contrast, the core of multiscale materials modeling stands on explicit microstructure representations and homogenization schemes. For elastic and nonlinear plastic materials with scale separation, theories for homogenization are well established. Most recent studies have been focusing on improving the efficiency and accuracy of microstructure modeling. However, for materials with strain localizations, such as damage and fracture, the homogenization condition is closely entangled with the microstructure's length-scale variations, making it challenging to define a consistent multiscale model. In the paper, we will address this gap within the framework of Deep Material Network (DMN) \cite{liu2019deep,liu2019exploring}, which is a machine learning model with physics-based building blocks.
	
	Meanwhile, the rise of Machine Learning (ML) has been continuously advancing the frontier of materials modeling and multiscale simulations. Although feedforward neural networks have been applied to materials modeling and characterization since the 1990s \cite{ghaboussi1991knowledge,hkdh1999neural}, their popularity was limited mainly due to the lack of computational resources. Recent advancements of computer hardware systems and open-source ML platforms stimulate the renaissance of deep learning. They allow researchers to effectively explore more sophisticated network architectures, such as convolutional neural networks \cite{lecun1995convolutional}, recurrent neural networks \cite{hochreiter1997long,cho2014properties}, and the transformers \cite{vaswani2017attention}. Importantly, the successes of these customized networks in computer vision and natural language processing also motivate mechanicians to design new ML architectures specialized for materials modeling and multiscale simulations, which could potentially discover hidden physical relationships and outperform existing knowledge-driven models in certain tasks.
	
	A category of knowledge-driven multiscale models is based on micromechanics theories, such as Mori-Tanaka method \cite{mori1973average,qu1993effect} and self-consistent methods \cite{hill1965self,christensen1979solutions}. Given the elegance of Eshelby's solutions \cite{eshelby1957determination,liu2008solutions} and mean-field schemes, they are widely adopted in current research and industrial applications. A major barrier of micromechanics theories in describing highly nonlinear materials is that the analytical solutions of idealized geometries cannot capture localized deformations, especially when it comes to failure analysis. Therefore, those predictions often lose the multiscale characteristics and require extra calibrations on the empirical model parameters. In contrast, the Representative Volume Element (RVE)-based methods simulate microstructures explicitly, and the discretized models are usually solved by Finite Element (FE) methods \cite{kouznetsova2002a,belytschko2008a} or fast Fourier transform (FFT)-based methods \cite{moulinec1998a}. Couplings between the microscale RVE and the macroscale structure yield the so-called concurrent multiscale simulations, referred to as FE$^2$ \cite{feyel2000a,feyel2003a} or FE-FFT \cite{kochmann2018efficient} depending on the model type at each scale.
	
	By formulating the microstructure as a boundary value problem, RVE-based methods offer more flexibility on the choices of material constitutive laws, such as anisotropic elasticity, plasticity with history dependency, and nonlinear hyperelasticity under large deformations. However, there has been less consensus on using RVE for failure analysis or other straining localization phenomena. Their uniqueness originates from the loss of ellipticity when local damage or fracture appears in the RVE. As the equilibrium condition becomes unstable, the deformations tend to localize into a single layer of elements. In order to ensure the objectivity with respect to mesh size and shape in the RVE analysis, numerical treatments are needed to regularize the ill-posed boundary value problem, and a few representatives are gradient-enhanced or non-local damage models \cite{de1995gradient,peerlings1996gradient,geers1998strain,kuhl2000anisotropic,bazant2002nonlocal}, phase-field fracture \cite{miehe2010thermodynamically,miehe2015phase}, crack band theories \cite{bavzant1983crack,gorgogianni2020mechanism}, and cohesive zone models \cite{camacho1996computational,ortiz1999finite,park2011cohesive}. In particular, the cohesive zone models have been coupled with extended/generalized finite element methods for crack growth modeling \cite{moes1999finite,moes2002extended}. The crack band model \cite{bavzant1983crack} regularizes the post-failure softening stiffness by the mesh size, and similar approaches have been widely adopted due to their simplicity and effectiveness.
	
	Although the spurious mesh dependency could be alleviated by the aforementioned numerical techniques at a single scale, key challenges emerge when one considers the multiscale coupling. As a strain localization band (e.g., a crack or a damage zone) cuts through the RVE and breaks the periodicity, prescribing proper microscale boundary conditions becomes much more difficult. Besides, the homogenization results depend on the microscale RVE size, which does not necessarily match the macroscale length scale of localization. This mismatch will result in energy inconsistency of scale transition when the homogenized bulk stress-strain responses are directly passed to the macroscale material point. Namely, if the RVE size is smaller than the macroscale length scale of localization (e.g., the mesh size), the macroscale energy dissipation would be over-predicted. In Section \ref{sec:issues}, we will discuss the issues of strain localization modeling in detail. 
	
	Another challenge facing RVE-based methods is the efficiency issue since the direct numerical simulations (DNS) often contain many degrees of freedom (DOF). It becomes the bottleneck for materials design and concurrent multiscale simulations, both of which require a recurrent assessment of the RVE model. In this regard, many ML methods and frameworks have been proposed to improve the materials models' efficiency by learning from experience or data. The training data can come from either physical experiments or simulations, while the simulation data are usually more obtainable in large amount. One way of mining the data is to perform model reduction on the full-field DNS results, such as proper orthogonal decomposition (POD) \cite{yvonnet2007a,fritzen2018two,rocha2020adaptive}, self-consistent clustering analysis (SCA) \cite{liu2016self,liu2018microstructural,gao2020predictive}. In comparison, methods have also been formulated based on the stress-strain data -– e.g., Gaussian process modeling \cite{chen2018multiscale,bostanabad2018uncertainty}, model-free approaches \cite{kirchdoerfer2016data,ibanez2018manifold,eggersmann2019model,he2020physics}, feedforward neural networks \cite{ghaboussi1991knowledge,le2015computational,bessa2017,fritzen2019fly,lu2018data}, recurrent neural networks \cite{wang2018multiscale,frankel2019predicting,mozaffar2019deep,logarzo373smart}. 
	
	We consider DMN \cite{liu2019deep,liu2019exploring} as a blend of both approaches: it finds a reduced-order representation of the DNS RVE model while the training only requires stress-strain data. \textcolor{black}{By construction, the analytical solutions of the physics-based building block guarantee an arbitrary material network to satisfy the frame indifference and the Hill-Mandel principle.} Gajek et al. \cite{gajek2020micromechanics} also studied its micromechanical principle and thermodynamic consistency. \textcolor{black}{DMNs have been applied to several representative material systems, including hyperelastic rubber composites under large deformation, polycrystalline materials with crystal plasticity, and different types of carbon fiber reinforced polymer composites \cite{liu2019exploring,liu2020deep}.} Furthermore, the transfer learning approach of DMN for creating unified microstructure database has been proposed in \cite{liu2019transfer} with the application to short-fiber reinforced composites\cite{liu2020intelligent}.
	
    A few ML or data-driven methods have dealt with multiscale failure analysis. Oliver et al. \cite{oliver2017reduced} performed the model reduction via the POD method for multiscale fracture problems. A domain decomposition strategy was proposed to treat the regular domain and the localization band separately in the RVE. Liu et al. \cite{liu2018microstructural} applied the SCA method to multiscale damage analysis, and the microscale damage parameters are calibrated to restore the energy consistency. 
    \textcolor{black}{Bessa et al. \cite{bessa2017} combined Gaussian processes and the SCA method \cite{liu2016self} for fracture toughness predictions based on the plastic strain field.}
    Wang et al. \cite{wang2018multiscale} used recurrent neural networks to generate cohesive laws for modeling localized physical discontinuities at different length scales of multi-permeability porous media. Recently, DMNs with enriching cohesive layers were developed for interfacial failure analysis \cite{liu2020deep} (e.g., debonding in composites). Although this earlier work has not tackled the localization issue, it establishes the basis of failure modeling within the DMN framework.
	
	\subsection{Localization across scales: Issues of RVE-based modeling}\label{sec:issues}
\begin{figure}[!t]
		\centering
		\graphicspath{{Figures/}}
		\subfigure[\textcolor{black}{Limitation on the RVE size}]{\includegraphics[clip=true,trim = 9.0cm 5.5cm 10.cm 5.5cm,width=0.48\textwidth]{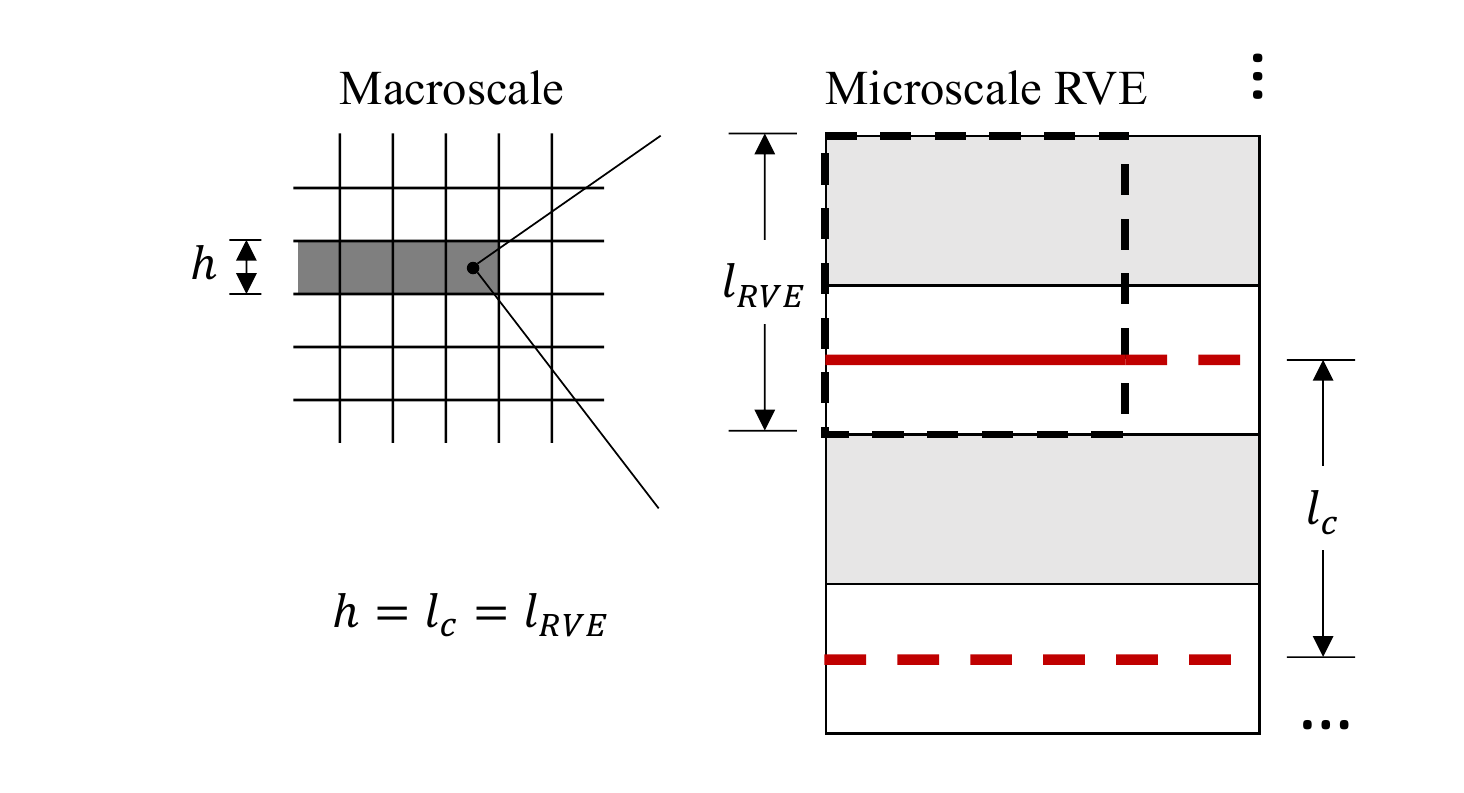}}
		\subfigure[\textcolor{black}{Difficulties of applying boundary conditions}]{\includegraphics[clip=true,trim = 9.0cm 5.5cm 10.0cm 5.5cm,width=0.48\textwidth]{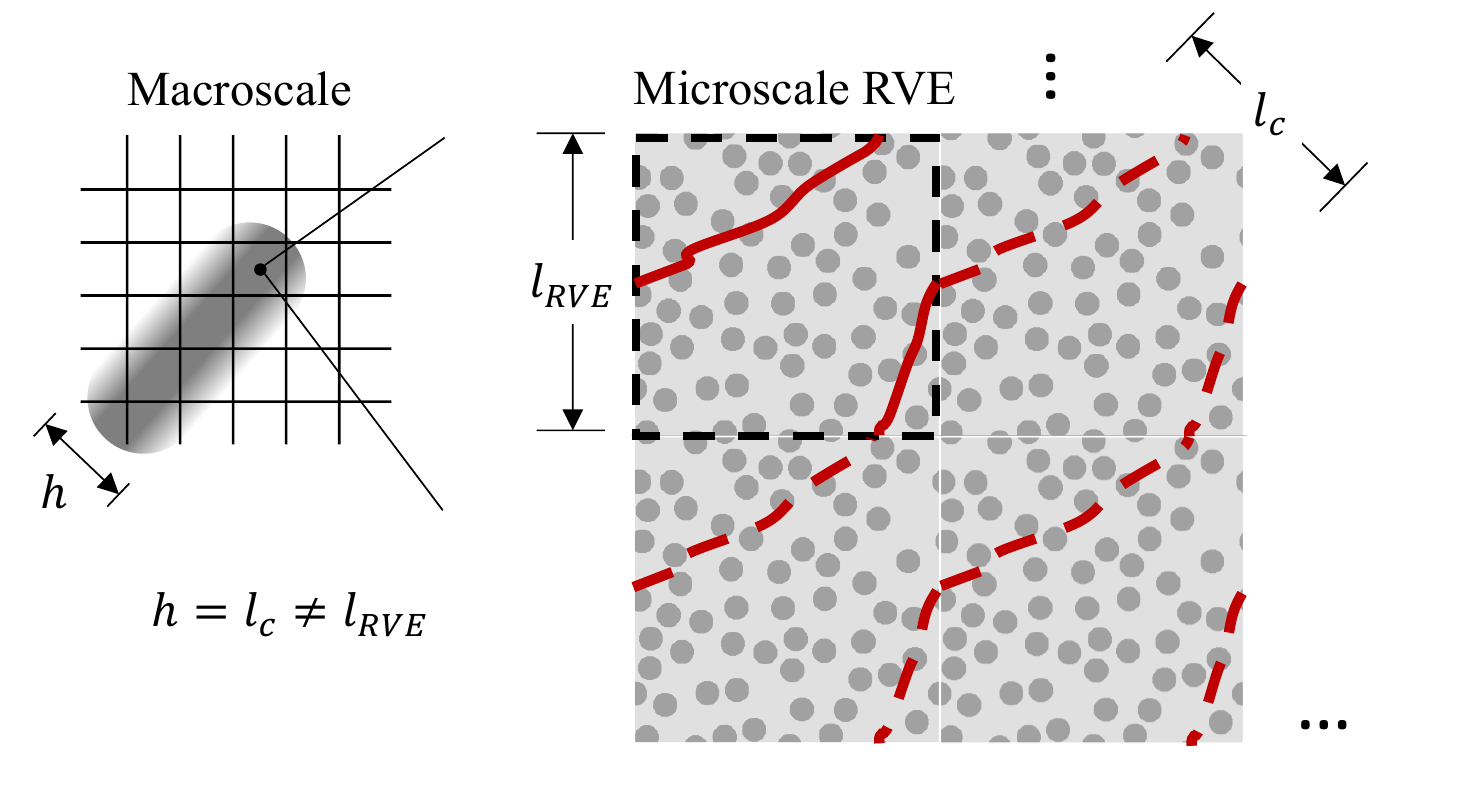}}
		\caption{Issues of RVE modeling when localization is presented. The macroscale length parameter, the RVE size, and the localization size are denoted by $h$, $l_{RVE}$, and $l_c$, respectively. \textcolor{black}{Energy consistency imposes $h=l_c$. In (a), $h$ is equal to the element size. In (b), $h$ is set by the size of the nonlocal regularization scheme. The localization size $l_c$ depends on the crack orientation, while the simulated crack path is distorted by the periodic boundary conditions applied on the RVE.}}
		\label{fig:crackissue}
	\end{figure}

	For failure analysis or strain localization modeling in general, RVE-based methods encounter certain difficulties that have long been recognized by researchers in the field \cite{belytschko2008a,belytschko2010coarse,bazant2010can,geers2010multi,bosco2014multiscale}. \textcolor{black}{In this paper, we will focus on the continuous multiscale framework 
		\footnote{\textcolor{black}{An alternative approach is to enrich the macroscale with a cohesive discontinuity model \cite{bosco2014multiscale,bosco2015multi,toro2016cohesive}. The responses of the macroscale discontinuity and its surrounding continuous material are coupled with localized and non-localized regions inside the RVE, respectively. The resulting continuous-discontinuous framework removes the limitations on the RVE size.}} 
	where the stress-strain responses at a macroscale integration/material point come from the bulk RVE homogenization.} To help explain the issues, we introduce three length parameters involved in a typical multiscale simulation based on RVE modeling: the macroscale length parameter $h$, the RVE size $l_{RVE}$, and the microscale localization size $l_{c}$.  The macroscale length parameter $h$ is usually determined by the element size because the softening region localizes into a single layer of elements, as shown in Figure \ref{fig:crackissue} (a). Alternatively, if one uses nonlocal or gradient-based regularization to reduce the mesh sensitivity, $h$ is set by the averaging size in the numerical scheme. The RVE size $l_{RVE}$ is the edge length of the square unit cell. The microscale localization size $l_{c}$ is the homogenization length scale of a localization band in the microscale model. For example, if the periodic boundary conditions are applied on the RVE model, $l_c$ is equal to the vertical distance between the repeating cracks. Therefore, depending on the modeling approach and the localization configuration, $l_c$ is not necessarily equal to the RVE size $l_{RVE}$.
	
	Here, we summarize three main issues of RVE analysis that limits its applications to large-scale multiscale simulations with localization:
	\begin{enumerate}
		\item The optimum RVE size changes with macroscale length parameter $h$ and localization configuration (e.g., crack orientation); otherwise, proper energy regularization must be introduced for a predefined RVE size. 
		\item The boundary conditions on a rectangular RVE model are not well defined due to the change of periodicity induced by localization.
		\item Solving the RVE with local material damage or softening can be time-consuming. 
	\end{enumerate}
	
	For energy consistency across scales, the localization size $l_{c}$ should be equal to the macroscale length parameter $h$. Regarding the crack configuration shown in Figure \ref{fig:crackissue} (a), $l_{c}$ is equal to the RVE size. As a result, the optimum RVE size $l_{RVE}$ is set by the macroscale length parameter $h$.
	However, in a real-world multiscale simulation, the element size of the macroscale model is usually much larger than the microstructural characteristic length, $h \gg l_{RVE}$. Imposing $l_{RVE} = h$ would make the RVE model too expensive to be solved, thus defeating the purpose of multiscale simulations. Meanwhile, a single RVE with the predefined size is usually not sufficient because the optimum $l_{RVE}$ may change with the element size in the macroscale model or even the crack orientation (see Figure \ref{fig:crackissue} (b)), which is unknown before running the simulation. 
	
	On the other hand, applying proper boundary conditions on a rectangular RVE model is questionable as the localization surface (or band in 2-D space) breaks the periodicity of RVE in the normal direction to the surface. 
	%For this reason, the existence of RVE under localization is often argued in the literature. As the scales are no longer separated, the homogenized stress-strain responses depend on the RVE size $l_{RVE} $, which is contradictory to the original definition of an RVE \cite{hill1965self}. 
	\textcolor{black}{Boundary conditions commonly adopted for RVEs without the loss of ellipticity would result in non-physical localization configurations.} As shown in Figure \ref{fig:crackissue} (b),  the crack is distorted numerically to satisfy the periodic boundary conditions on the RVE, while physically, it should propagate straightly in $30^\circ$.  \textcolor{black}{To resolve this issue, one needs to adapt the aspect ratio, rotate the rectangular RVE domain \cite{hirschberger2009computational}, or even change the periodicity definitions \cite{bosco2014multiscale} according to the localization orientation.} Nonetheless, similar to the choice of RVE size, this becomes a ``Chicken-and-Egg" problem since the localization orientation is oftentimes unknown before analyzing the RVE.
	
	Last but not least, the appearance of material softening puts extra burdens on the numerical solver, while RVE models based on DNS can be rather time-consuming even for well-posed problems. If one intends to solve the RVE problem implicitly, numerical treatments like viscous regularization are required to make local tangent stiffness matrices positive-definite. In general, it demands more iterations for solving the RVE problem in the course of localization, so that improving the model efficiency becomes more prominent. 
	
	\subsection{Design of the paper}
	
	We will introduce a cell division scheme for consistent scale transition in the DMN framework to address the aforementioned issues in strain localization modeling. New algorithms of crack activation and evolution will be developed for the implicit failure analysis. By coupling DMNs with macroscale finite element models, we will demonstrate the applications of concurrent multiscale simulations to a broad range of material systems. 
	
	The remainder of this paper is organized as follows. In Section \ref{sec:scale}, we propose the idea of cell division for tracking the length scales inside a DMN. Analytical solutions of the two-layer building block are derived under the scenario of dividing an ellipsoidal (or elliptic in 2-D space) cell. The cell-division results on the binary-tree networks are demonstrated for two representative microstructures. Section \ref{sec:failure} focuses on DMN-related algorithms for failure analysis, including the cohesive law with viscous regularization, activations of potential crack surfaces, and the overall implicit solution scheme. Section \ref{sec:examples} provides examples ranging from single material point studies to concurrent multiscale simulations of nonlinear anisotropic composites. Advantages and limitations are discussed in Section \ref{sec:ad}. Conclusions and future work are summarized in Section \ref{sec:conclusion}.
	
	\section{Scale transition in DMN: Concept of cell division} \label{sec:scale}
	\subsection{Preliminary}
	Deep Material Network (DMN) is a mechanistic machine learning method for modeling materials across different scales. Its key features include the physics-based building block with interpretable fitting parameters, extrapolation capability for material and geometric nonlinearities, and efficient inference with a small number of DOF. The data-driven framework of DMN starts from the offline stage, where linear elastic DNS are performed to generate the training data. Gradient-based optimization algorithms are then used to find the optimum fitting parameters. In the online prediction stage, the network with trained fitting parameters can be extrapolated to consider nonlinear material responses based on the mechanistic building block's analytical solutions. A more detailed description of the data-driven framework is provided in \ref{ap:a1}. 
	
	DMN has been applied to several representative multiscale material systems, including particle-reinforced rubber composites, short or continuous fiber-reinforced composites, and polycrystalline materials. Formulated based on the network structure, DMN takes a different route of describing multiscale material behaviors from RVE analysis, which solves a boundary value problem of partial differential equations. Importantly, it offers the potentials of overcoming the aforementioned issues in RVE-based methods (see Section \ref{sec:issues}): the analytical solutions of DMN building block avoid the difficulties of defining proper boundary conditions on a rectangular unit cell, and the reduced-order DMN model is more efficient than a full-field RVE analysis. However, to fully tackle the multiscale localization problem, a consistent scale transition scheme still needs to be developed within the framework.
	
	This motivates us to introduce the concept of ``cell division". Essentially, each node in the network is associated with a so-called ``cell" to encode the corresponding node's length scales. The cell of the top node of DMN represents macroscale material point, so that its dimensions are defined by the macroscale length parameters $h$. \textcolor{black}{Cells are divided backward following the geometric structure of each building block in the network. Eventually, the length scales of microscale DOFs are encoded in the micro-cells at the bottom layer. One can regard the micro-cell as an equivalent microscale material point. When a localization band develops through the material point, its physical dimensions are set by the corresponding micro-cell.} 

	To simplify the mathematical formulation of the cell division process, we assume each cell to be an ellipsoid in 3-D space, or an ellipse in 2-D space. The implicit equation of an ellipsoidal cell can be expressed in a 3-D Cartesian coordinate system $(x,y,z)$ as
	\begin{equation}
	\begin{bmatrix}x&y&z\\\end{bmatrix}\begin{bmatrix}
	{A}_{11}&{A}_{12}&{A}_{13}\\
	&{A}_{22}&{A}_{23}\\
	sym&&{A}_{33}\\
	\end{bmatrix}
	\begin{bmatrix}
	x\\y\\z\\
	\end{bmatrix} = 1,
	\end{equation}
	where we refer $\textbf{A}$ as the ``scale tensor" in the paper. The second-order tensor $\textbf{A}$ is positively definite and symmetric. Specifically, if a node possesses a spherical domain with diameter $h$, its scale tensor can be written as
	\begin{equation}
	\textbf{A} = \begin{bmatrix}
	4/h^2&0&0\\
	&4/h^2&0\\
	sym&&4/h^2\\
	\end{bmatrix}.
	\end{equation}
	
	\begin{figure}[!t]
		\centering
		\graphicspath{{Figures/}}
		\includegraphics[clip=true,trim = 0.2cm 2.0cm 0.5cm 2.0cm, width = 1.00\textwidth]{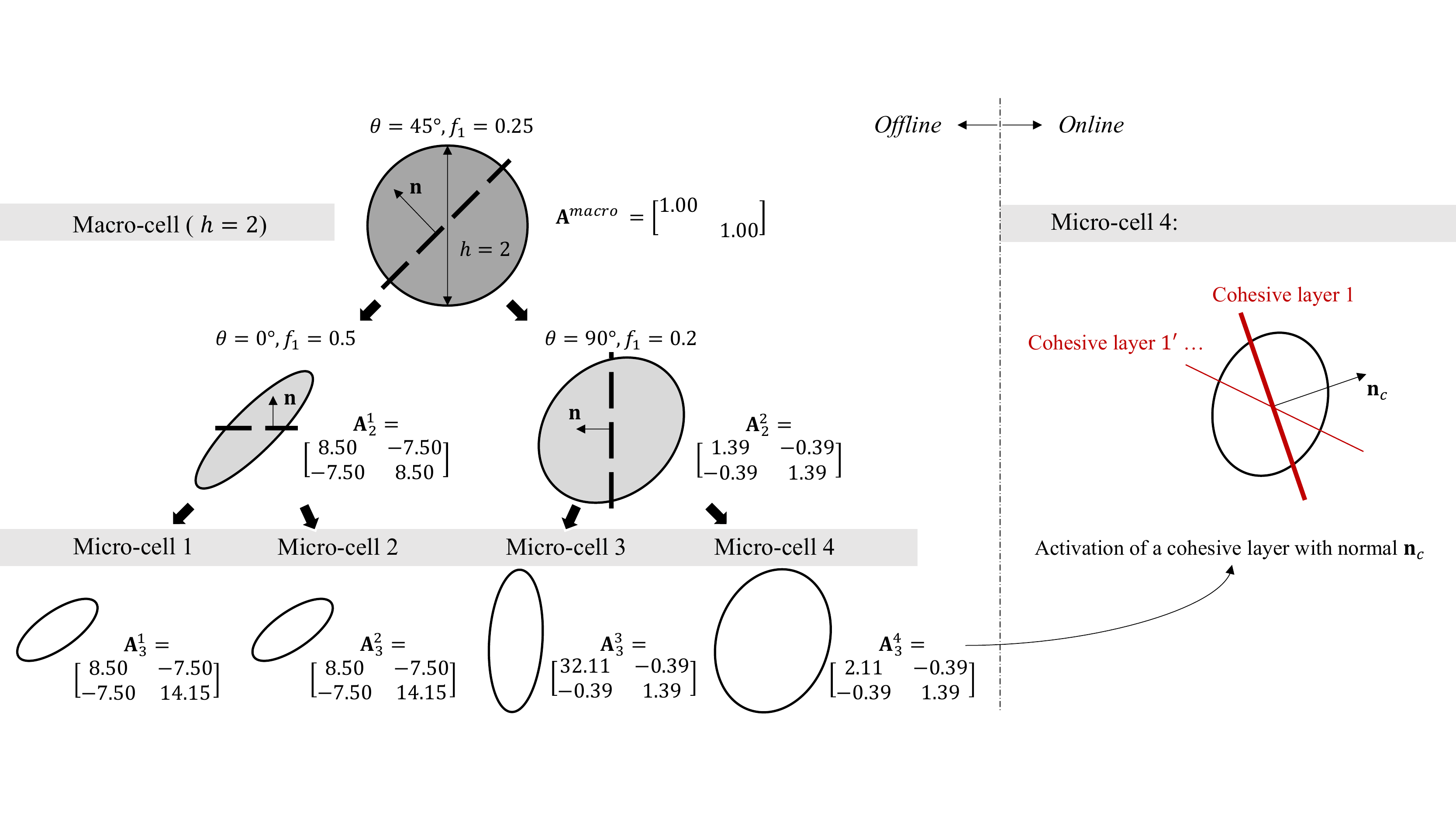}
		\caption{Illustration of the cell division process for scale transition in 2-D space. The macroscale cell is represented by an ellipse defined by the ``scale tensor" $\textbf{A}^{macro}$, which is related to the macro scale parameter $h$. \textcolor{black}{In the offline back-propagation process, each cell is divided based on the interface orientation $\theta$ (or the normal $\textbf{n}$) and the first child node's volume fraction $f_1$ of the trained two-layer building block.} The nodes in the bottom layer are regarded as the DOF of DMN, whose length scales are defined by the micro-cells. \textcolor{black}{In the online stage, the cohesive layer can be activated within a micro-cell to model the failure, while its normal $\textbf{n}_c$ changes with the micro-cell's deformation state during the online simulation.}}
		\label{fig:cellDivision}
	\end{figure}
	
	\textcolor{black}{Figure \ref{fig:cellDivision} illustrates the cell division process on the scale tensors for a 2-D DMN with three network layers ($N=3$).} The top node, which represents the macroscale material point, has a scale tensor $\textbf{A}^{macro}$, which is commonly determined by the element size or the nonlocal averaging size in the macroscale model. \textcolor{black}{We divide the mother node's cell into the two child nodes' cells according to the interface's orientation and the phase fraction of each building block, which are determined from the offline training.}  As information of the scale tensor propagates from the top layer to the bottom layer, $\textbf{A}^{macro}\rightarrow \textbf{A}_2^1, \textbf{A}_2^2 \rightarrow \textbf{A}_3^1, \textbf{A}_3^2,\textbf{A}_3^3, \textbf{A}_3^4$, the scales of the micro-cells can be properly tracked.  \textcolor{black}{The cell division's mathematical formulation in a generic two-layer building block and the overall scheme will be presented in Section \ref{sec:division}.}
	
	\textcolor{black}{In the online extrapolation stage, crack planes can be activated in each micro-cell to model microscale material failure, as shown for ``Micro-cell 4" in Figure \ref{fig:cellDivision}.} Specifically, we model a crack plane by the cohesive-layer approach. The cohesive layer's orientation is determined by the base material's current deformation state and the associated failure criteria. Once a cohesive layer is activated and attached to the DMN node, its constitutive behaviors will be described by a traction-separation law. The scales of the micro-cell come into play when one needs to determine the``reciprocal length scale parameter" \cite{liu2020deep}, ${v}$, which can be interpreted as the inverse of the effective thickness of the base material normal to the cohesive layer. Physically, it characterizes the contribution of the separation displacement to the overall strain of the micro-cell. We will show how to derive the expression of ${v}$ as a function of the micro-cell's scale tensor $\textbf{A}_N^j$ and the normal of crack plane $\textbf{n}_c$ in Section \ref{sec:activation}.
	
	\begin{remark}
		In our previous work for materials with predefined deformable interfaces \cite{liu2020deep}, the orientations and reciprocal length parameters of the cohesive layers are the fitting parameters optimized based on DNS training data, and they will be kept constant in the prediction stage after training. By contrast, the cohesive layers in this work are utilized to model the crack planes whose orientations and length scales depend on the base material's current deformation state during the online analysis. Therefore, these geometric parameters are not pre-determined from offline training.
	\end{remark}
	
	Once the rotation angles and the reciprocal length parameters of the cohesive layers are determined, the enriched DMN can be solved through implicit analysis.  At each iteration of Newton's method, the stiffness tensor $\textbf{C}_{N}$ and residual strain $\delta\boldsymbol{\sigma}_N$ of an enriched bottom-layer node/cell can be computed by
	\begin{equation}\label{eq:Dcoh}
	\textbf{C}_{N}= \left[\textbf{D}^{base} + \sum_{p=1}^{N_c}{v}_c^p\textbf{R}_c^p\tilde{\textbf{G}}^p(\textbf{R}_c^p)^{-1}\right]^{-1}
	\end{equation}
	and
	\begin{equation}\label{eq:decoh}
	\delta\boldsymbol{\sigma}_N =  -\textbf{C}_{N}\left[\delta\boldsymbol{\varepsilon}^{base} + \sum_{p=1}^{N_c}{v}_c^p\textbf{R}_c^p\delta\tilde{\textbf{d}}^p\right],
	\end{equation}
	where the compliance tensor $\textbf{D}^{base}$ and the residual strain $\delta\boldsymbol{\varepsilon}^{base}$ are obtained by evaluating the constitutive model of the base material. $\tilde{\textbf{G}}^p$ and $\tilde{\textbf{d}}^p$ are the compliance matrix and residual displacement vector come from the traction-separation law of the $p$-th cohesive layer.  Note that these residual quantities appear only when the system experiences geometric or material nonlinearities, such as plasticity in the base material and softening in the cohesive layer. While this work will mainly focus on small-strain formulation, a short discussion on the finite-strain formulation with geometric nonlinearity is provided in \ref{ap:a2}. Moreover, $N_c$ denotes the total number of enriching cohesive layers in the base material. If the cohesive layer is assumed to behave isotropically in the crack plane (see Section \ref{sec:coh}), the rotation matrix $\textbf{R}^p_c$ for the $p$-th cohesive layer can be solely determined from its normal vector $\textbf{n}_c^p$.
	
	The information of stiffness tensors and residual stresses at the bottom layer is then propagated through the network all the way to the top node, which stores the macroscopic quantities: $\textbf{C}^{macro}$ and $\delta\boldsymbol{\sigma}^{macro}$. Afterward, the macroscale boundary conditions are analyzed at the top node to fill any undefined macroscopic stress or strain components. For example, in a 3-D concurrent multiscale simulation, the top node receives all six strain components of a material/integration point $\Delta\boldsymbol{\varepsilon}^{macro}$ in the macroscale FE model, and the missing stress components can be obtained as
	\begin{equation}
	\Delta \boldsymbol{\sigma}^{macro} =\textbf{C}^{macro}\Delta\boldsymbol{\varepsilon}^{macro} + \delta \boldsymbol{\sigma}^{macro}.
	\end{equation}
	After the stress and strain information are propagated back to the base material and enriched cohesive layers of each micro-cell at the bottom layer, the convergence is checked by comparing the updated microscale incremental strains and displacement vectors with ones from the last iteration. Outside the loop for Newton's method, we also need algorithms to track the crack configuration and evaluate its convergence accordingly. The complete description of the solution scheme will be provided in Section \ref{sec:scheme}. 
	
	\subsection{Division in two-layer building block}\label{sec:division}
	In a generic two-layer building block, we denote the scale tensor of the mother node as $\textbf{A}^0$, and ones of the child nodes as $\textbf{A}^1$ and $\textbf{A}^2$. The unit normal to the two child materials' interface is $\textbf{n}$. The volume fraction of the first child node is $f_1$, and the other child node has $f_2 = 1-f_1$. For a given ellipsoidal cell with a shape tensor $\textbf{A}$, we define the cutting surface, $\Lambda(\textbf{A},\textbf{n})$, as the intersection between the ellipsoid and a cutting plane, which has a normal $\textbf{n}$ and passes through the center of the ellipsoid. Meanwhile, $V(\textbf{A})$ is the volume of the ellipsoid. 
	
	In our formulation, three conditions need to be satisfied during the cell division process:
	\begin{enumerate}
		\item As centered at the origin, child cells are always within the mother cell.
		\item The mother and child cells share identical cutting surfaces with $\textbf{n}$ normal to the building block's interface,
		\begin{equation}\label{eq:cond1}
		\Lambda(\textbf{A}^0,\textbf{n})\cong\Lambda(\textbf{A}^1,\textbf{n})\cong\Lambda(\textbf{A}^2,\textbf{n}).
		\end{equation}
		\item Consistency of volume fractions:
		\begin{equation}\label{eq:cond2}
		f_1 = V(\textbf{A}^1)/V(\textbf{A}^0),\quad \text{with }V(\textbf{A}^0)=V(\textbf{A}^1)+V(\textbf{A}^2).
		\end{equation}
	\end{enumerate}
	As we will show later in this section, the simplicity of the two-layer building block not only enables analytical homogenization of the mechanical quantities \cite{liu2019deep,liu2019exploring} but also allows us to derive analytical functions of the cell-division process for scale transition under the above conditions.
	
\begin{figure}[!t]
		\centering
		\graphicspath{{Figures/}}
		\subfigure[][$\textbf{A}^0 = \begin{bmatrix}8.5&-7.5\\-7.5&8.5\end{bmatrix}$.]{\includegraphics[clip=true,trim = 1.0cm 0cm 1.7cm 0cm,width=0.212\textwidth]{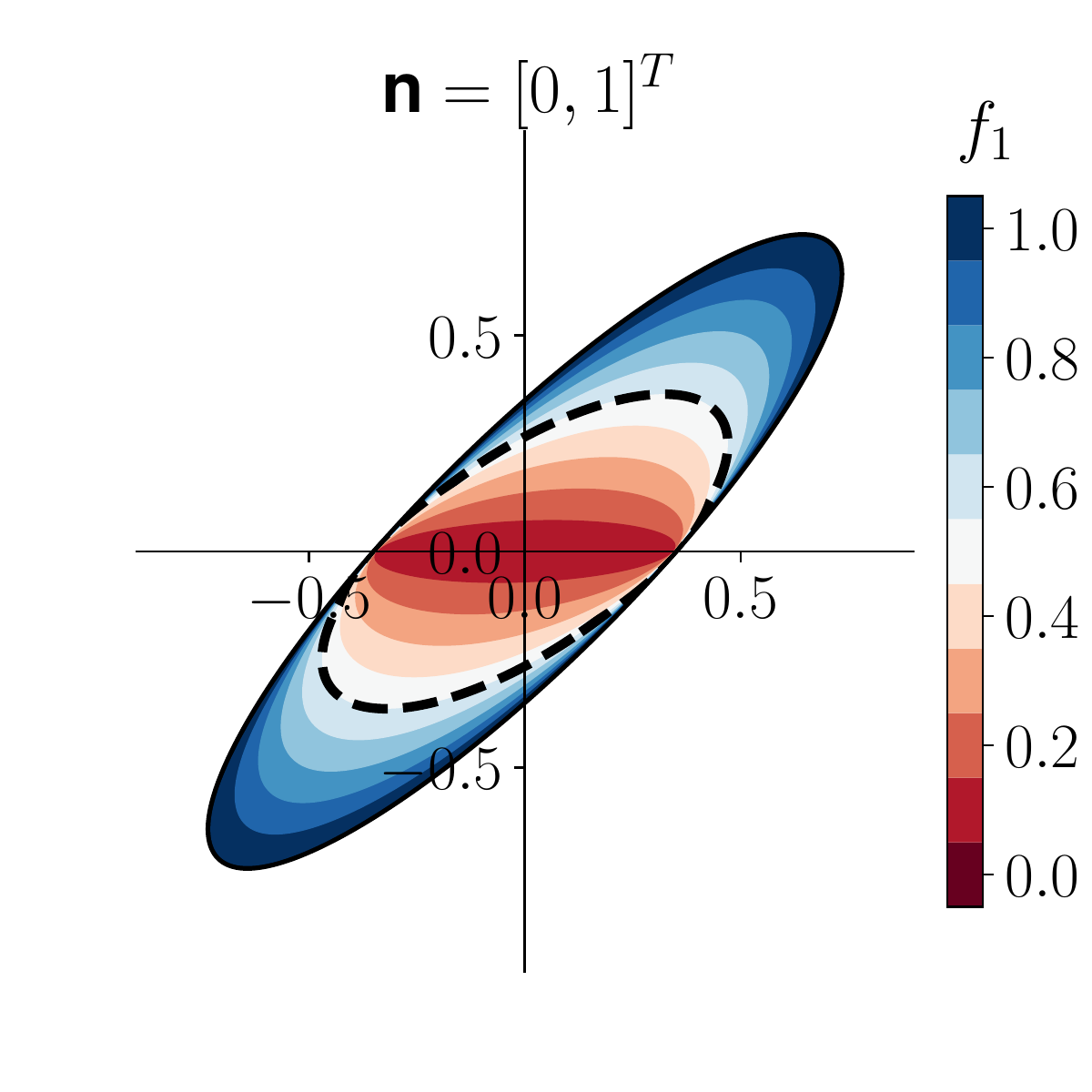} \includegraphics[clip=true,trim = 1.0cm 0cm 0cm 0cm,width=0.25\textwidth]{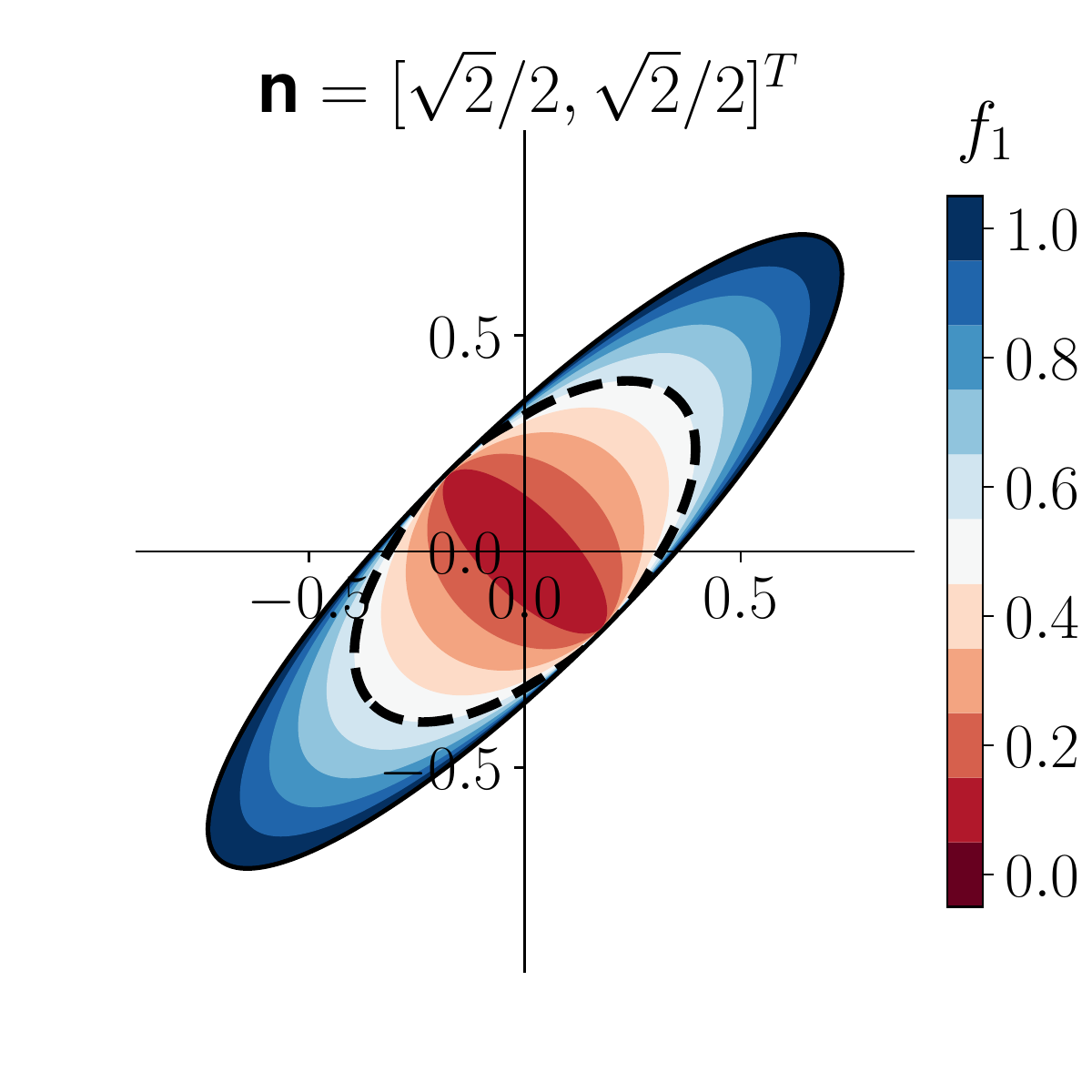}}
		\quad
		\subfigure[][$\textbf{A}^0 = \begin{bmatrix}1.39&-0.39\\-0.39&1.39\end{bmatrix}$.]{\includegraphics[clip=true,trim = 1.0cm 0cm 1.7cm 0cm,width=0.212\textwidth]{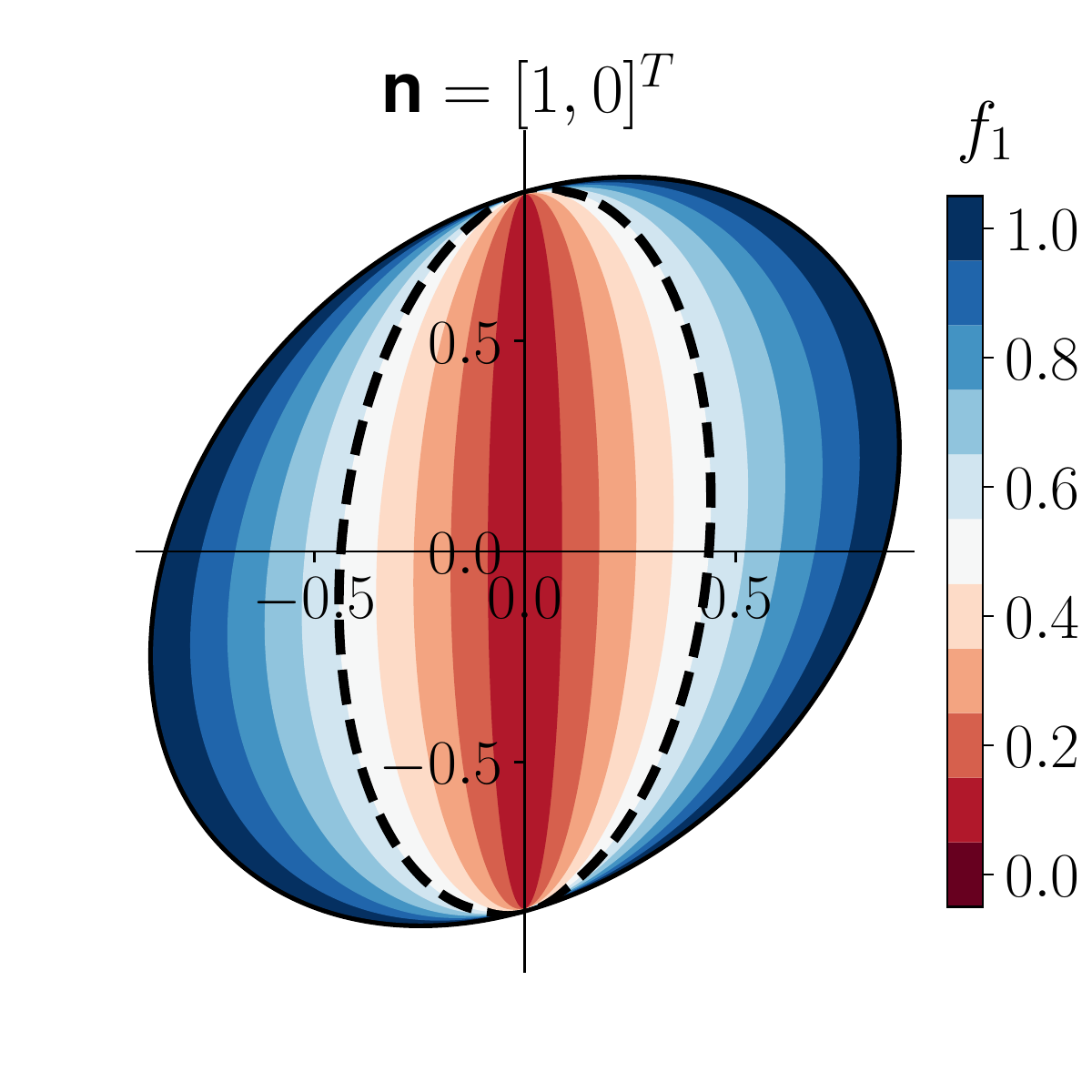} \includegraphics[clip=true,trim = 1.0cm 0cm 0cm 0cm,width=0.25\textwidth]{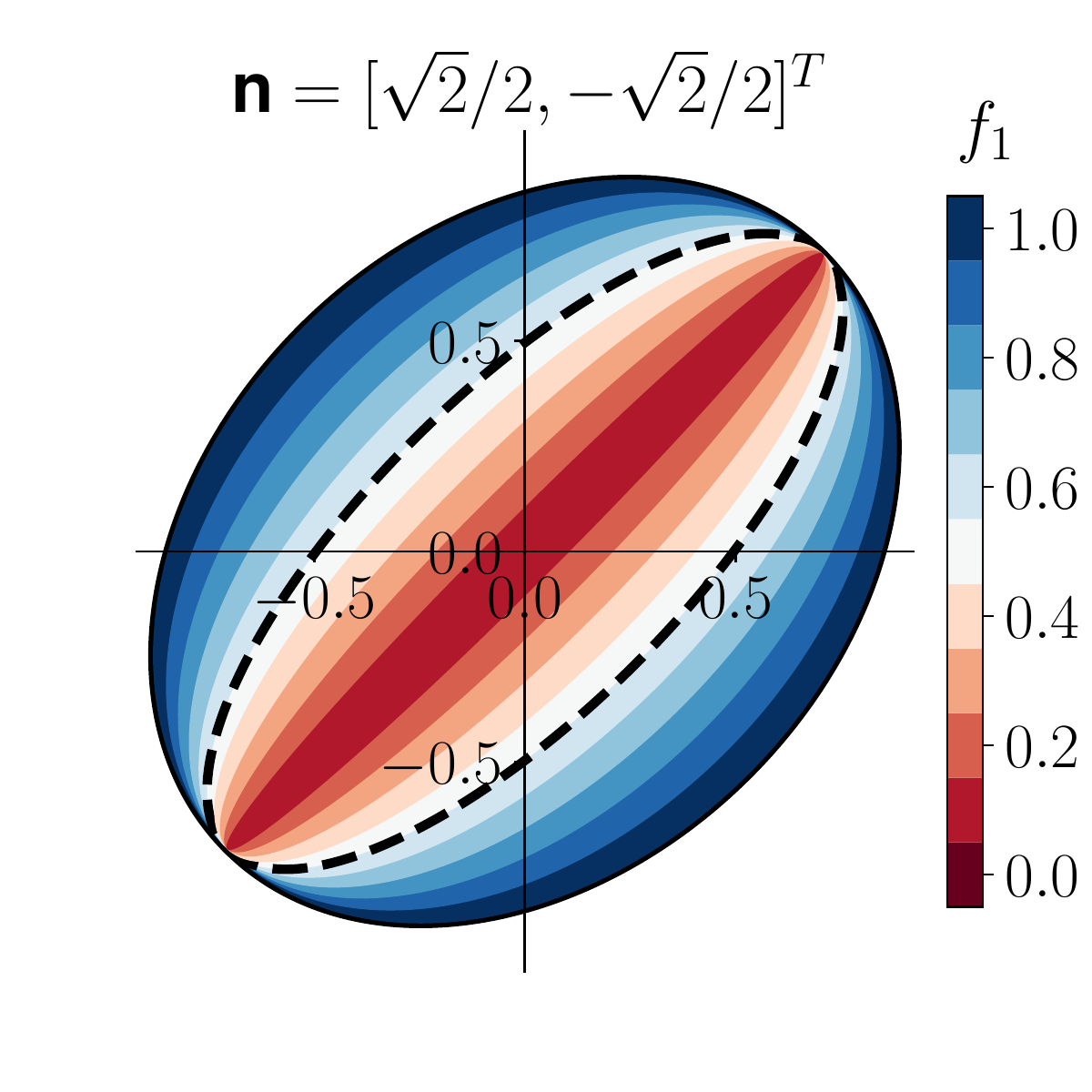}}
		\caption{Divisions of the mother node in a 2-D building block for different combinations of the scale tensors $\textbf{A}^0$ and the normal of the cutting plane $\textbf{n}$. Specifically, $\textbf{A}^0$s are picked from the two nodes at the second network layer in Figure \ref{fig:cellDivision}. The cell of the first child node with $f_1=0.5$ (or an equal cut) is highlighted by the dashed line in each plot.}
		\label{fig:split}
	\end{figure}

	Figure \ref{fig:split} presents the cell-division results under four sets of $\textbf{A}^0$  and $\textbf{n}$ in 2-D space. For each case, the elliptic cells of the first child node are plotted for $f_1$ varying from 0.0 to 1.0, with color ranging from red to blue. As $f_1$ approaches 0, the cell of the first child node essentially falls onto the cutting surface and has an infinitesimal thickness in the normal direction. One can also see that the child cells are always contained inside their mother cell, which is ensured by the first division condition.
	
	\begin{remark}
		The cell division for scale transition is analogical to the de-homogenization process for back-propagating physical quantities, such as stress and strain tensors. \textcolor{black}{In the DMN computation, it is sufficient to set the normal $\textbf{n}$ of cutting plane in the cell-division model to be $\{0,0,1\}^T$ in 3-D or $\{0,1\}^T$ in 2-D space, and the solutions can be generalized to an arbitrary normal direction through a rotation operation}.  In Section \ref{sec:activation}, the approach will be reused to determine the reciprocal length parameter of a crack surface.
	\end{remark}
	
	Driving the formulations of $\textbf{A}^1$ and $\textbf{A}^2$ for $\textbf{A}^0$ of an arbitrary ellipsoidal cell is not trivial. However, the process becomes much more straightforward for a unit sphere, because two principal axes of its child ellipsoid are in the cutting plane, while the third axis is normal to the plane. Based on this observation, we propose the following procedure for the derivation:
	\begin{enumerate}
		\item Rotate and scale the mother ellipsoid with $\textbf{A}^0$ to a unit sphere with $\hat{\textbf{A}} ^0=\textbf{I}$. The transformation also applies to the cutting plane, resulting in a new normal $\hat{\textbf{n}}$.
		\item In the transformed configuration, cut the unit sphere based on $\hat{\textbf{n}}$. Compute the scale tensors $\hat{\textbf{A}} ^1$ and $\hat{\textbf{A}}^2$.
		\item Transform $\hat{\textbf{A}} ^1$ and $\hat{\textbf{A}} ^2$ back to $\textbf{A}^1$ and $\textbf{A}^2$ in the original configuration.
	\end{enumerate}
	
	We will derive the expressions of $\textbf{A}^1$ and $\textbf{A}^2$ for the 3-D building block, while the same procedure can also be applied to the 2-D building block \cite{liu2019deep}. Through the eigenvalue analysis of $\textbf{A}^0$, we can find a rotation matrix $\boldsymbol{{\mathcal{R}}}$ which aligns the principal axes of the rotated ellipsoid $\textbf{A}'$ with the coordinate axes,
	\begin{equation}\label{eq:R}
	\textbf{A}'= \begin{bmatrix}
	\lambda_1&&\\
	&\lambda_2&\\
	&&\lambda_3\\
	\end{bmatrix} = \boldsymbol{{\mathcal{R}}} ^T \textbf{A}^0 \boldsymbol{{\mathcal{R}}},
	\end{equation}
	where $\lambda_1$, $\lambda_2$, and $\lambda_3$ are the eigenvalues of $\textbf{A}^0$. For the rotated ellipsoid, the square root of $\lambda_i$ is equal to the inverse of the length of its $i$-th semi-axis. Therefore, another scaling operation is prescribed on the rotated ellipsoid to transform it to a unit sphere, and the scaling matrix is 
	\begin{equation}\label{eq:T}
	\boldsymbol{{\mathcal{T}}} = \begin{bmatrix}
	1/\sqrt{\lambda_1}&&\\
	&1/\sqrt{\lambda_2}&\\
	&&1/\sqrt{\lambda_3}\\
	\end{bmatrix}\quad
	\text{s.t. } \hat{\textbf{A}}^0=\boldsymbol{{\mathcal{T}}}\boldsymbol{{\mathcal{R}}} ^T \textbf{A}^0 \boldsymbol{{\mathcal{R}}}\boldsymbol{{\mathcal{T}}} = \textbf{I}.
	\end{equation}
	
	Let $\textbf{n}^s$ and $\textbf{n}^t$ denote two orthogonal vectors in the original cutting plane, where we have $\textbf{n}^s \times \textbf{n}^t=\textbf{n}$. After the transformation, the new vectors in the plane are
	\begin{equation}
	\hat{\textbf{n}}^s = \boldsymbol{{\mathcal{T}}}^{-1}\boldsymbol{{\mathcal{R}}} ^T\textbf{n}^s, \quad \hat{\textbf{n}}^t = \boldsymbol{{\mathcal{T}}}^{-1}\boldsymbol{{\mathcal{R}}} ^T\textbf{n}^t.
	\end{equation}
	The new normal to the transformed cutting plane is
	\begin{equation}\label{eq:normal}
	\hat{\textbf{n}} = \dfrac{\hat{\textbf{n}}^s\times\hat{\textbf{n}}^t}{|\hat{\textbf{n}}^s\times\hat{\textbf{n}}^t|} =\dfrac{\boldsymbol{{\mathcal{T}}}\boldsymbol{{\mathcal{R}}}^T\textbf{n}}{|\boldsymbol{{\mathcal{T}}}\boldsymbol{{\mathcal{R}}}^T\textbf{n}|}. 
	\end{equation}
	
	After the unit sphere is cut by the plane with normal $\hat{\textbf{n}}$, the scale tensors of the two child nodes are obtained as
	\begin{equation}\label{eq:unitcut}
	\hat{\textbf{A}}^1 = \textbf{I}-\left(1-\dfrac{1}{(f_1)^2}\right)\hat{\textbf{n}}\otimes\hat{\textbf{n}},\quad
	\hat{\textbf{A}}^2 = \textbf{I}-\left(1-\dfrac{1}{(1-f_1)^2}\right)\hat{\textbf{n}}\otimes\hat{\textbf{n}},
	\end{equation}
	which satisfy all three division conditions in the transformed configuration. 
	
	The inverse transformations of $\hat{\textbf{A}}^1$ and $\hat{\textbf{A}}^2$ return the expressions for $\textbf{A}^1$ and $\textbf{A}^2$ in the original configuration,
	\begin{equation}\label{eq:A1A2}
	\textbf{A}^1 = \boldsymbol{{\mathcal{R}}}\boldsymbol{{\mathcal{T}}}^{-1}\hat{\textbf{A}}^1\boldsymbol{{\mathcal{T}}}^{-1}\boldsymbol{{\mathcal{R}}}
	^T,\quad
	\textbf{A}^2 = \boldsymbol{{\mathcal{R}}}\boldsymbol{{\mathcal{T}}}^{-1}\hat{\textbf{A}}^2\boldsymbol{{\mathcal{T}}}^{-1}\boldsymbol{{\mathcal{R}}}
	^T.
	\end{equation}
	By substituting terms in Eq. (\ref{eq:A1A2}) with Eq. (\ref{eq:normal}) and (\ref{eq:unitcut}), we arrive at
	\begin{equation}\label{eq:A1}
	\textbf{A}^1 = \boldsymbol{\mathcal{A}}^1(\textbf{A}^0,f_1,\textbf{n}) = \textbf{A}^0-\left(1-\dfrac{1}{(f_1)^2}\right)\dfrac{\textbf{n}\otimes\textbf{n}}{|\boldsymbol{{\mathcal{T}}}\boldsymbol{{\mathcal{R}}}^T\textbf{n}|^2}
	\end{equation}
	and
	\begin{equation}\label{eq:A2}
	\textbf{A}^2 = \boldsymbol{\mathcal{A}}^2(\textbf{A}^0,f_1,\textbf{n}) =  \textbf{A}^0-\left(1-\dfrac{1}{(1-f_1)^2}\right)\dfrac{\textbf{n}\otimes\textbf{n}}{|\boldsymbol{{\mathcal{T}}}\boldsymbol{{\mathcal{R}}}^T\textbf{n}|^2}.
	\end{equation}
	Since both of the inverse rotation and scaling operations are affine transformations, $\textbf{A}^1$ and $\textbf{A}^2$ also satisfy all three division conditions in the original configuration.
	
	Meanwhile, the area of $\Lambda(\textbf{A}^0,\textbf{n})$ can be calculated as
	\begin{equation}\label{eq:Area}
	S(\textbf{A}^0,\textbf{n}) = \dfrac{\pi}{|\hat{\textbf{n}}^s \times \hat{\textbf{n}}^t|} = \dfrac{\pi}{\sqrt{\det{\textbf{A}^0}}|\boldsymbol{{\mathcal{T}}}\boldsymbol{{\mathcal{R}}}^T\textbf{n}|}.
	\end{equation}
	Implied by the second division condition in Eq \ref{eq:cond1},
	\begin{equation}
	S(\textbf{A}^0,\textbf{n}) = S(\textbf{A}^1,\textbf{n}) = S(\textbf{A}^2,\textbf{n}).
	\end{equation}
	
	\color{black}
	\bigskip
	\noindent\fbox{\begin{minipage}{46em}\label{mp:3}
			\medskip
			\centering\textbf{Box 2.2.1 Cell-division scheme of DMN in global coordinate system}		
			\begin{enumerate}[itemsep=0mm]
				\item Compute the volume fractions $(f_1)_{i=1,...,N-1}^{k=1,2,...,2^{i-1}}$ for all the building blocks based on Eq. (\ref{eq:f1})
				\item Initialization at the top node: $\boldsymbol{\mathcal{O}}_1^1\leftarrow\textbf{I}$, $\textbf{A}_1^1\leftarrow\textbf{A}^{macro}$
				\item Backward propagation of scale tensors:\newline
				\begin{algorithm}[H]
					\For{$(i = 1;\ i < N;\ i = i + 1)$}{
						\For{$(k = 1;\ k < 2^{i-1}+1;\ k = k + 1)$}{
							\begin{enumerate}[itemsep=0mm]
								\item Compute local rotation matrix $\textbf{R}_i^k(\alpha_i^k,\beta_i^k,\gamma_i^k)$
								\item Update global orientation matrix: $\boldsymbol{\mathcal{O}}_i^k\leftarrow (\textbf{R}_i^k)\boldsymbol{\mathcal{O}}_i^k$
								\item Interface's normal vector in global coordinate system: $\textbf{n}_i^k=(\boldsymbol{\mathcal{O}}_i^k)^T\textbf{n}_0$
								\item Cell division in the building block based on Eq. (\ref{eq:A1}) and (\ref{eq:A2}):
								$
								\textbf{A}_{i+1}^{2k-1} = \boldsymbol{\mathcal{A}}^1\left(\textbf{A}_i^k,(f_1)_i^k,\textbf{n}_i^k\right), \quad
								\textbf{A}_{i+1}^{2k} = \boldsymbol{\mathcal{A}}^2\left(\textbf{A}_i^k,(f_1)_i^k,\textbf{n}_i^k\right)
								$
								\item Update child nodes' global orientation matrix:  $\boldsymbol{\mathcal{O}}_{i+1}^{2k-1}\leftarrow\boldsymbol{\mathcal{O}}_i^k$, $\boldsymbol{\mathcal{O}}_{i+1}^{2k}\leftarrow\boldsymbol{\mathcal{O}}_i^k$
							\end{enumerate}
						}
					}
				\end{algorithm} 
			\end{enumerate}
			\par
	\end{minipage}}
	\bigskip
	
	To summarize, we describe the cell-division scheme for a 3-D material network with depth N in Box \hyperref[mp:3]{2.2.1}. The fitting parameters are the activations $z^{j=1,2,...,2^{N-1}}$ in the bottom layer, and rotation angles ($\alpha_{i=1,...,N}^{k=1,2,...,2^{i-1}}$,$\beta_{i=1,...,N}^{k=1,2,...,2^{i-1}}$,$\gamma_{i=1,...,N}^{k=1,2,...,2^{i-1}}$) of all the active nodes. Physically, $\max(0,z^j)$ (or the ReLU activation function) returns the weight of $j$-th node in the bottom layer. Through forward propagation, the weights of the $k$-th node in the $i$-th layer can be obtained as
	\begin{equation}
	w^k_i = \sum_{j=2^{N-i}(k-1)+1}^{2^{N-i}k} \max(0,z^j).
	\end{equation}
	The volume fraction of the first child node for the $k$-th building block in the $i$-th layer is
	\begin{equation}\label{eq:f1}
	(f_1)_i^k =  w_{i+1}^{2k-1}/w_{i}^{k}.
	\end{equation}
	The interface's normal of the two-layer building block before the rotation operation is $\textbf{n}_0=\{0,0,1\}^T$. The rotation matrix of the building block for a 1-D vector is given by $\textbf{R}_i^k(\alpha_i^k,\beta_i^k,\gamma_i^k)$ \cite{liu2019exploring}. For demonstration purpose, we present all the scale tensors and normal vectors in the global coordinate system, and the matrix $\boldsymbol{\mathcal{O}}_i^k$ is introduced to track the global orientation of the corresponding building block. Note that the scheme in Box \hyperref[mp:3]{2.2.1} is presented for a perfect binary-tree network architecture. In practice, the binary tree will be compressed during the training \cite{liu2019deep,liu2019exploring}. Nevertheless, the main structure of the scheme will essentially stay the same.
	
	\color{black}
	
	\subsection{Two microstructures}
	This paper mainly focuses on two representative microstructures: a 3-D particle-reinforced composite and a 2-D composite with identical circular inclusions embedded in the matrix phase. For the 2-D composite, we will rely on the physical interpretations of DMN fitting parameters to directly transfer it to a 3-D unidirectional-fiber composite, so that no extra training is required for generating the 3-D model. For both cases, periodic boundary conditions are used in the RVE analysis based on finite element methods. 
	
	\begin{figure}[!t]
		\centering
		\graphicspath{{Figures/}}
		\includegraphics[clip=true,trim = 3cm 5cm 3cm 4cm,width=0.98\textwidth]{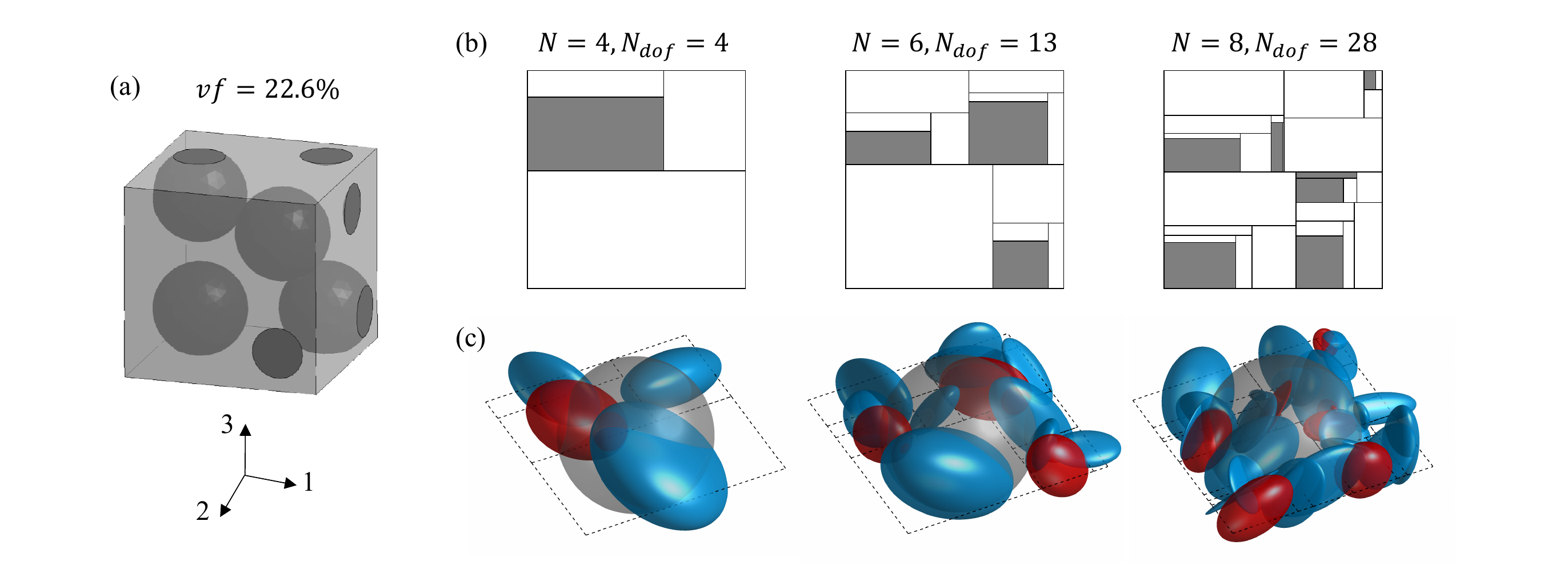}
		\caption{3-D particle-reinforced composite: (a) RVE geometry for finite element analysis, the volume fraction of the particle phase is 22.6\%; (b) Tree-map plots of trained DMNs with depth $N=4$, 6, and 8. The number of DOF, $N_{dof}$, is shown on top of each plot; (c) Micro-cells arranged by the corresponding treemap plot with $\textbf{A}^{macro}=\textbf{I}$. Micro-cells of the particle phase are colored by red,  and ones of the matrix phase blue. The macro-cell is plotted in gray at the center of each plot. }
		\label{fig:geo1}
	\end{figure}
	
	The geometry of the 3-D particle-reinforced composite is shown in Figure \ref{fig:geo1} (a), and there are four identical spherical particles embedded in the matrix. The volume fraction of the particle phase is 22.6\%. To improve the DNS training data's accuracy, we mesh the RVE by 10-node tetrahedron finite elements. The resulting DNS model has 84,693 nodes and 59,628 elements. Tree-map plots of the DMNs with different depth after training are given in Figure \ref{fig:geo1} (b). The number of DOF $N_{dof}$ is counted as the number of active nodes in the bottom layer. For $N=4$, 6, and 8, we have $N_{dof} = 4$, 13, and 28, respectively. More details about the data generation and DMN training process can be found in \ref{ap:a1} and our previous paper \cite{liu2019exploring}. The micro-cell of each DOF can be obtained from the cell division process, and we place each micro-cell at the center of its corresponding block in the tree-map plot as shown in Figure \ref{fig:geo1} (c). For each case, the macro-cell has a scale tensor $\textbf{A}^{macro}=\textbf{I}$, which appears as the gray sphere in the plot.
	
	\begin{figure}[!t]
		\centering
		\graphicspath{{Figures/}}
		\includegraphics[clip=true,trim = 3cm 4cm 2.5cm 3.5cm,width=0.98\textwidth]{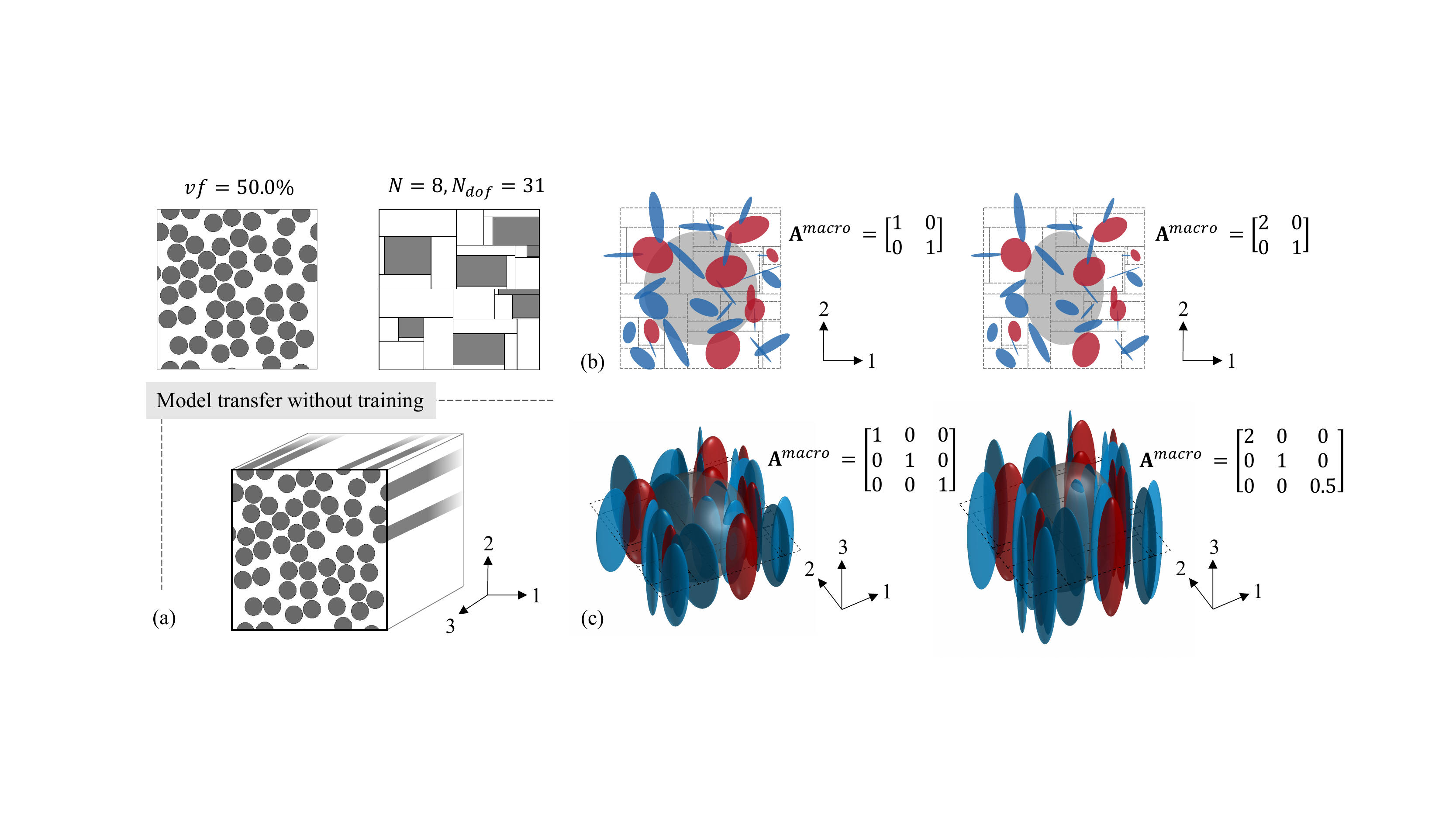}
		\caption{3-D unidirectional fiber composite from DMN model transfer. (a) The 2-D RVE with circular inclusions is modeled by finite elements, and the volume fraction of the inclusion phase is equal to 50.0\%. After training, the 2-D DMN with 8 layers has 31 DOF as shown in the treemap plot. The 3-D DMN models of the unidirectional fiber composite are created by transferring the fitting parameters without extra training. (b) 2-D micro-cell configurations for two different macro scale tensors. (c) 3-D micro-cell configurations for two different macro scale tensors after model transfer.}
		\label{fig:geo2}
	\end{figure}
	
	The 2-D composite has identical circular inclusion embedded in the matrix with the volume fraction equal to 50\%, as shown in Figure \ref{fig:geo2} (a). After discretization, the DNS model has 199,014 FE nodes and 198,212 4-node 2-D plane strain elements, which guarantees that there are at least three layers of elements between any two inclusions. The depth of DMN is set to 8, and there are 31 DOF after training. The 2-D configurations for two $\textbf{A}^{macro}$ tensors are provided in Figure \ref{fig:geo2} (b). The 3-D configurations of micro-cells after the model transfer are shown in Figure \ref{fig:geo2} (c).
	
	Before diving into the details of model transfer from 2-D to 3-D composites, let us first look at the fitting parameters of the 2-D DMN with depth $N$: activations $z^{j=1,2,...,2^{N-1}}_{(2D)}$ and rotation angles $\theta^{k=1,2,...,2^{i-1}}_{i=1,2,...,N}$. Physically, $\max(0,z^j_{2D})$ returns the weight of the $j$-th node at the bottom layer, and the angle $\theta^{k}_{i=1}$ controls the rotation or interface orientation of the $k$-th building block in the $i$-th layer. As the 3-D unidirectional fiber composite can be generated by stretching the 2-D model in the out-of-plane direction, there is no need to change the network topology in the plane. All the activations remain the same:
	\begin{equation} \label{eq:transfer}
	z^j_{(3D)} = z^j_{(2D)}, \quad \text{for } j = 1, 2,...,2^{N-1}.
	\end{equation} 
	For the 3-D rotation angles, we have
	\begin{equation}
	\alpha^k_i = \theta^k_i, \beta^k_i=0, \gamma^k_i=0\quad \text{for } i=2,3,..,N \text{ and } k = 1, 2,...,2^{i-1}.
	\end{equation}
	To align the fibers along the third axis as shown in Figure \ref{fig:geo2}(a) , we treat the rotation at the top node ($i=1$) differently:
	\begin{equation}
	\alpha^1_1 = \theta^1_1, \beta^1_1=\dfrac{\pi}{2}, \gamma^1_1=0.
	\end{equation}
	
	\begin{remark}
		It is possible to train the 3-D DMNs directly from a 3-D DNS model of the unidirectional fiber composite. In fact, we have demonstrated this approach in \cite{liu2019exploring}. However, in general, the 2-D DNS model takes less time for data generation, and the 2-D DMN also trains faster as it has fewer fitting parameters given the same depth. We realize it helpful to present this nice feature enabled by the physics-based building block of DMN, while a complete discussion of model transfer is beyond the scope of this paper.
	\end{remark}
	
	By the nature of model transfer in Eq. (\ref{eq:transfer}), the network configuration and weights have not been changed. Therefore, the 2-D and 3-D networks share the same treemap plots in Figure \ref{fig:geo2} (b). Meanwhile, since the cutting planes in the cell-division process are always parallel to the third axis, one can see from Figure \ref{fig:geo2} (c) that all the micro-cells have the same dimension in the fiber direction as the macro-cell.
	
	\section{Failure Algorithms}\label{sec:failure}
	\subsection{Cohesive law with viscous regularization}\label{sec:coh}
	As discussed in Section \ref{sec:division}, crack planes are activated in the micro-cells to simulate the failure behaviors.
	The crack plane is modeled by a cohesive-layer model based on one-dimensional effective traction-separation law \cite{camacho1996computational,ortiz1999finite}. The total opening displacement vector $\textbf{d}$ and the traction $\textbf{t}$ can be written as
	\begin{equation}\label{eq:td}
	\textbf{d} = d_n \textbf{n}_c + \textbf{d}_S, \quad \textbf{t} = t_n \textbf{n}_c + \textbf{t}_S,
	\end{equation}
	where $\textbf{n}_c$ is the unit normal to the crack plane. \textcolor{black}{Different from $\textbf{n}$ of the interface in the two-layer building block, $\textbf{n}_c$ is not determined from offline training.} Instead, as will be shown in Section \ref{sec:activation}, it depends on the deformation state of the corresponding micro-cell during the online analysis. An effective opening displacement $d_m$ is further introduced to simplify the mixed-mode cohesive law, and the free energy density per unit undeformed area $\phi$ becomes a function of $d_m$,
	\begin{equation}\label{eq:free}
	\phi = \phi\left(\textbf{d},\textbf{q}\right) = \phi\left(d_m,\textbf{q}\right),
	\end{equation}
	where $\textbf{q}$ are some internal variables for describing the irreversible processes and material states. Another simplification arises from the assumption that the cohesive law is isotropic in the crack plane. This indicates that the resistance to sliding is independent of the direction of sliding. and the effective displacement $d_m$ only depends on $d_n$ and the magnitude $|\textbf{d}_S|$. Furthermore, the effective traction can be written as
	\begin{equation}
	t_m = \dfrac{\partial \phi}{\partial d_m}(d_m,\textbf{q}), 
	\end{equation}
	and it will be used to formulate the crack initiation criteria for activating the cohesive layers.
	
	The compressive normal stress should not cause the cohesive layer to fail, so that it does not contribute to the free energy density. In addition, no friction effect has been included in our cohesive model. The tensile and compressive cases are considered separately as below,
	\begin{enumerate}[itemsep=0mm]
		\item For the tensile case $d_n\geq0$, the effective opening displacement $d_m$ is
		\begin{equation}\label{eq:t-tension}
		d_m = \sqrt{{d_n}^2 + \beta^2 |\textbf{d}_S|^2},%= \sqrt{\beta^2|\textbf{d}|^2 + (1-\beta^2)(\textbf{d}\cdot \textbf{n}_c)^2},
		\end{equation}
		where the positive parameter $\beta$ defines the ratio of effects from normal and shear displacements. The cohesive law becomes
		\begin{equation}
		\textbf{t} = \dfrac{\partial\phi}{\partial \textbf{d}} =\dfrac{\partial\phi}{\partial {d}_m}\dfrac{\partial{d}_m}{\partial \textbf{d}}= \dfrac{t_m}{d_m}\left(d_n\textbf{n}_c+\beta^2\textbf{d}_S\right).
		\end{equation}
		Based on Eq. (\ref{eq:td}) and (\ref{eq:t-tension}), the effective traction stress can be written as
		\begin{equation}\label{eq:tm1}
		t_m(\textbf{t},\textbf{n}_c) = \sqrt{{t_n}^2+\beta^{-2}|\textbf{t}_S|^2}.
		\end{equation}
		\item For the compressive case $d_n<0$, the effective opening displacement $d_m$ is
		\begin{equation}
		d_m = \beta |\textbf{d}_S|.%=\beta \sqrt{|\textbf{d}|^2 - (\textbf{d}\cdot \textbf{n}_c)^2},
		\end{equation}
		The cohesive law becomes
		\begin{equation}\label{eq:t-compress}
		\textbf{t} = t_n\textbf{n}+\dfrac{\partial\phi}{\partial \textbf{d}_S} =t_n\textbf{n}+\dfrac{\partial\phi}{\partial {d}_m}\dfrac{\partial{d}_m}{\partial \textbf{d}_S} = t_n\textbf{n}+ \dfrac{t_m}{d_m}\beta^{2}\textbf{d}_S.
		\end{equation}
		Based on Eq. (\ref{eq:td}) and (\ref{eq:t-compress}), the effective traction stress can be written as
		\begin{equation}\label{eq:tm2}
		t_m(\textbf{t},\textbf{n}_c) = \beta^{-1}|\textbf{t}_S|.
		\end{equation}
	\end{enumerate}
	
	\begin{figure}[!t]
		\centering
		\graphicspath{{Figures/}}
		\subfigure[][Effective cohesive law.]{\includegraphics[clip=true,trim = 12cm 6cm 12cm 6cm,width=0.4\textwidth]{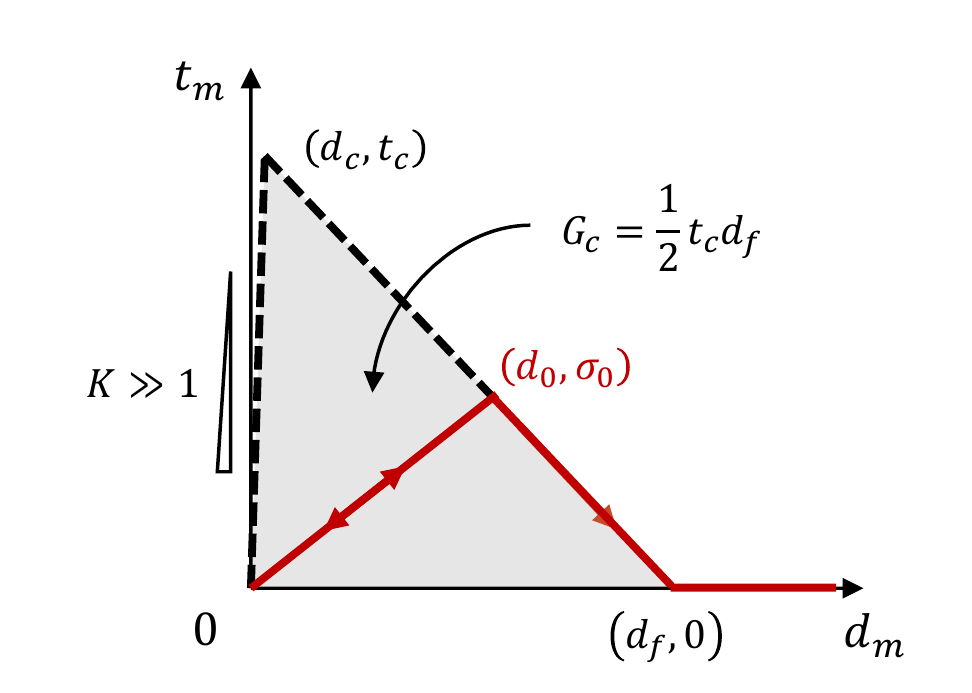}}
		\subfigure[][Mohr's circle in 3-D space.]{\includegraphics[clip=true,trim = 12cm 6cm 12cm 6cm, width = 0.4\textwidth]{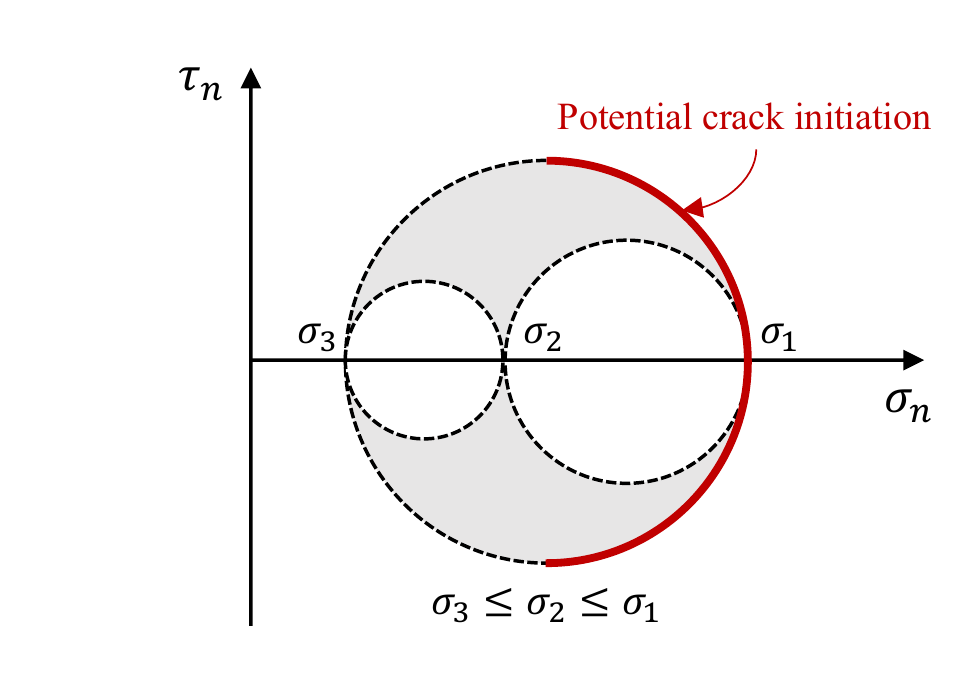}}
		\caption{Cohesive layer for crack modeling. (a) The bilinear cohesive law for effective opening displacement $d_m$ and the effective traction $t_m$. The initial stiffness $K$ should be large enough to mimic perfect bonding before failure, and the critical energy release rate $G_c$ is equal to the area under the curve. An unloading-loading path during softening is highlighted as the red solid line.  (b) Mohr's circle for a 3-D stress state $\boldsymbol{\sigma}$. The gray area covers the permissible normal and shear stress states, while the red solid line highlights potential states for crack-plane initiation. }
		\label{fig:Mohr}
	\end{figure}
	
	As shown in Figure \ref{fig:Mohr}(a), a 1-D bilinear $t_m-d_m$ effective cohesive law is adopted in this work. To simulate a perfectly bonded crack surface before the onset of failure, the initial stiffness of the elastic regime $K$ should be set to a large value. The critical effective traction is denoted as $t_c$, and the corresponding opening displacement as $d_c$. The internal variables $(d_0, t_0)$ store the feasible state with the maximum effective traction. The cohesive layer fails completely at $d_m = d_f$ with $t_m=0$. However, to avoid singularity in the analytical DMN calculation, we add a tiny stiffness $\kappa$ to the entire $t_m-d_m$ curve:
	\begin{equation}
	t_m \leftarrow t_m + \kappa d_m.
	\end{equation}
	For all the numerical examples to be studied in Section \ref{sec:examples}, we have
	\begin{equation}\label{eq:K}
	K = 1\times 10^8 \text{ GPa}/\text{mm}, \quad \kappa = 1\times 10^{-4}\text{ GPa}/\text{mm}.
	\end{equation}
	Meanwhile, the normal component of traction $\textbf{t}$ in Eq. (\ref{eq:t-compress}) for $\textbf{d}_n<0$ needs to be defined separately,
	\begin{equation}
	t_n = K d_n, 
	\end{equation}
	which represents a nearly impenetrable crack surface. More details on the bilinear curve and the derivation of the tangent stiffness tensor for implicit analysis can be found in \cite{liu2020deep}.
	
	Softening in the post-failure region is known to cause convergence difficulties in the implicit analysis. When the total strain energy to release is higher than the fracture toughness of the homogenized material, the solution is hard to find during unstable crack propagation. To overcome these issues, viscous regularization is introduced in the DMN framework. In \cite{liu2020deep}, we put the viscous effect directly on the opening displacement, which suits well in our debonding analysis with limited interfacial effects. However, for failure analysis interested in this paper, the separation displacement often dominates the overall strain due to localization, which causes nonphysical viscous stress even at the full failure of the material.
	
	Instead, we apply viscous regularization on the damage parameter of an inviscid backbone cohesive model, which is defined as
	\begin{equation}
	D = \dfrac{\sigma_0}{\sigma_c + K_h(d_0-d_c)},
	\end{equation}
	where $K_h$ is a positive hardening stiffness of the reference undamaged material. Its magnitude is chosen based on the base material's stiffness and the cohesive layer's characteristic length. Without viscosity, this hardening modulus does not affect the response of the cohesive layer. However, a properly selected $K_h$ will help improve the convergence of the implicit algorithm with viscous regularization. Specifically, we let the effective stiffness of the cohesive layer ($K_h/{v}_c$) be equal to Young's modulus of the base material $E^{base}$, so that
	\begin{equation}
	K_h = E^{base} {v}_c,
	\end{equation}
	Here, $v_c$ is the reciprocal length parameter of the cohesive layer, and its expression will be provided in Eq. (\ref{eq:vc}) when we discuss the activation of crack surfaces.
	
	The viscous damage parameter $\dot{D}_v$ is controlled by the following evolution equation:
	\begin{equation}
	\dot{D}_v = \dfrac{1}{\tau}(D-D_v),
	\end{equation}
	where $\tau$ is the viscosity coefficient representing the relaxation time of the viscous system. Here we use a backward Euler method to update $D_v$. Finally, the effective traction stress after viscous regularization is
	\begin{equation}
	t_v = \begin{cases}
	D_v \left[\dfrac{\sigma_c + K_h(d_0-d_c)}{d_0} d_m\right], & \text{if } d_m < d_0 \text{ or } \dot{d}_m<0; \\
	D_v\left[\sigma_c + K_h(d_m-d_c) \right],& \text{otherwise.}
	\end{cases}
	\end{equation}
	
	In summary, the cohesive layers' material behaviors are governed by four parameters: the critical effective traction $t_c$, the critical energy release rate $G_c$,  the ratio $\beta$ for defining the effective opening displacement, and the relaxation time $\tau$ for viscous regularization.
	
	\subsection{Activation of crack surfaces} \label{sec:activation}
	
	Assume the stress state of the base material for an arbitrary micro-cell is $\boldsymbol{\sigma}$. To decide whether a crack surface should be activated, we will first search the plane(s) with the maximum effective traction, and its normal $\textbf{n}_c$ is 
	\begin{equation}\label{eq:activationlaw}
	\textbf{n}_c = \argmax_{\textbf{n}'} \,t_m(\boldsymbol{\sigma}\cdot \textbf{n}', \textbf{n}'),
	\end{equation}
	where the expressions of the effective traction $t_m$ are given in Eq. (\ref{eq:tm1}) and (\ref{eq:tm2}) for tension and compression, respectively. 
	
	As shown in Figure \ref{fig:Mohr}(b), the Mohr circle of a three-dimensional stress state can be utilized to guide the searching process. To arrive at the Mohr circle,  we need to determine the values of principal stresses and the principal directions by the eigenvalue analysis of the stress tensor $\boldsymbol{\sigma}$. Without loss of generality, we order the three eigenvalues as
	\begin{equation}
	\sigma_1 \geq \sigma_2 \geq \sigma_3.
	\end{equation}
	By the definition of $t_m$, the optimum point should lie on the right half of the outer circle $C_2$, as highlighted by the red curve in the plot. For any point inside the permissible area, one can always find a point on $C_2$ that has a larger effective traction. Searching on $C_2$ also indicates a rotation along the second principal axis. The center and radius of $C_2$ are 
	\begin{equation}
	\bar{\sigma} = \dfrac{1}{2}(\sigma_1 + \sigma_3), \quad \bar{\tau} = \dfrac{1}{2}(\sigma_1 - \sigma_3).
	\end{equation}
	Here we use the angle $\theta$ to denote the rotation from the state with $\sigma_n = \sigma_1$, so that the normal and shear stresses can be expressed as
	\begin{equation}
	\sigma_n = \bar{\sigma} + \bar{\tau}\cos 2\theta \quad\text{and}\quad \tau_n = \bar{\tau}\sin 2\theta,
	\end{equation}
    with $\theta \in [-\pi/4,\pi/4]$ corresponding to the right half of $C_2$
	
	Then, let us consider the effective traction in Eq. (\ref{eq:tm1}) for the tension loading case, or $\sigma_n > 0$,
	\begin{equation}
	t_m = \sqrt{(\bar{\sigma} + \bar{\tau}\cos 2\theta)^2+\beta^{-2}(\bar{\tau}\sin 2\theta)^2}.
	\end{equation}
	The stationary points in terms of $\theta$ are
	\begin{equation}
	\theta^* = \pm\dfrac{1}{2}\arccos(\dfrac{\bar{\sigma}}{\bar{\tau}(\beta^{-2}-1)})
	\quad \text{ with }
	t_m''(\theta^*) = (1 - \beta^{-2})\bar{\tau}^2\sin^2 2\theta^*.
	\end{equation}
	Therefore, the solution $\theta^*$ exists as a potential global maximum point only if
	\begin{equation}
	0<\dfrac{\bar{\sigma}}{\bar{\tau}(\beta^{-2}-1)}<1 \text{ and } \bar{\sigma}>0 \text{ and } \beta < 1.
	\end{equation}
	
	On the other hand, the effective traction in Eq. (\ref{eq:tm2}) for the compressive case, or $\sigma_n < 0$, can be written as
	\begin{equation}
	t_m = \beta^{-1}|\bar{\tau}\sin 2\theta|,
	\end{equation}
	which reaches maximum when $\theta = \pm \pi/4$.
	
	In summary, for any given $\boldsymbol{\sigma}$ and $\beta$, we only need to check the following candidates for the maximum effective traction:
	\begin{equation}\label{eq:potentialpoints}
	\theta = 0, \quad \theta = \pm \dfrac{\pi}{4},
	\end{equation}
	\begin{equation*}
	\theta = \pm\dfrac{1}{2}\arccos(\dfrac{\bar{\sigma}}{\bar{\tau}(\beta^{-2}-1)}) \quad\text{ if } 0<\dfrac{\bar{\sigma}}{\bar{\tau}(\beta^{-2}-1)}<1 \text{ and } \bar{\sigma}>0 \text{ and } \beta < 1.
	\end{equation*}
	
	\begin{remark}
		The failure algorithms are primarily designed for the specific cohesive law based on the effective traction-separation behavior in Section \ref{sec:coh}. Despite its simplicity, a limitation of the one-dimensional model is that the fracture energy is constant regardless of the fracture mode. However, sometimes it becomes necessary to consider the variances of fracture energy under different modes, e.g., Mode I and Mode II fractures. In those situations, cohesive law models based on more general potentials can be used \cite{park2011cohesive}, while the viscous regularization scheme and the activations criteria should be revised accordingly.
	\end{remark}
	
	When the maximum effective traction exceeds the critical value $t_c$, 
	\begin{equation}
	t_m(\boldsymbol{\sigma}\cdot \textbf{n}_c, \textbf{n}_c) > t_c,
	\end{equation}
	the crack surface with normal $\textbf{n}_c$  will be activated in a micro-cell. Depending on the solutions in Eq. (\ref{eq:potentialpoints}), two crack surfaces might be activated at the same time. To differentiate the two crack surfaces numerically, a small perturbation is added to their $t_c$ values. Note that the displacement vector in the new cohesive layers at loading step $n$ is set to be 
	\begin{equation}
	\textbf{d}^{(n-1)} = \dfrac{\boldsymbol{\sigma}^{(n-1)}\cdot \textbf{n}_c}{K},
	\end{equation}
	where the superscript $(n-1)$ labels a quantity from previous load step $n-1$. This guarantees the equilibrium condition after inserting the cohesive layers to the base material. Since the stiffness $K$ is a large value (see Eq. (\ref{eq:K})), the mismatch in the kinematic constraints does not affect the overall accuracy.
	
	Finally, we need to determine the reciprocal length parameter, as illustrated in Figure \ref{fig:cellDivision}. Assume the scale tensor of the microcell is $\textbf{A}_N^j$. Similar to Eq. (\ref{eq:Area}), the area of the crack surface is 
	\begin{equation}\label{eq:area}
	S(\textbf{A}_N^j,\textbf{n}_c) = \dfrac{\pi}{\sqrt{\det{\textbf{A}_N^j}}|\boldsymbol{{\mathcal{T}}}\boldsymbol{{\mathcal{R}}}^T\textbf{n}_c|}.
	\end{equation}
	where $\boldsymbol{{\mathcal{T}}}$ and $\boldsymbol{{\mathcal{R}}}$ are defined in Eq. (\ref{eq:R}) and (\ref{eq:T}), respectively. Meanwhile, the volume of the micro-cell with $\textbf{A}_N^j$ is 
	\begin{equation}
	V = \dfrac{4\pi}{3\sqrt{\det{\textbf{A}_N^j}}}.
	\end{equation}
	For an isotropic spherical micro-cell,  the reciprocal length parameter $v_c$ can be naturally set equal to the inverse of the diameter of the sphere, with $v_c=2S/3V$. Extending this formula to a general ellipsoidal micro-cell yields
	\begin{equation}\label{eq:vc}
	{v}_c(\textbf{A}_N^j,\textbf{n}_c) = \dfrac{2S}{3V} = \dfrac{1}{2|\boldsymbol{{\mathcal{T}}}\boldsymbol{{\mathcal{R}}}^T\textbf{n}_c|}.
	\end{equation}
	
	\subsection{Solution scheme for implicit DMN failure analysis}\label{sec:scheme}
	\bigskip
	
	\noindent\fbox{\begin{minipage}{46em}\label{mp:1}
			\medskip
			\centering\textbf{Box 3.3.1 Solution scheme for implicit DMN analysis and crack activations}
			
			\begin{enumerate}[itemsep=0mm]
				\setcounter{enumi}{-1}
				\item Initialization. $M^{(0)}=0$, $\boldsymbol{\mathcal{L}}^{(0)}=\varnothing$. Given $\textbf{A}^{macro}$, perform cell division to get the scale tensor $\textbf{A}^j_N$ for each micro-cell.
				\item For load step $n$, the macroscale strain increment is $\Delta \boldsymbol{\varepsilon}^{macro}$. Initialize the global crack configuration use the converged one from the last load step $n-1$: $M^{new}\leftarrow M^{(n-1)}$, $\boldsymbol{\mathcal{L}}^{new} \leftarrow \boldsymbol{\mathcal{L}}^{(n-1)}$
				\item Newton's method:
				\begin{enumerate}[itemsep=0mm]
					\item If the number of iterations exceeds the limit, go to 5.
					\item Evaluate the constitutive laws of the base material and enriched cohesive layers of each micro-cell to get the stiffness tensor and residual stress of the corresponding bottom-layer node
					\item Forward propagation of stiffness tensors and residual stresses to the top node with $\textbf{C}^{macro}$ and $\delta \boldsymbol{\sigma}^{macro}$
					\item Compute the macroscale stress increment $\Delta \boldsymbol{\sigma}^{macro} =\textbf{C}^{macro}\Delta\boldsymbol{\varepsilon}^{macro} + \delta \boldsymbol{\sigma}^{macro}$
					\item Backward propagation of stress and strain tensors from the top node to the bottom layer. Compute new incremental strain tensors of base materials and new incremental displacement vectors of cohesive layers
					\item Check convergence. If not converged, go to 2(a)
				\end{enumerate}
				\item Crack activations:
				\begin{enumerate}[itemsep=0mm]
					\item Find the micro-cell index $I_c$ and surface normal $\textbf{n}_c$, which satisfy the allowances (see Remark \ref{rem:5}) and maximize
					\begin{equation*}
					\Delta t = t_m(\boldsymbol{\sigma}^{I_c}\cdot\textbf{n}_c, \textbf{n}_c)-t_c
					\end{equation*}
					\item If $\Delta t\leq0$, no more new crack surfaces, go to 4.
					\item If $\Delta t> 0$, compute ${v}_c$ and initialize the internal variables $\textbf{q}_c$ of the new enriched cohesive layer(s). Update the list $M^{new}\leftarrow M^{new}+1$, $\boldsymbol{\mathcal{L}}^{new}\leftarrow \boldsymbol{\mathcal{L}}^{new} + \{I_{c}, {v}_{c}, \textbf{n}_c, S\}$
					\item Go to 2(a) to start a new round of iterations
				\end{enumerate}
				
				\item Global crack configuration reaches convergence. Update the internal variables, $\boldsymbol{\mathcal{L}}^{(n)}\leftarrow\boldsymbol{\mathcal{L}}^{new}$, $M^{(n)}\leftarrow M^{new}$
				\item Invoke time step refinement,  restore $M^{(n-1)}$, $\boldsymbol{\mathcal{L}}^{(n-1)}$, and all internal variables.
			\end{enumerate}
			\par\smallskip
	\end{minipage}}
	\bigskip
	
	To improve the overall convergence of the multiscale failure analysis, we propose a solution scheme which decouples the DMN implicit solver and the activation of new crack surfaces. First of all, let us define the global crack configuration $\boldsymbol{\mathcal{L}}$ of DMN,
	\begin{equation}
	\boldsymbol{\mathcal{L}} = \{\mathcal{L}^1, \mathcal{L}^2, ..., \mathcal{L}^{{M}}\},
	\end{equation}
	where ${M}$ is the total number of crack surfaces. The list $\mathcal{L}^i$ stores the information of the $i$-th crack surface (or cohesive layer in DMN formulation),
	\begin{equation}
	\mathcal{L}^i = \{I_{c}^i, {v}_{c}^i, \textbf{n}_c^i,S^i\},
	\end{equation}
	where $I_c^i$ is the index of micro-cell where the crack surface is located, ${v}_c^i$ is the reciprocal length parameter, $\textbf{n}_c^i$ is the unit normal, and $S^i$ is the area of the crack surface. Note that once the cohesive layer for a crack surface is activated, we also need to keep track of its internal variables $\textbf{q}_c$. At the beginning of analysis, or load step $0$, no crack surface exists in the network for an undamaged material, so we have
	\begin{equation}
	\boldsymbol{\mathcal{L}}^{(0)}=\varnothing. 
	\end{equation}
	
	In Box \hyperref[mp:1]{3.3.1}, we list the solution scheme for a material point (or integration point) in a concurrent multiscale simulation, where the macro strain increment $\Delta \boldsymbol{\varepsilon}^{macro}$ is applied at the top node of DMN. \textcolor{black}{In the initialization step, the scale tensors $\textbf{A}^{j=1,2,...,2^{N-1}}_N$ for all the micro-cells are computed once based on the trained network architecture, and they will be fixed during the online simulation. We determine the minimum network depth $N$ by assessing the DMN test error (typically less than 1\%). Since the two-layer building block has a mixture of equilibrium and kinematic conditions, both stress and strain tensors are back-propagated in Step 2(e) to determine all the components inside the stress-strain relationship of each node.}
	
	In all our simulations, the relative tolerances of the incremental strain tensors and the displacement vectors in Newton's method (Step 2(f) in Box 4.1) are both set to $1.0\times 10^{-6}$ for convergence, and the maximum iteration number is set to 40. 
	
	\begin{remark}\label{rem:5}
		In Step 3(a), we introduce two allowances of crack activation in each micro-cell to improve the solution scheme's robustness: (1)The total number of crack surfaces does not exceed 4; (2) The cosine similarity between the normals of a new crack surface and any existing one is less than $\sqrt{2}/2$.
	\end{remark}
	
	The first allowance limits the memory cost of an enriched DMN, though adaptive memory allocation could be implemented in the future to help relieve this constraint. The second allowance based on cosine similarity is introduced to suppress redundant crack activations. Due to viscous regularization, the stress softening is delayed, and the effective traction may exceed the critical value. As a result, new cracks may be invoked at a similar orientation as the existing cracks, which is not desired for the analysis. 
	
	If convergence cannot be reached for the current loading step, adaptive load step refinement is enabled to ease solving the nonlinear system. Essentially, the original load step is divided equally into two sub-steps. Note that the time increment should also be refined for the consistency of viscous regularization. The maximum number of load step refinements is set to 10 in this paper. Detailed descriptions of the adaptive load step refinement algorithm are provided in Box \hyperref[mp:2]{C.1}.
	
	\section{Examples}\label{sec:examples}
	In concurrent multiscale structural simulations, the macroscale FEA model can be either explicit or implicit, since both the stress and the tangent stiffness tensor are available at the top node of DMN. However, in multiscale failure analyses with strain softening and localization, we will focus on the explicit dynamics in the macroscale FEA model, while the microscale DMN models are always solved implicitly.
	
	We begin with a single material point modeled by DMN in Section \ref{sec:single}. The failure behaviors of the particle-reinforced composite shown in Figure \ref{fig:geo1} will be evaluated under tension, compression, shear, and cyclic loadings. Meanwhile, we will study the effects of the macroscale length parameter $h$, the viscous relaxation time $\tau$, and the DMN depth $N$. \textcolor{black}{The primary purpose of this work is to advocate DMN with the proposed scale transition scheme as an advanced framework for strain localization modeling. Instead of benchmarking it to DNS RVE model (e.g. FEM), which encounters certain issues as discussed in Section \ref{sec:issues}, we will focus on DMN's extrapolation capability of representing various failure modes and loading paths, the energy consistency with the macroscale length scale, and its performance for online multiscale simulations validated by experiments.}
	
	Three representative examples of the concurrent multiscale simulation based on microscale DMNs will be presented in the following sections: 1) Section \ref{sec:crushtube}, the dynamic crush of a composite tube meshed by 3-D thin-shell elements; 2) Section \ref{sec:3point},  three-point bending test under 2-D plane strain condition; 3) Section \ref{sec:coupon}, composite off-axis coupon test modeled by 3-D solid elements. Moreover, predictions from the off-axis coupon model of a unidirectional carbon fiber reinforced polymer composite will be validated against experimental data from the literature \cite{gao2020predictive}.
	
	\subsection{Single material point: parametric study}\label{sec:single}
	
	In this section, the 3-D particle-reinforced composite is first trained by DMN and tested for its online extrapolation as a single material point. The particle phase is assumed to be linearly elastic, and the matrix phase is elasto-plastic with failure. All the material parameters are provided in Table \ref{table:para1}. The default value of the relaxation time $\tau$ is $1.0\times 10^{-4}$ ms, while we will study its effect on the material responses in Section \ref{sec:hyper}. Additionally, the strain rate on the single material point is set to $1.0\text{ ms}^{-1}$.
	
	\begin{table}[htb!]
		\captionabove{\textcolor{black}{Microscale material parameters of the particle reinforced composite with matrix failure.}} % title of Table
		\centering % used for centering table
		\label{table:para1} % is used to refer this table in the text
		{\tabulinesep=1.0mm
			\begin{tabu}{c c c c c} % centered columns (4 columns)
				\hline
				\multirow{2}{*}{Particle}& $E_p$ (GPa) & $\nu_p$ &  &  \\
				& 500.0 & 0.3 &  &  \\ 
				\hline
				\multirow{4}{*}{Matrix} & $E_m$ (GPa) & $\nu_m$ & $\sigma^Y$ (GPa) & Hardening   \\ 
				&100.0 & 0.30 & 0.1 & Eq. (\ref{eq:piecewise_hardening}) \\ 
				\cline{2-5}
				&$t_c$ (GPa) & $G_c$ (GPa$\cdot$mm)& $\beta$ & $\tau$ (ms) \\ 
				&0.15 & $6\times10^{-4}$ & 1.0 & $1.0 \times 10^{-4}$\\
				\hline
				%inserts single line
		\end{tabu}}
	\end{table}
	
	In terms of the matrix plasticity, von Mises plasticity with isotropic hardening surface is used,  and the yield stress $\sigma^Y$ is described as a function of the effective plastic strain $\varepsilon_p$,
	\begin{equation}\label{eq:piecewise_hardening}
	\sigma^Y(\varepsilon_p)=
	\begin{cases}
	0.1 + 10\cdot{\varepsilon}_{p}, & \varepsilon_p \in [0,0.01) \\
	0.18 + 2\cdot{\varepsilon}_{p}, & \varepsilon_p \in [0.01,\infty)
	\end{cases}
	\text{GPa}.
	\end{equation} 
	
	The macroscale material point is assumed to have the same length scale in all directions, so that the spherical macro-cell is represented by a isotropic scale tensor,
	\begin{equation}
	\textbf{A}^{macro} = \dfrac{4}{h^2} \textbf{I}.
	\end{equation}
	By default, the macro length scale $h$ is set to 2 mm. The micro-cells after the cell division process based on the trained network structure are shown in Figure \ref{fig:geo1} (c). Later in Section \ref{sec:h}, we will also examine the effects of $h$ on the overall material responses.
	
\begin{figure} [!t]
		\centering
		\graphicspath{{Figures/}}
		\subfigure[Tension loadings.]{\includegraphics[clip=true,trim = 8.9cm 2.5cm 8.8cm 0.0cm,width=0.45\textwidth]{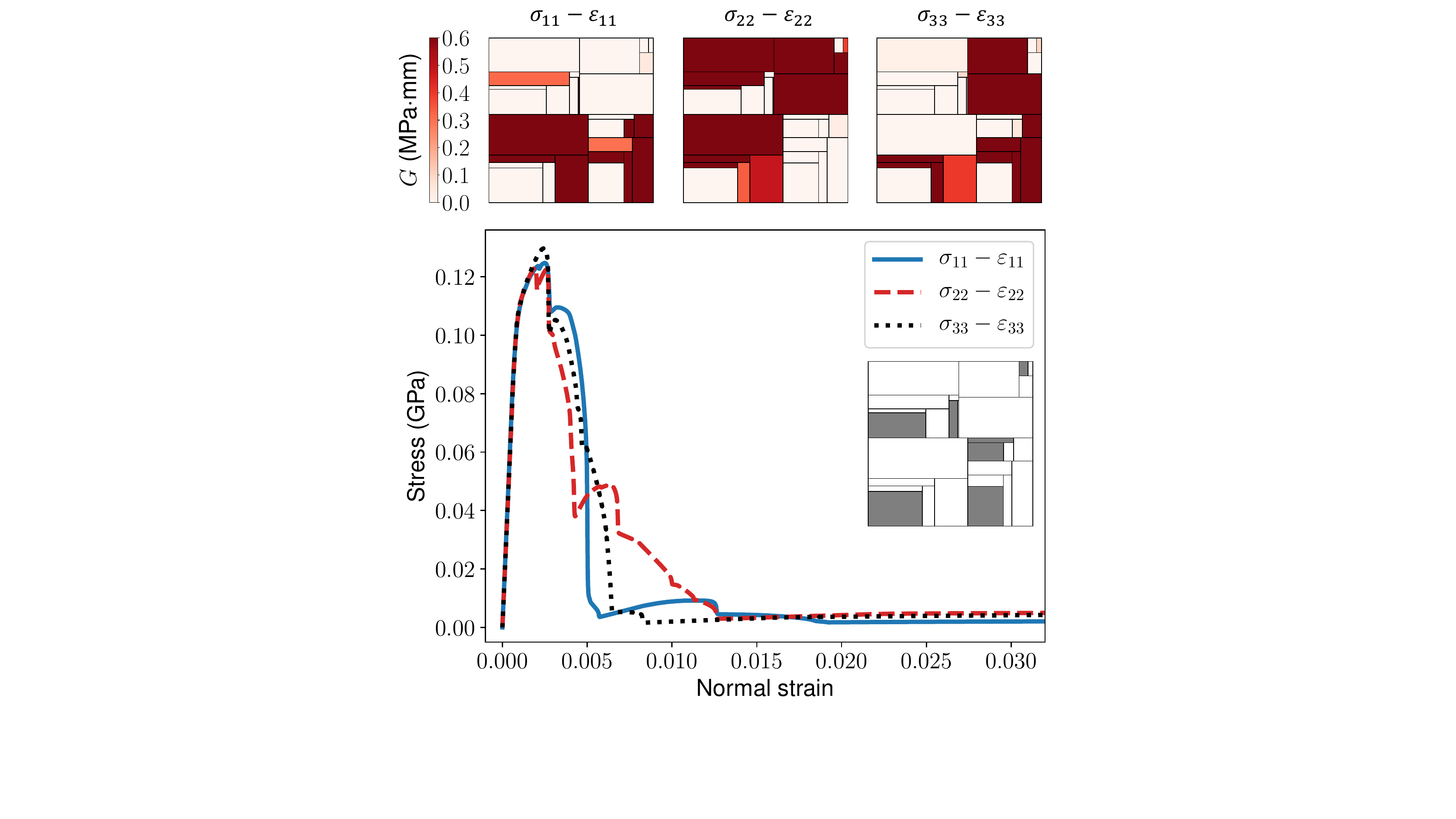}}
		\subfigure[Shear loadings.]{\includegraphics[clip=true,trim = 8.9cm 2.5cm 8.8cm 0.0cm,width=0.45\textwidth]{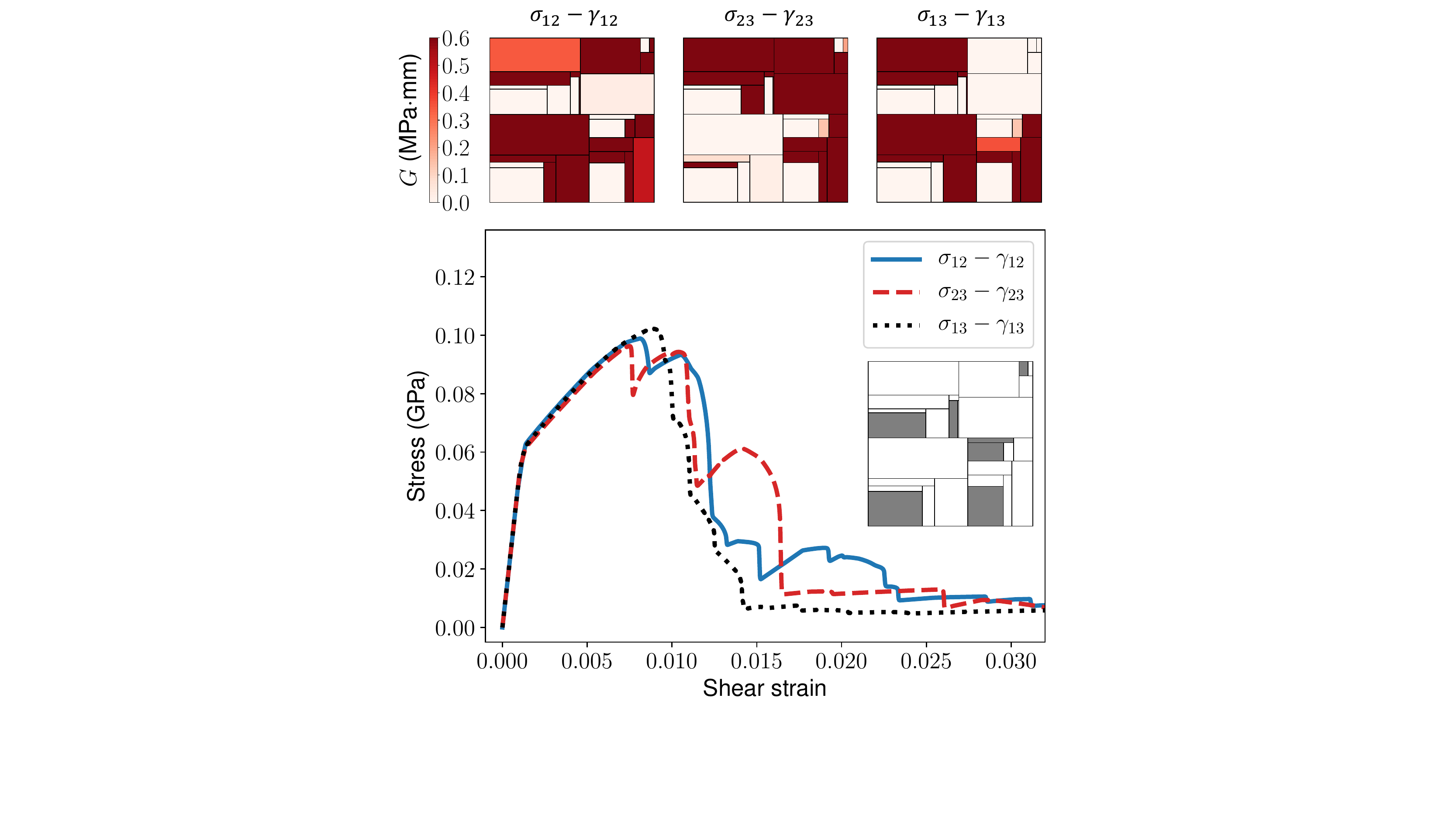}}
		\caption{DMN predictions for the particle-reinforced composite under three uniaxial tension loadings and three shear loadings. The number of layers $N$ is 8, and the number of microscale DOF or micro-cells in the DMN is $N_{dof}=33$. For each loading case, the contour treemap plot of the released energy per crack area $G$ at $\varepsilon^{macro}=0.03$ is placed above the stress-strain plot. In a micro-cell, $G$ vanishes if no crack surface is activated.}
		\label{fig:ex1-1}
	\end{figure}

	Figure \ref{fig:ex1-1} shows the stress-strain curves under six loading directions for the DMN with $N=8$, which has 33 active DOF in the bottom layer. Because the particle-reinforced composite is nearly isotropic, it is expected to behave similarly in three uniaxial tension loadings and in three shear loadings. For the uniaxial tension loadings, the composite reaches the maximum stress around 0.125 GPa, which is less than the critical effective traction $t_c$ of the matrix material due to stress concentration induced by the particle reinforcement. For the shear loadings, the plastic hardening effect is more prominent, and the softening starts at a lower stress level of 0.10 GPa. In addition, treemap plots colored by the released energy per crack surface area $G$ in each DOF at $\varepsilon^{macro}=0.03$ are presented above the stress-strain curves. A micro-cell with fully separated crack surfaces has $G$ equal to $G_c$ ($G_c=0.6$ MPa$\cdot$mm, see Table \ref{table:para1}), while $G$ vanishes if there exists no crack surface. Essentially, this contour treemap plot reflects the global crack configuration of DMN, which is shown to vary with the applied macroscale boundary conditions.
	
	\begin{figure} [!t]
		\centering
		\graphicspath{{Figures/}}
		\subfigure[Monotonic loadings in tension and compression.]{\includegraphics[clip=true,trim = 8.9cm 2.5cm 8.8cm 5.0cm,width=0.45\textwidth]{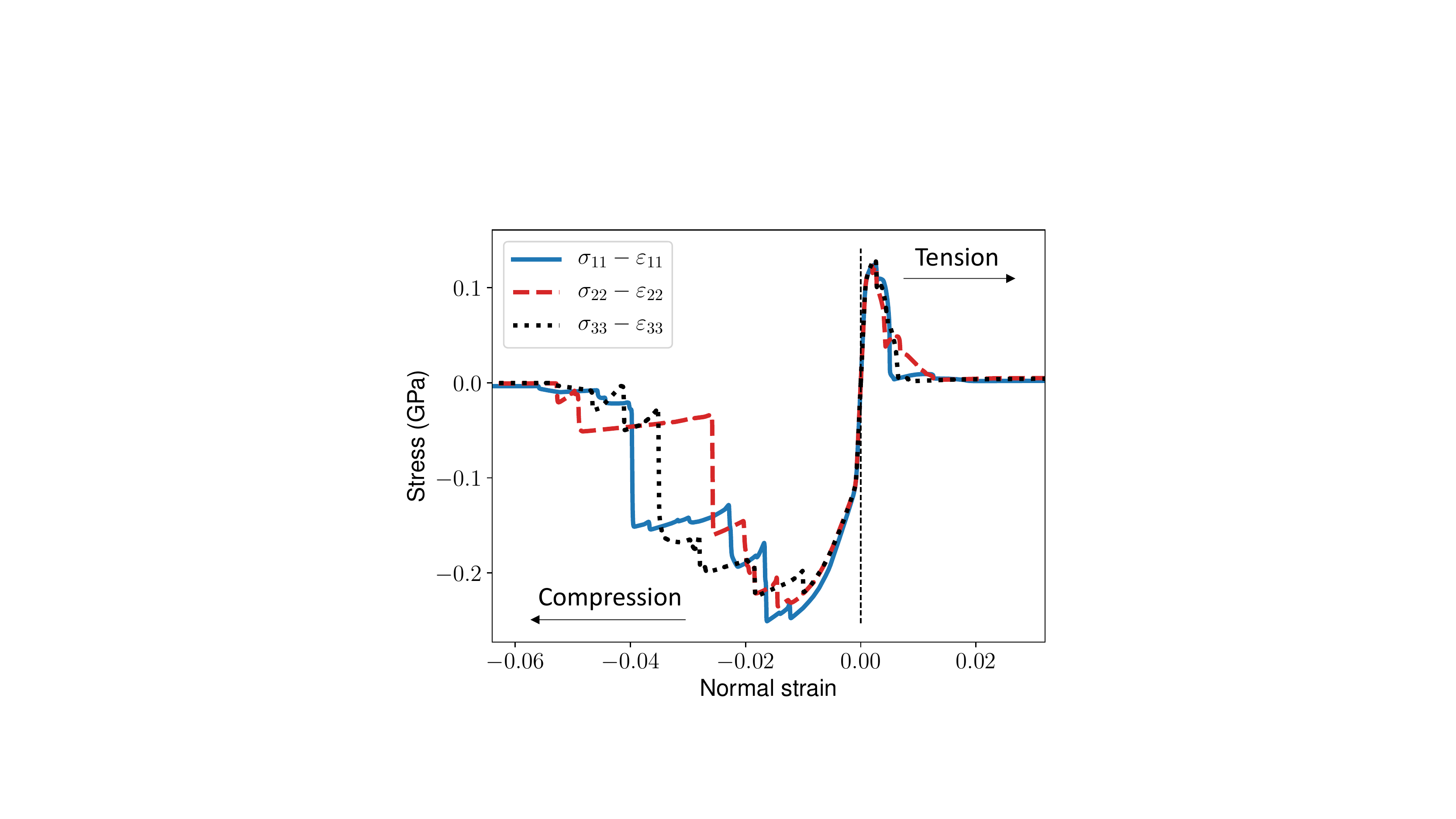}}
		\subfigure[Cyclic loadings with different strain magnitudes.]{\includegraphics[clip=true,trim = 8.9cm 2.5cm 8.8cm 5.0cm,width=0.45\textwidth]{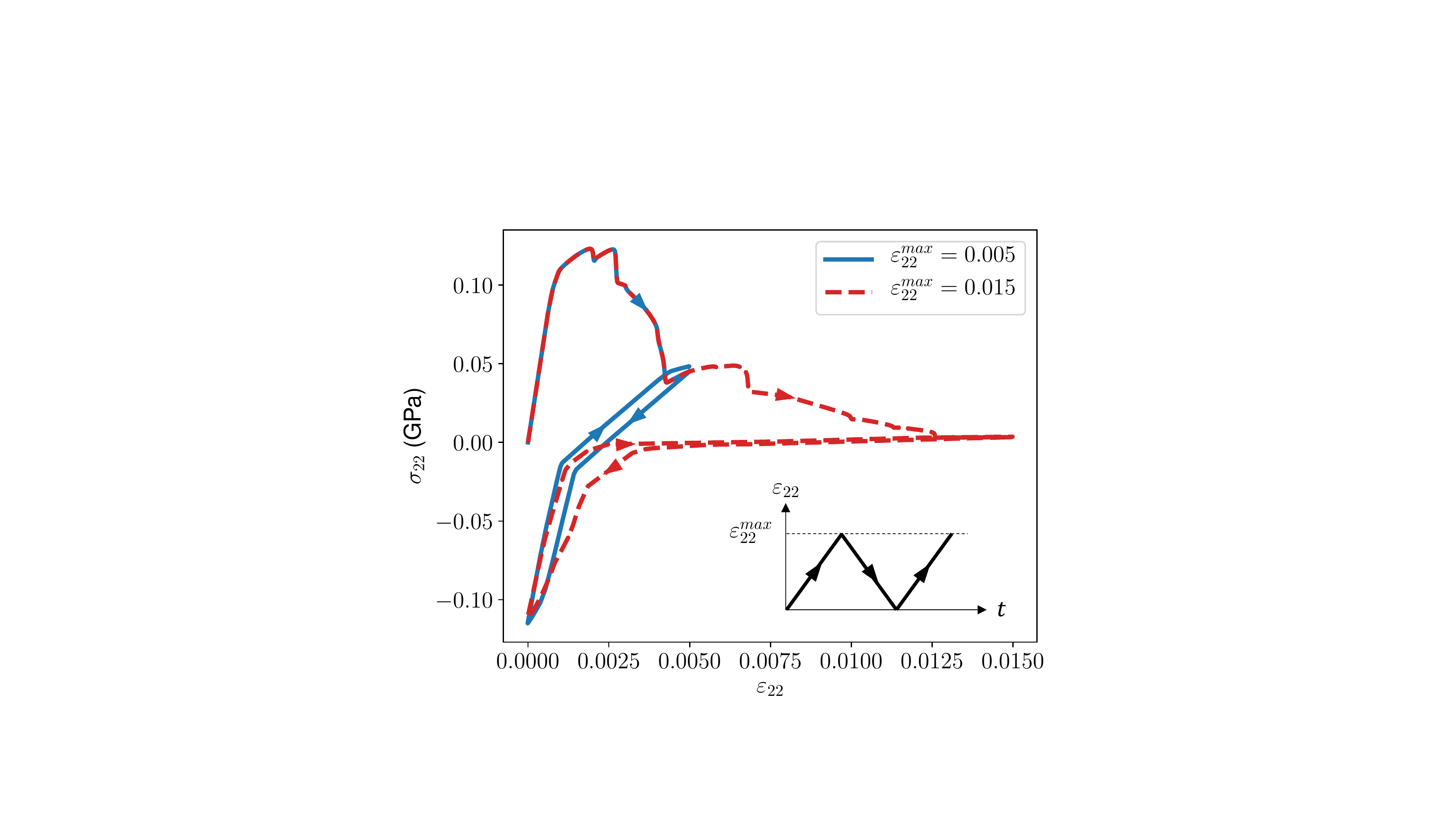}}
		\caption{Stress-strain curves of the particle reinforced composite under more general loading conditions. The number of layers $N$ is 8. In (a), the uniaxial tension and compression loadings are evaluated independently. In (b), each uniaxial cyclic loading path contains two loading and one unloading processes between 0 and  $\varepsilon_{22}^{max}$, as shown in the embedded strain history plot.}
		\label{fig:tensioncompression}
	\end{figure}
	
	To demonstrate the physics introduced by the cohesive layers, we compare the material responses under uniaxial tension and compression loadings in Figure \ref{fig:tensioncompression} (a). As the compressive stress does not contribute to the effective traction $t_m$ of a potential crack surface in the matrix phase, the composite material reaches a higher stress magnitude of $0.25$ GPa under compression. For all the cases, the DMNs fail completely with negligible residual stresses. 
	
	\begin{remark}
		Although not necessary for predicting the multiscale failure responses, it is useful to define a global damage indicator for tracking the failure process and informing discontinuous crack algorithms in the macroscale model. However, in the concurrent multiscale simulations to be shown in Section \ref{sec:3point} and \ref{sec:coupon}, we get around potential numerical issues like element distortions by using a volumetric strain-based criterion to trigger the element deletion in the macroscale model.
	\end{remark}
	
	The cyclic loading behaviors with different strain magnitude $\varepsilon_{22}^{max}$ are shown in Figure \ref{fig:tensioncompression} (b). When the DMN material is unloaded at $\varepsilon_{22}=0.005$ in the middle of the softening process, it first goes through a linear elastic regime with a degraded stiffness, and then transits to a compressive state with almost the same stiffness as the undamaged material. The reloading curve closely follows the unloading curve but differs slightly due to plasticity induced at the end of compression. For the second case, the material is close to complete failure at $\varepsilon_{22}=0.015$. Therefore, the first part of the unloading curve has nearly zero stiffness, but it recovers the original stiffness as the opening displacements of microscale crack surfaces reduce to zero.
	
	In essence, DMN allows modelers to describe complex nonlinear multiscale material behaviors using simple microscale material laws. For various multiscale systems without strain localization, like plasticity and hyperelasticity, we have validated its accuracy against DNS in \cite{liu2019deep,liu2019exploring}, and we will also revisit the DNS validation for the particle-reinforced composite in Section \ref{sec:hyper}. More importantly, the scale transition of DMN enabled by the proposed cell division process overcomes some critical issues of RVE-based models with strain localization, achieving a consistent and efficient framework for multiscale failure analysis.
	
	\subsubsection{Effects of macro length scale}\label{sec:h}
	As discussed in Section \ref{sec:issues}, the macro length scale $h$ is usually predetermined by the macroscale element size or the nonlocal regularization size. The cell division process consistently tracks the scale transition in DMN without retraining the network or introducing extra energy regularization for different macro length scales.
	
	\begin{figure} [!t]
		\centering
		\graphicspath{{Figures/}}
		\subfigure[Uniaxial tension $\sigma_{22}-\varepsilon_{22}$.]{\includegraphics[clip=true,trim = 0.0cm 0.0cm -0.5cm 0.0cm,width=0.44\textwidth]{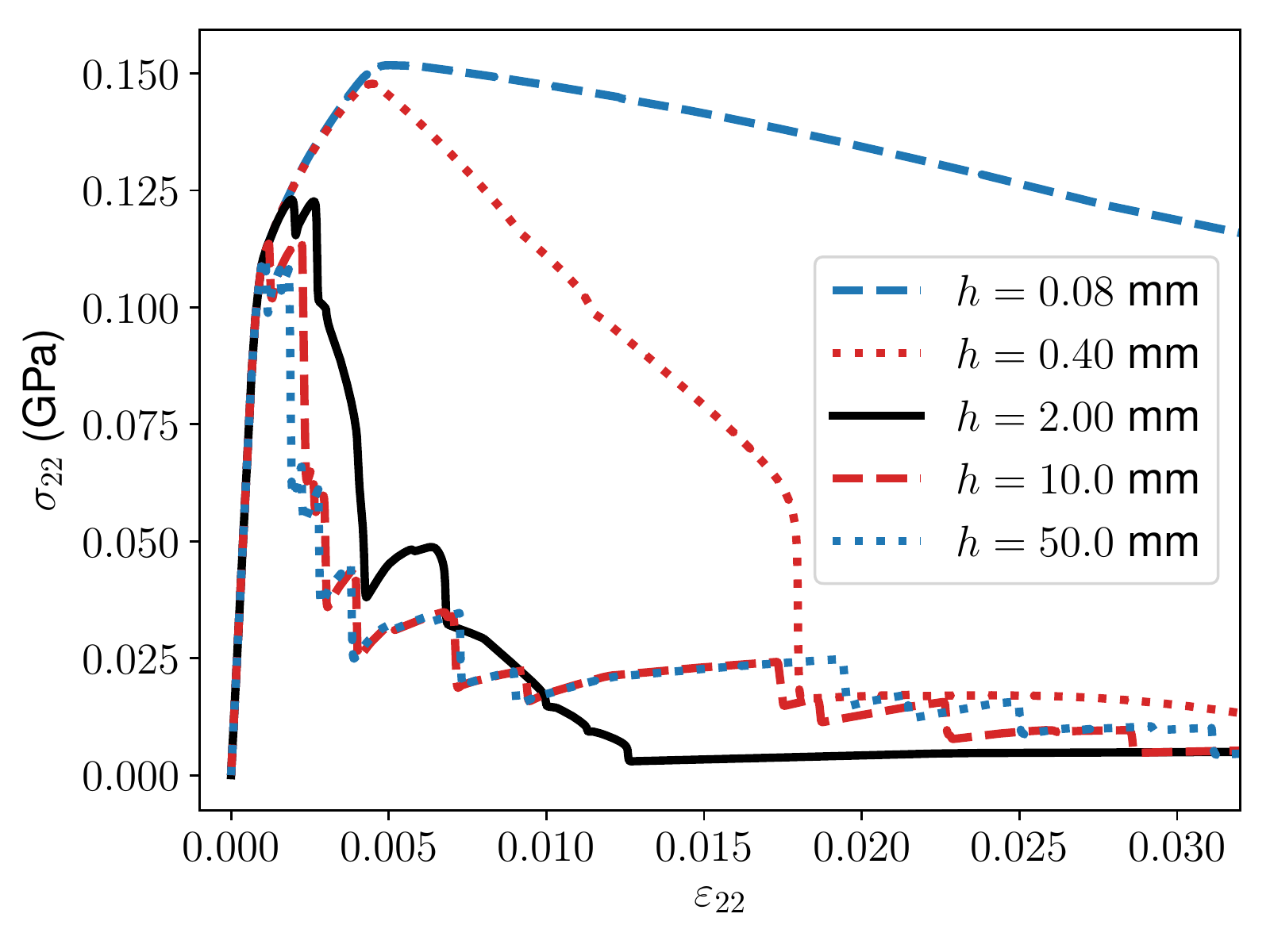}}
		\subfigure[Maximum stress and released energy.]{\includegraphics[clip=true,trim = 8.9cm 2.5cm 8.8cm 5.0cm,width=0.45\textwidth]{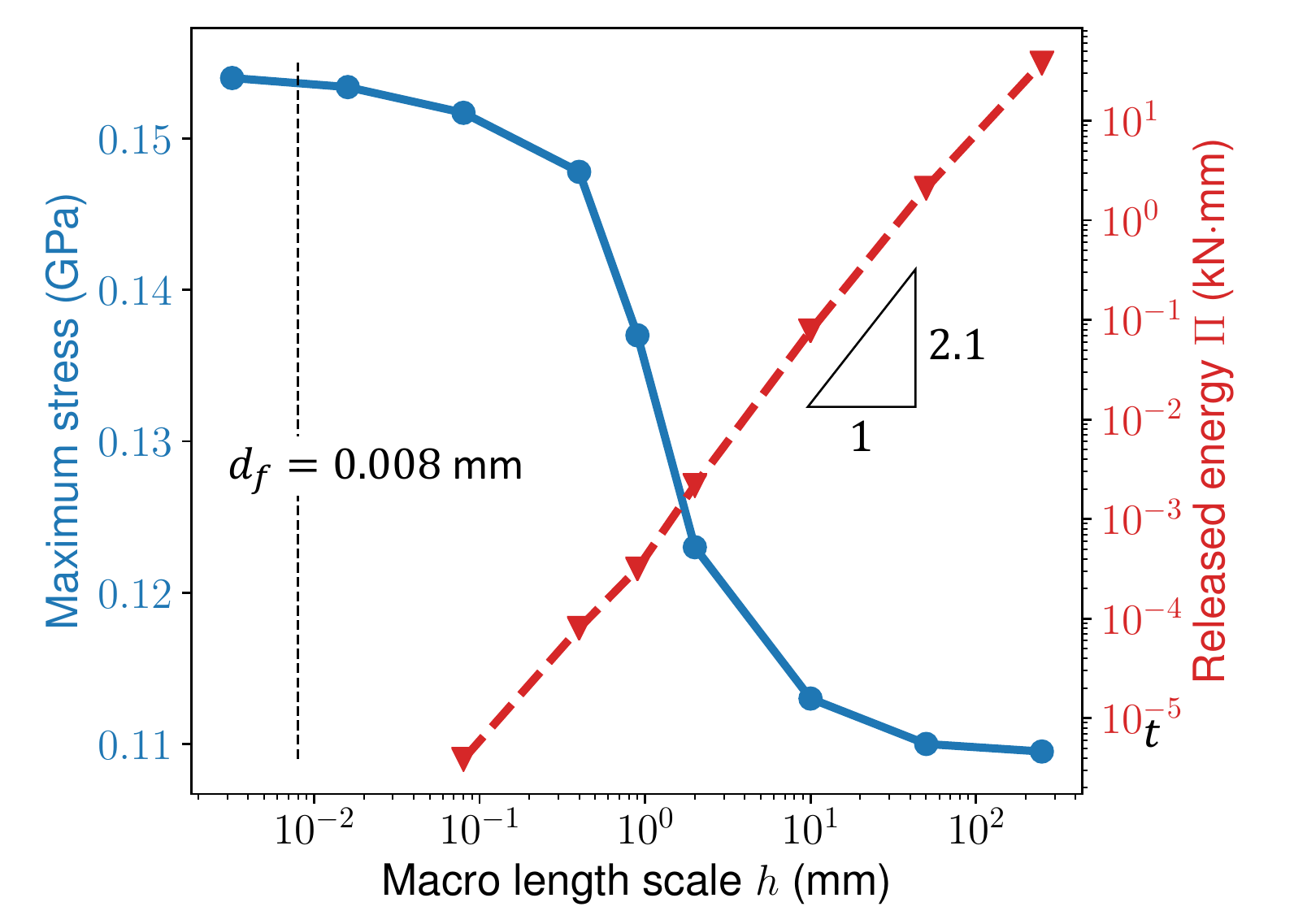}}
		\caption{Effects of the macro length scale $h$. The macro scale tensor is $\textbf{A}^{macro}={4}/{h^2\cdot\textbf{I}}$, and the number of layers $N$ is 8. In (b), the maximum stress is extracted from the corresponding $\sigma_{22}-\varepsilon_{22}$ curve in (a). The released energy $\Pi$ is the total free energy $\phi$ of all the activated crack surfaces in the network. The dashed line highlights the full separation displacement of the cohesive layer $d_f=0.008$ mm.}
		\label{fig:length}
	\end{figure}
	
	Figure \ref{fig:length} (a) presents $\sigma_{22}-\varepsilon_{22}$ curves of $h$ varying from 0.08 mm to 50.0 mm. As we can see from the plots, the magnitude of the macroscale softening stiffness increases with $h$.  When $h$ is larger than 10.0 mm, the stress-strain curve has a sharp drop and does not change much as $h$ increases. In these cases, the activated crack surfaces' overall fracture energy is not sufficient to release the total strain energy in the bulk material, and the excess energy is mainly dissipated via viscosity.
	
	Figure \ref{fig:length} (b) plots the maximum stress and released energy against the macro length scale $h$ under uniaxial tension loadings. The maximum stress is directly extracted from the stress-strain curves in Figure \ref{fig:length} (a), while the released energy $\Pi$ is computed as the total free energy $\phi$ (see Eq. (\ref{eq:free})) of all the crack surfaces in $\boldsymbol{\mathcal{L}} = \{\mathcal{L}^1,\mathcal{L}^1,...,\mathcal{L}^M\}$:
	\begin{equation}\label{eq:pi}
	\Pi = \sum_{i=1}^{M} \phi^i S^i,
	\end{equation}
	where $S^i$ is the surface area of the $i$-th crack as defined in Eq. (\ref{eq:Area}). Note that this released energy is not equal to the area under the stress-strain curve, due to the existence of matrix plasticity and viscous dissipation. The energy predictions for $h<0.08$ mm are not shown as the material point does not fail completely at $\varepsilon_{22} = 0.1$.
	
	As $h$ approaches the characteristic length of the cohesive layer $d_f = 0.008$ mm, the material point's maximum stress (or strength) reaches a plateau close to the critical effective traction of the cohesive layer $t_c=0.15$ GPa. When the material point has a small macro length scale, the reciprocal length parameters $v_c$ of the cohesive layers become so large that the global softening is delayed. Eventually, the overall strength is governed by the cohesive layers in the matrix phase.
	
	Another important observation is that the released energy in the material point is approximately proportional to $h^2$. This is physically sound since the material point has an isotropic scale tensor and the global (effective) crack surface is a two-dimensional manifold cutting through the material point. If one uses volumetric damage mechanics in the matrix phase, the resulting energy would be proportional to $h^3$, which requires post-regularizations based on $h$ (e.g., element size) to restore the energy consistency \cite{bazant1983crack,liu2018microstructural}. However, in the DMN framework, the property $\Pi\propto h^{2}$ is reproduced naturally from the cell division process and the cohesive-layer enrichment.
	
	\subsubsection{Effects of relaxation time and network depth}\label{sec:hyper}
	The purpose of putting the viscous regularization on the cohesive layer's damage parameter is to overcome the convergence difficulties in an implicit analysis. A larger relaxation time $\tau$ induces more viscosity or damping in the system. On the one hand, we want to keep the added viscosity small enough to limit its effects on the material responses. On the other hand, a more damped system tends to be more stable and requires fewer time-step refinements to converge. Therefore, $\tau$ should be chosen properly to balance accuracy and efficiency.
	
	\begin{figure} [!t]
		\centering
		\graphicspath{{Figures/}}
		\subfigure[Relaxation time $\tau$.]{\includegraphics[clip=true,trim = 0.0cm 0.0cm -0.5cm 0.0cm,width=0.44\textwidth]{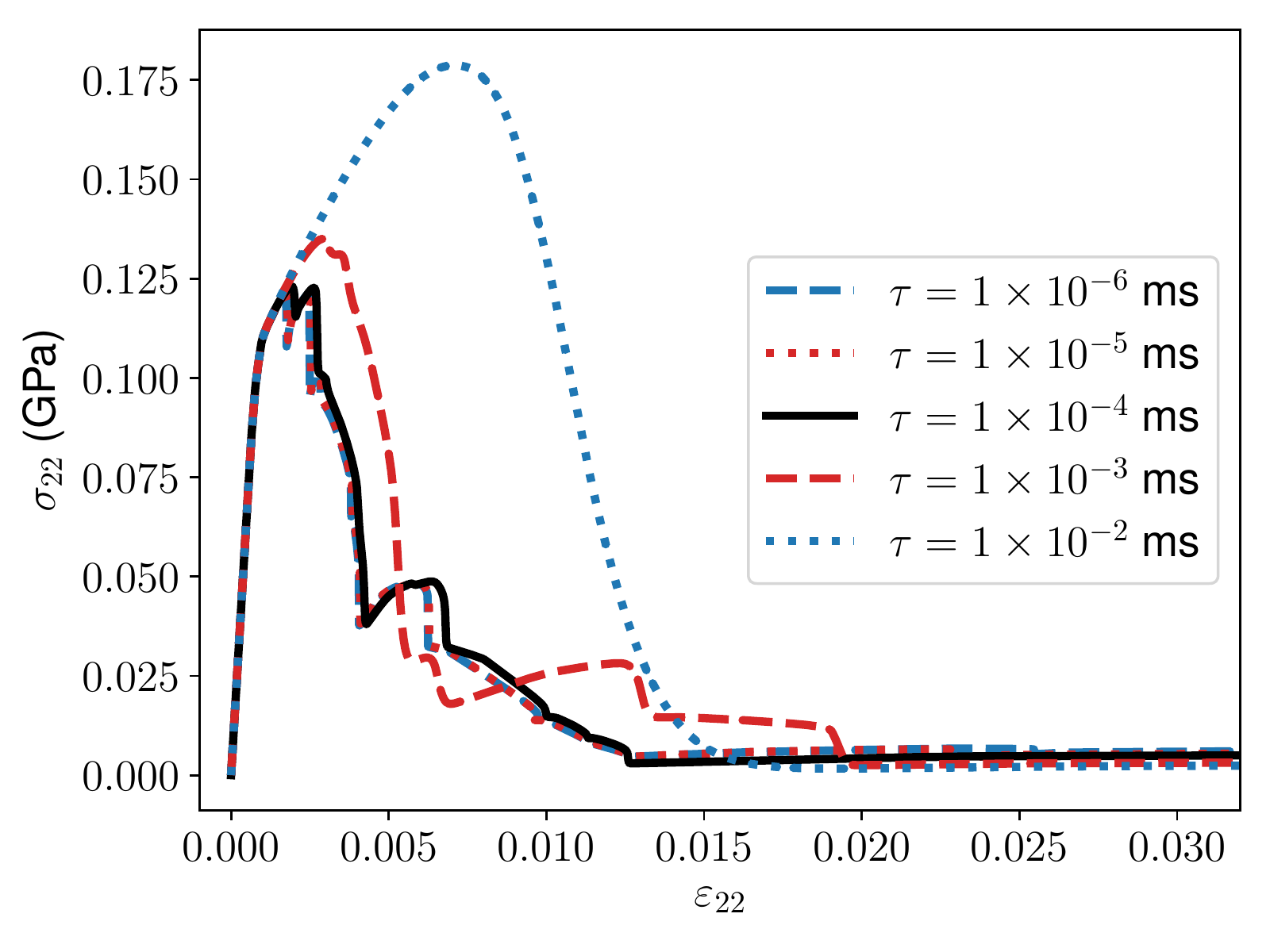}}
		\subfigure[Number of DMN layers $N$.]{\includegraphics[clip=true,trim = 0.0cm 0.0cm -0.5cm 0.0cm,width=0.44\textwidth]{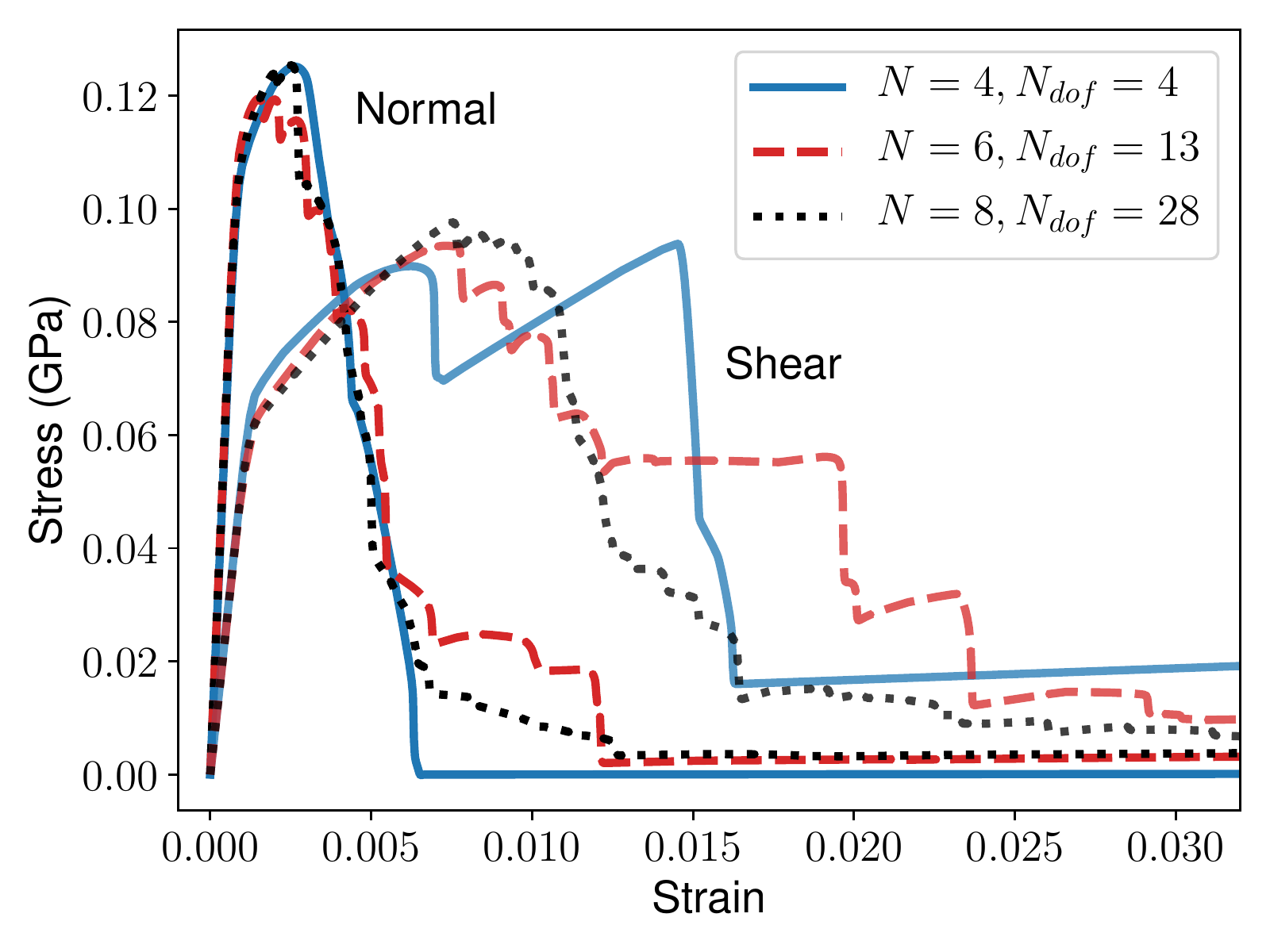}}
		\caption{Effects of model parameters to be selected by modelers. In (b), each stress-strain curve is averaged from all three uniaxial tension (or shear) loadings. }
		\label{fig:hyperparameter}
	\end{figure}
	
	Figure \ref{fig:hyperparameter} (a) shows $\sigma_{22}-\varepsilon_{22}$ curves for $\tau$ ranging from $1.0\times 10^{-6}$ to $1.0\times 10^{-2}$ ms. For the prescribed strain rate $\dot{\varepsilon}_{22} = 1.0 \text{ ms}^{-1}$, we observe little viscous effect when $\tau$ is less than $1.0\times 10^{-4}$ ms. However, smaller $\tau$ sacrifices the efficiency due to more time-step refinements in the implicit analysis. As $\tau$ increases to $1.0\times 10^{-2}$ ms, the onset of softening along the hardening curve is greatly delayed, and more strain energy is dissipated through the artificial viscosity. Although the macroscale strain rate we applied in this study seems very high at first glance, it can often appear in the strain localization region, especially when the macro length scale $h$ (e.g., element size) is tiny.
	
	Moreover, the number of DMN layers $N$, also referred to as the ``network depth", is an important hyper-parameter that controls the network's complexity. Figure \ref{fig:hyperparameter} (b) provides the stress-strain curves under tension and shear loadings for $N=4$, 6, and 8. Since the network with $N=4$ only has 4 DOF (see Figure \ref{fig:geo1}), we observe that its responses differ a lot in three orthogonal normal or shear loadings, while the network with $N=8$ shows much more isotropic failure behaviors as we can see from Figure \ref{fig:ex1-1}. To better demonstrate the results, we only plot the average stress-strain curves. Overall, the maximum stress and softening stiffness predictions are consistent across different network depths, while deeper networks with more DOF yield smoother responses. 
	
		\begin{figure} [!t]
		\centering
		\graphicspath{{Figures/}}
		\subfigure[Uniaxial tension $\sigma_{22}-\varepsilon_{22}$.]{\includegraphics[clip=true,trim = 0.0cm 0.0cm -0.5cm 0.0cm,width=0.44\textwidth]{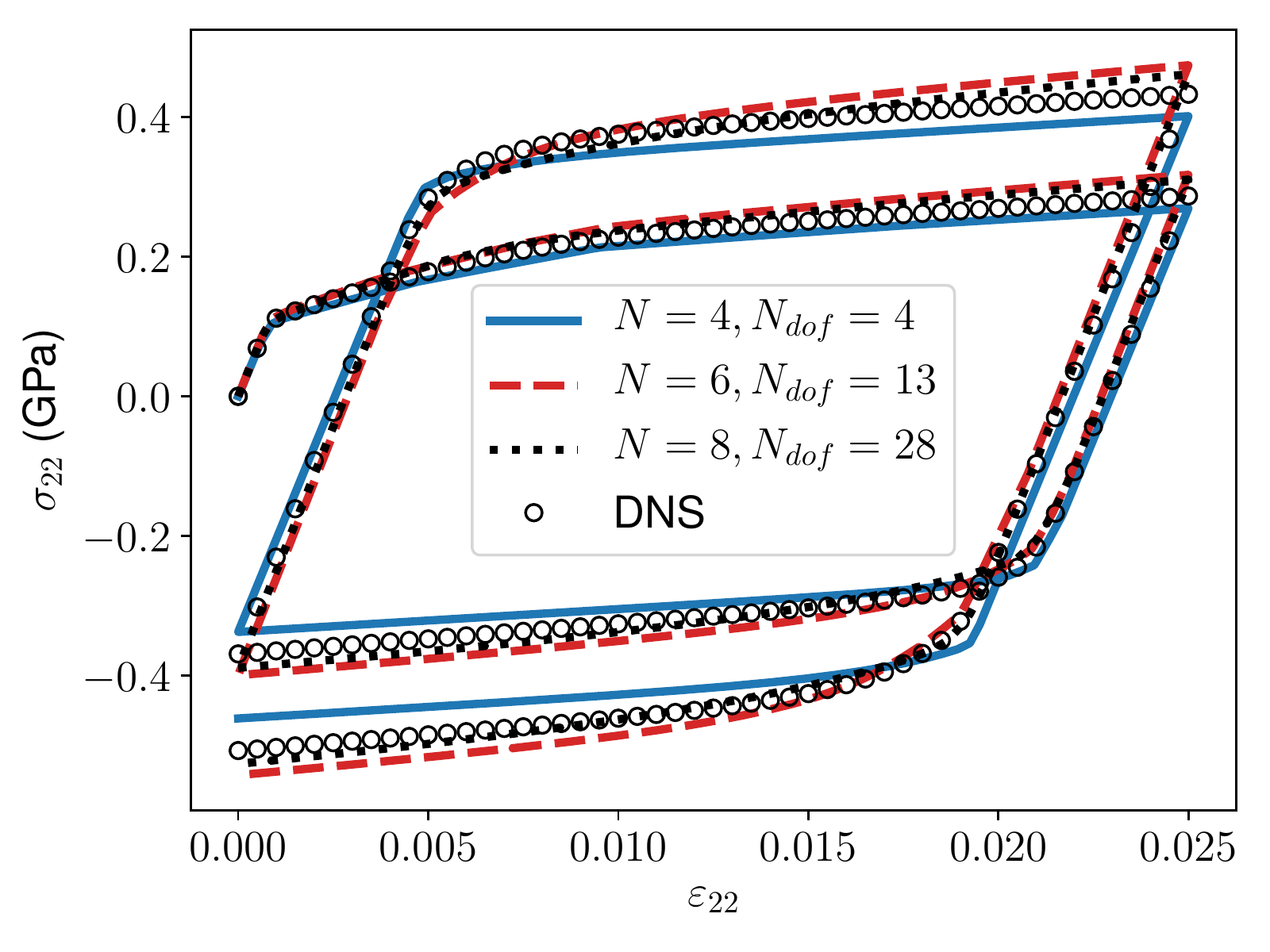}}
		\subfigure[Shear $\sigma_{12}-\gamma_{12}$.]{\includegraphics[clip=true,trim = 0.0cm 0.0cm -0.5cm 0.0cm,width=0.44\textwidth]{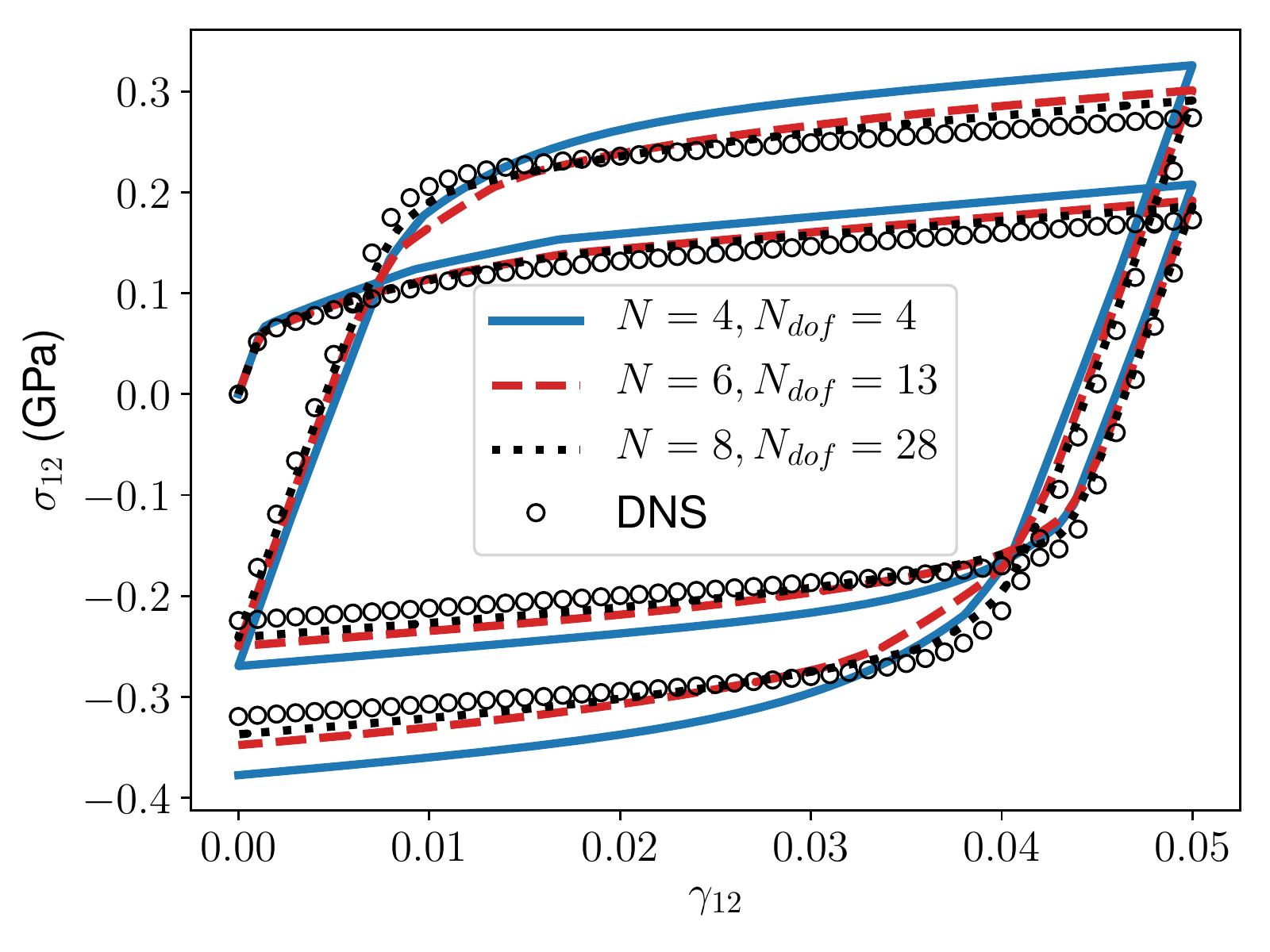}}
		\caption{Results of the particle-reinforced composite without matrix failure. The network depths are $N = 4$, 6, and 8. The DNS results are denoted by circles.}
		\label{fig:plas}
	\end{figure}
	
	DNS results of RVE models are available for the nonlinear elastoplastic composite without matrix failure, so we will use them to validate the accuracy of DMN. Figure \ref{fig:plas} shows the stress-strain curves from DMN and DNS under cyclic uniaxial tension and shear loadings. Good convergence of DMN to DNS results is observed. For $N=6$ and $N=8$, the networks can predict the hardening behaviors very well for both loading cases, although only linear elastic data are used for offline training.
	
	\subsection{Dynamic crush of a composite tube}\label{sec:crushtube}
\begin{figure}[!t]
		\centering
		\graphicspath{{Figures/}}
		\includegraphics[clip=true,trim = 3cm 2cm 2cm 2cm,width=0.9\textwidth]{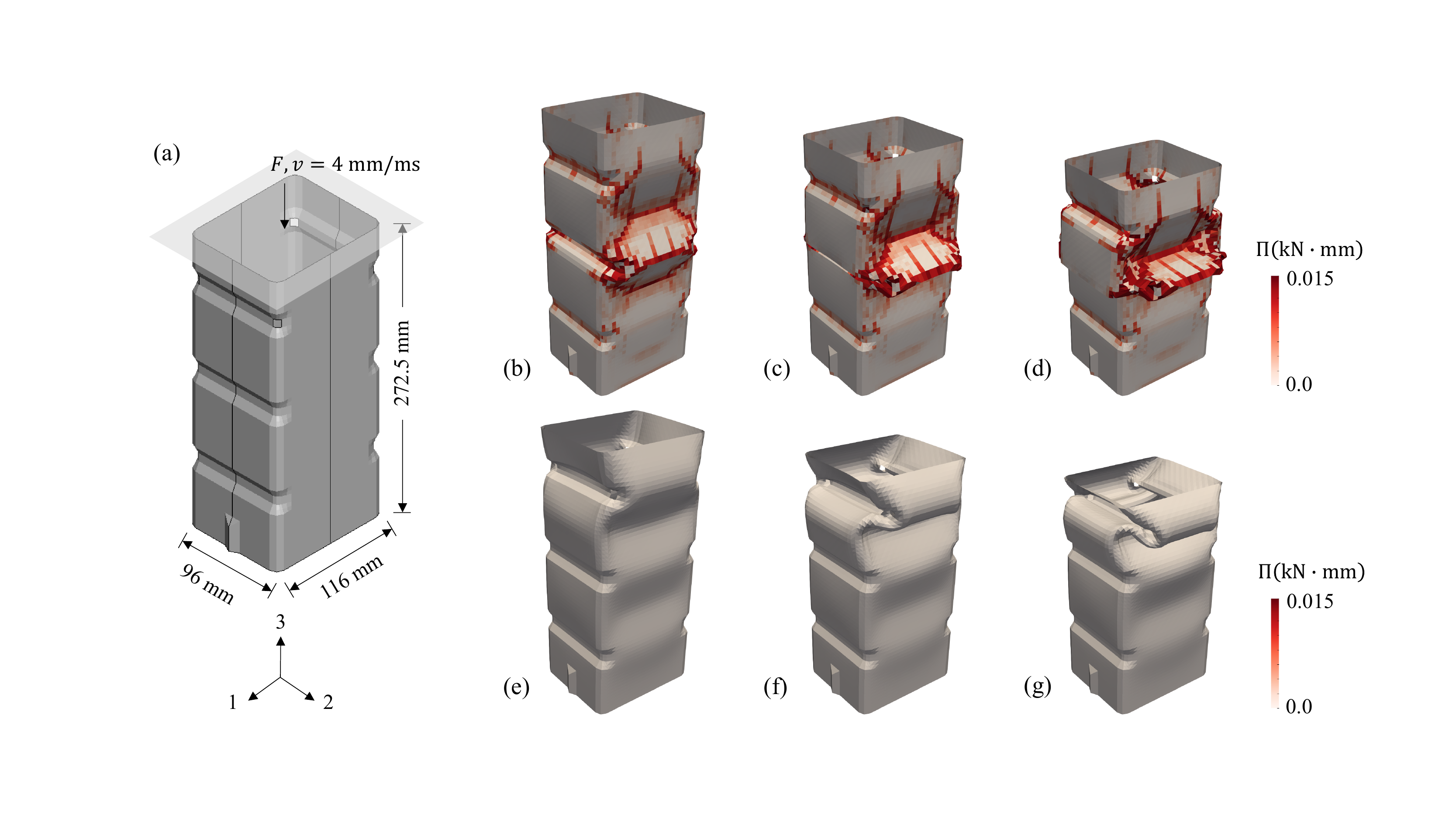}
		\caption{The crush tube modeled by the particle-reinforced composite. (a) The geometry of the tube. The rigid wall is moving downwards at 4 mm/ms. Only a quarter of the tube is simulated due to symmetry. (b-d) Snapshots of the tube with matrix failure at $T=3$, 9, 15 ms.  (e-f) Snapshots of the tube without matrix failure at $T=3$, 9, 15 ms. The mesh size of the tube is 4.16 mm. The deformed plots are colored by the released energy $\Pi$ of each element.}
		\label{fig:example2}
	\end{figure}

	In this section, we apply the microscale DMNs of the particle reinforced composite to concurrent multiscale simulations. In the macroscale, a symmetric crush tube is impacted by a moving wall on the top surface. The tube dimensions are provided in Figure \ref{fig:example2} (a), and the thickness of the tube is 2 mm. Due to the model symmetry, only a quarter of the tube is simulated. The velocity of the rigid wall is $v=4$ mm/ms. 
	
	The macroscale crush tube is meshed by the Belytschko-Tsay shell elements with thickness stretch, and 3 integration points are used across the thickness direction for each element. For the mesh refinement study, three element sizes are investigated: 8.32 mm, 4.16 mm, and 2.08 mm. Accordingly, the numbers of shell elements are 467, 1866, and 7464, respectively.  As the tube is mainly under compressive loads, no element deletion will be triggered to sustain physical self-contacts in the model.
	
	Each integration/material point in the crush tube is coupled to a DMN with $N=8$ and $N_{dof}=28$. The overall density of the DMN material is  $\rho = 5\times10^{-6} \text{ kg}/\text{mm}^3$. Other material parameters of the particle and matrix phases can be found in Table \ref{table:para1}. As non-local or gradient-based regularization is not used, we treat the element size as the in-plane macro length scale. Moreover, the material axis 3 shown in Figure \ref{fig:geo1} (a) is always normal to the shell plane, and the out-of-plane macro length scale is defined by the shell thickness. Given the thickness equal to 2 mm, the macro scale tensor $\textbf{A}^{macro}$ of each integration point is
	\begin{equation}
	\textbf{A}^{macro} = \begin{bmatrix}4/h^2&&\\&4/h^2&\\&&1.0\end{bmatrix} \text{mm}^{-2}.
	\end{equation}
	
	Figure \ref{fig:example2} (b-d) show contour plots of the released energy on the deformed tube at $T=3$, 9, and 15 ms, The definition of the released energy in DMN is given in Eq. (\ref{eq:pi}), while the element-wise value is averaged over all three integration points. The mesh size of the tube model is 4.16 mm. As we can see from the snapshots, a hexagonal crack pattern is formed under the impact, and the failures mostly appear in the middle section. We also simulate an identical tube model except no matrix failure is considered in the composite. The snapshots of the deformed tube at $T=3$, 9, and 15 ms are provided in Figure \ref{fig:example2} (e-g). As no crack surface is activated, the released energy is always zero in these plots. After the initial buckling, the plastic deformations are concentrated at the top section of the tube.
	
	\begin{figure}[!t]
		\centering
		\graphicspath{{Figures/}}
		\includegraphics[clip=true,trim = 3cm 0.5cm 2cm 0.0cm,width=0.8\textwidth]{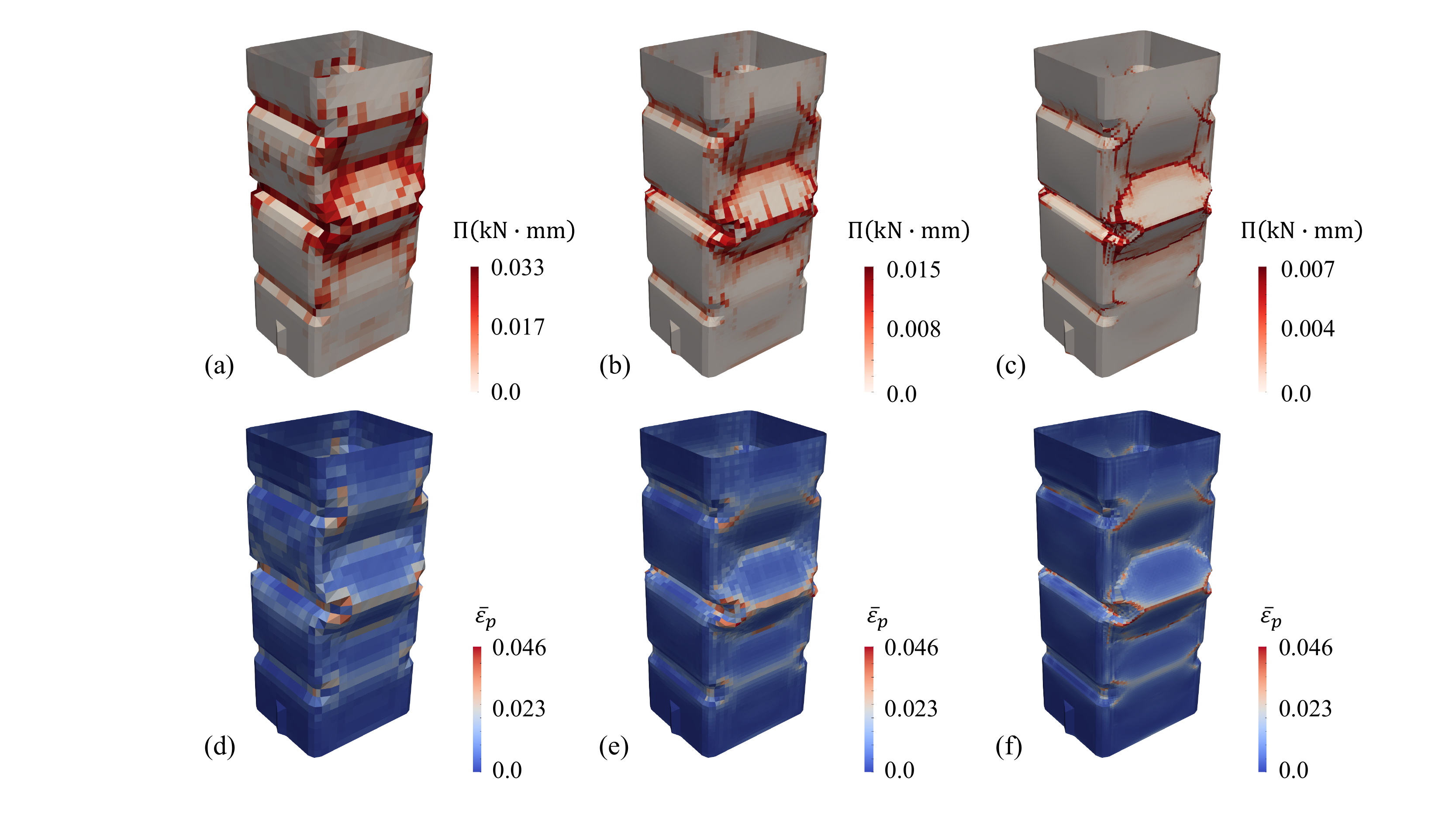}
		\caption{Snapshots at $T=3$ ms of failed tubes with different macroscale mesh sizes. (a-c) Contour plots of the released energy $\Pi$ for mesh sizes 8.32 mm, 4.16 mm, and 2.08 mm. (d-f) Contour plots of the average effective plastic strain at the inner surface for mesh sizes 8.32 mm, 4.16 mm, and 2.08 mm.}
		\label{fig:e2-length}
	\end{figure}
	
	Furthermore, we perform the mesh refinement study on the crush tube. First, the contour plots of the released energy are shown in Figure \ref{fig:e2-length} (a-c) for mesh sizes 8.32 mm, 4.16 mm, and 2.08 mm, respectively. The crack patterns predicted by the three models are similar in terms of the overall shape but differ slightly in some local regions. One may also observe that the magnitude of the released energy is nearly proportional to the mesh size. In contrast, the released energy of the single material point shown in Figure \ref{fig:length} (b) is proportional to the square of the macro length scale $h$. This is because the DMN macro-cell of the shell element is anisotropic, with its dimensions changing only in the shell plane upon mesh refinement.
	
	The average effective plastic strain $\bar{\varepsilon}_p$ is another physical quantity to characterize the overall material state. It is defined as
	\begin{equation}\label{eq:aplas}
	\bar{\varepsilon}_p = \sum_{i=1}^{N_{dof}} \varepsilon_p^i f^i,
	\end{equation}
	where $\varepsilon_p^i$ is the effective plastic strain of the $i$-th DOF, and $f^i$ is its volume fraction in the network. Typically, $\varepsilon_p$ vanishes if the phase is linear elastic. The contour plots of $\bar{\varepsilon}_p$ on the inner surface for different element sizes are provided in Figure \ref{fig:e2-length} (d-f). More than the plastic deformation patterns, the magnitudes of the average effective plastic strain are around the same, indicating that the onset of failure is not sensitive to the mesh size.
	
	\begin{figure} [!t]
		\centering
		\graphicspath{{Figures/}}
		\subfigure[Composite tube with matrix failure.]{\includegraphics[clip=true,trim = 0.0cm 0.0cm -0.5cm 0.0cm,width=0.44\textwidth]{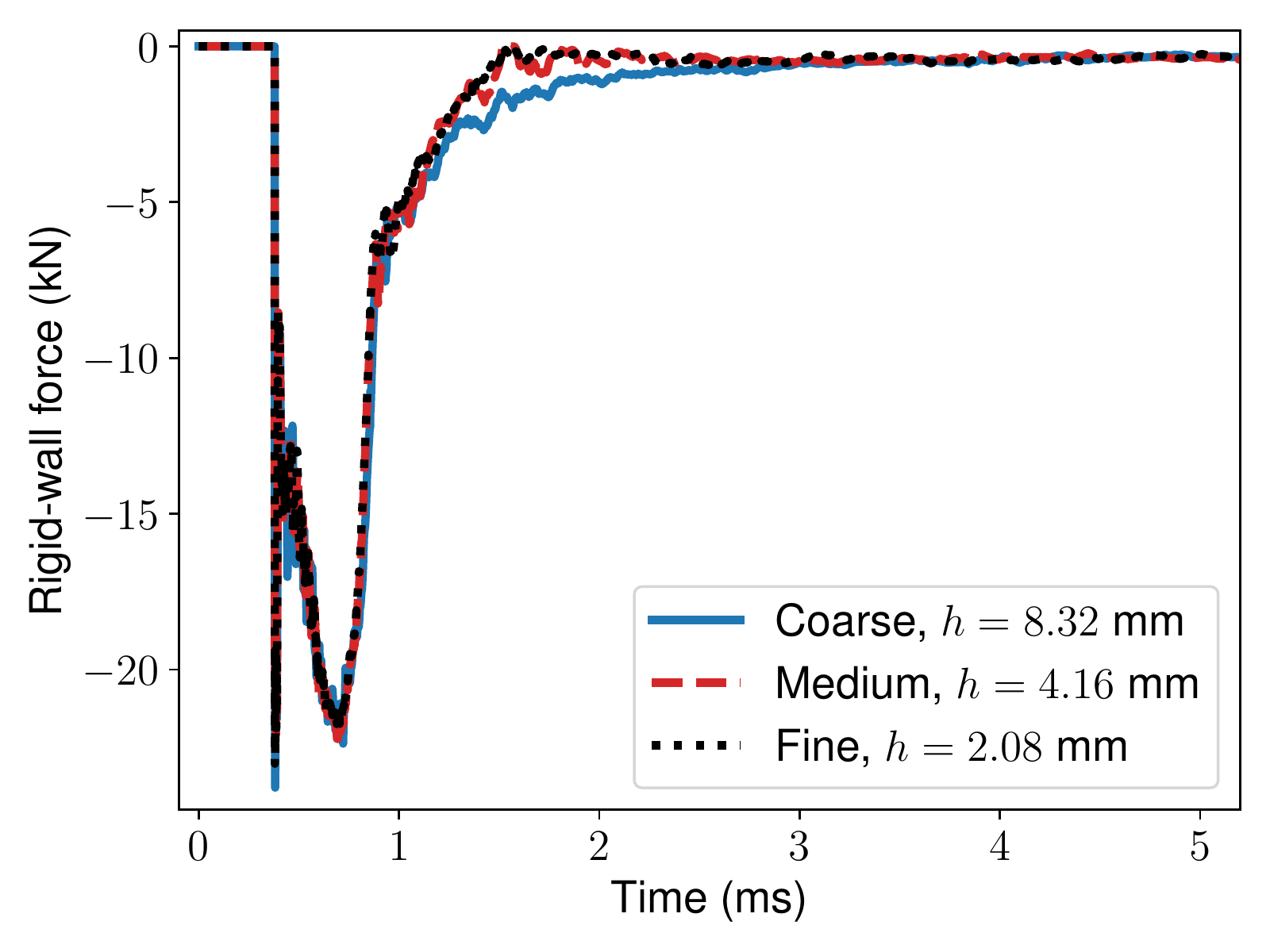}}
		\subfigure[Composite tube without matrix failure.]{\includegraphics[clip=true,trim = 0.0cm 0.0cm -0.5cm 0.0cm,width=0.44\textwidth]{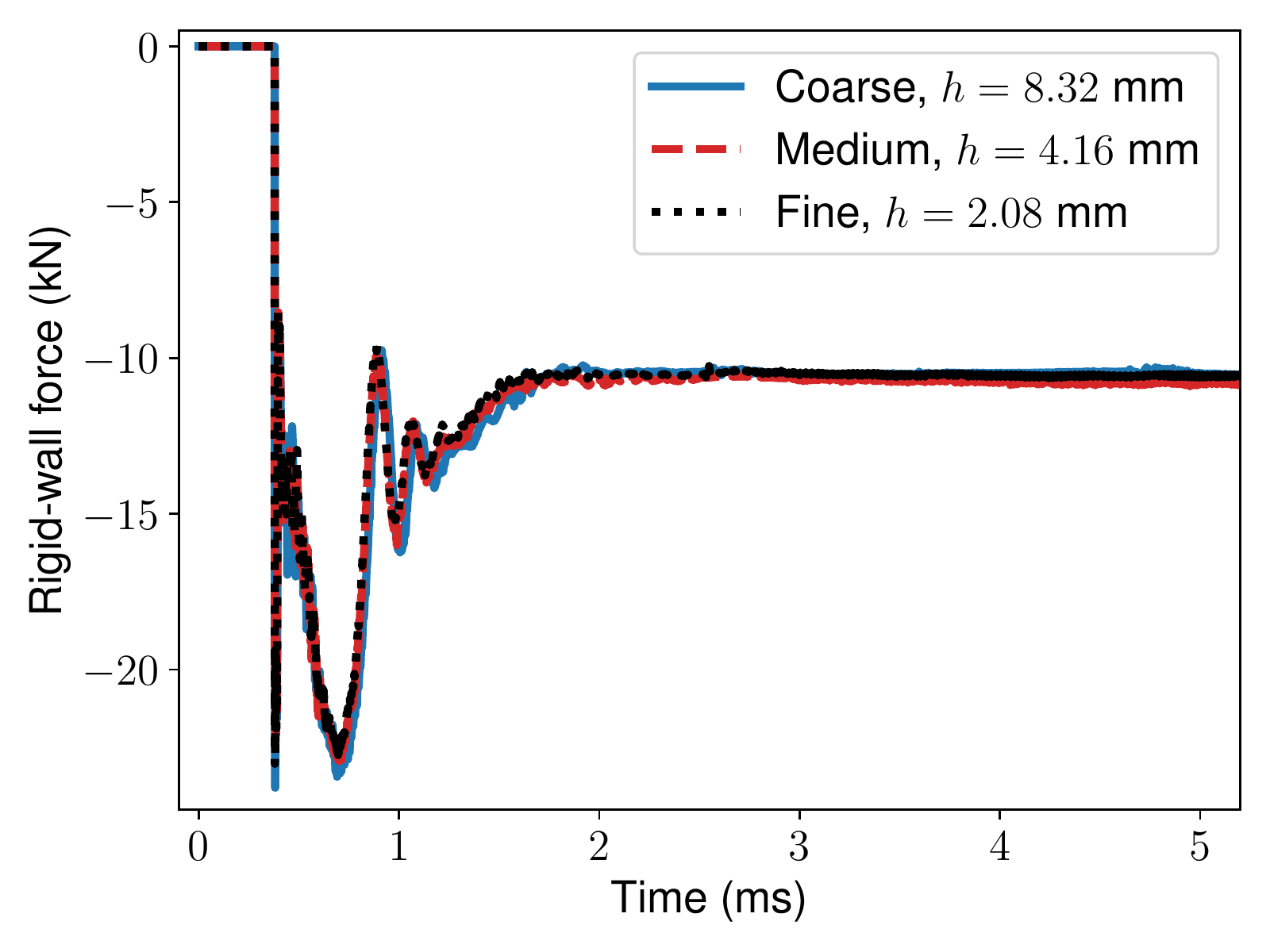}}
		\caption{Histories of the rigid-wall force for mesh sizes 8.32 mm (Coarse), 4.16 mm (Medium), and 2.08 mm (Fine).}
		\label{fig:ex2-force}
	\end{figure}
	
	Finally, we summarize the histories of rigid-wall force with and without matrix failure in Figure \ref{fig:ex2-force}. The first two spikes appear similarly in the two cases, which results from the initial elastic buckling of the tube. Afterward, the force on the composite tube with matrix failure drops to zero, while the one without matrix failure stables at a finite value around -11 kN. Regarding the convergence on the mesh size, the medium and fine meshes predict very close results for the first case. While for the second case with only matrix plasticity, all three meshes output similar force history curves.
	
	\subsection{Three-point bending tests with microstructure anisotropy}\label{sec:3point}
	From this section, we switch the focus to the unidirectional-fiber microstructure shown in Figure \ref{fig:geo2} (a) for modeling carbon fiber reinforced polymer composites. The volume fraction of the fiber phase is 50\%. The number of layers is $N=8$. After the offline training, 31 active DOF remains in the network, and 8 of them belong to the fiber phase. The carbon fibers are modeled by an orthotropic linear elastic material with large stiffness in the fiber direction, and no fiber failure is considered in this paper. The epoxy matrix is modeled by isotropic von Mises plasticity with an exponential hardening law,
	\begin{equation}
	\sigma^Y({\varepsilon}_{p}) = (\sigma^y-\sigma^u)\exp(-a{\varepsilon}_{p})+ E^h{\varepsilon}_{p} + \sigma^u,
	\end{equation}
	where $\sigma^y$ represents the yielding strength, $\sigma^u$ is the ultimate yield stress for large effective plastic strain, $E^h$ is a linear hardening stiffness, and $a$ is a dimensionless constant. The carbon fibers and epoxy matrix are assumed to be perfectly bonded. The overall density of the UD composite is set to
	\begin{equation}
	\rho = 1.6\times10^{-6} \text{ kg}/\text{mm}^3.
	\end{equation}
	All the microscale material parameters are provided in Table \ref{table:udpara}. Note that the elastic constants are measured directly from experiments on single-phase materials. The hardening parameters $(\sigma^y,\sigma^u, E^h, a)$ are fitted based on the uniaxial tension and shear tests on the epoxy matrix. The critical energy release rate is set to $3\times10^{-4}$ GPa$\cdot$mm. Specifically, the critical effective traction $t_c$ of the matrix phase is determined inversely by matching the transverse tension curve of the UD composite. 
	
	\begin{table}[htb!]
		\captionabove{Material parameters in the unidirectional-fiber microstructure for the CFRP composite. No fiber failure is considered in this work. The critical effective traction $t_c$ is calibrated inversely by the transverse tension data of the UD composite.} % title of Table
		\centering % used for centering table
		\label{table:udpara} % is used to refer this table in the text
		{\tabulinesep=1.0mm
			\begin{tabu}{c c c c c c c} % centered columns (4 columns)
				\hline
				\multirow{4}{*}{Carbon fiber}& $E_1$ (GPa) & $E_2$ (GPa) & $E_3$ (GPa) & $G_{12}$ (GPa) & $G_{13}$ (GPa)&$G_{23}$ (GPa)  \\
				& 245.0 & 19.8 & 19.8 & 29.2 & 29.2&5.9 \\ 
				\cline{2-7}
				&$\nu_{21}$ & $\nu_{31}$ & $\nu_{32}$ & &&\\
				&0.023 & 0.023 & 0.67 &  &&\\
				\hline
				\multirow{4}{*}{Epoxy} & $E_m$ (GPa) & $\nu_m$ & $\sigma^y$ (GPa) & $\sigma^u$ (GPa) &$E^h$ (GPa)& $a$ \\ 
				&3.8 & 0.387 & 0.025 & 0.115 & 0.01&140 \\ 
				\cline{2-7}
				&$t_c$ (GPa) & $G_c$ (GPa$\cdot$mm)& $\beta$ & $\tau$ (ms) & \\ 
				&$0.10$& $3\times10^{-4}$ & 1.0 & $1\times10^{-3}$ &\\
				\hline
				%inserts single line
		\end{tabu}}
	\end{table}
	
\begin{figure} [!t]
		\centering
		\graphicspath{{Figures/}}
		\subfigure[Transverse tension with different $h$.]{\includegraphics[clip=true,trim = 0.0cm 0.0cm -0.5cm 0.0cm,width=0.44\textwidth]{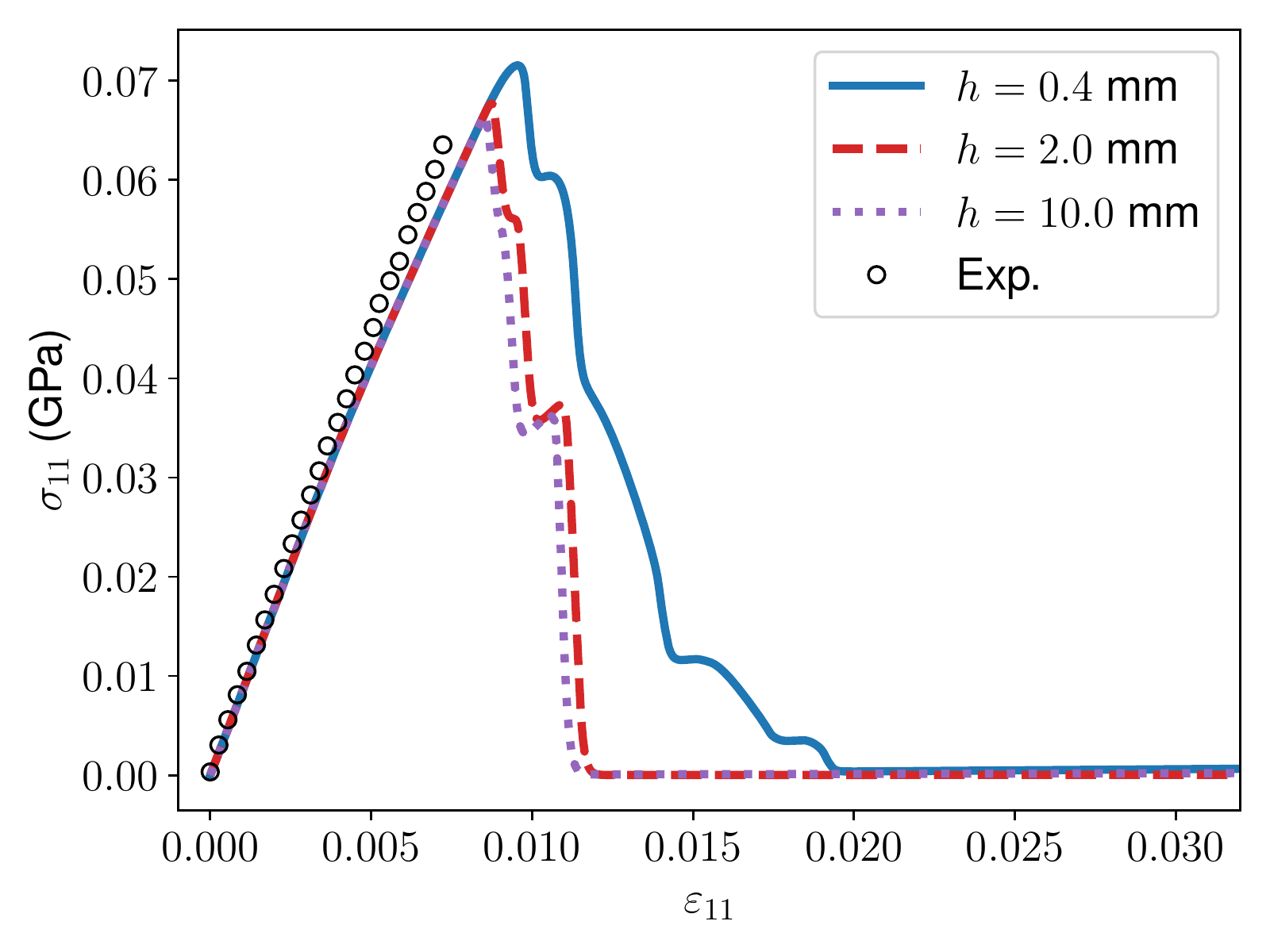}}
		\subfigure[Various loading directions.]{\includegraphics[clip=true,trim = 0.0cm 0.0cm -0.5cm 0.0cm,width=0.44\textwidth]{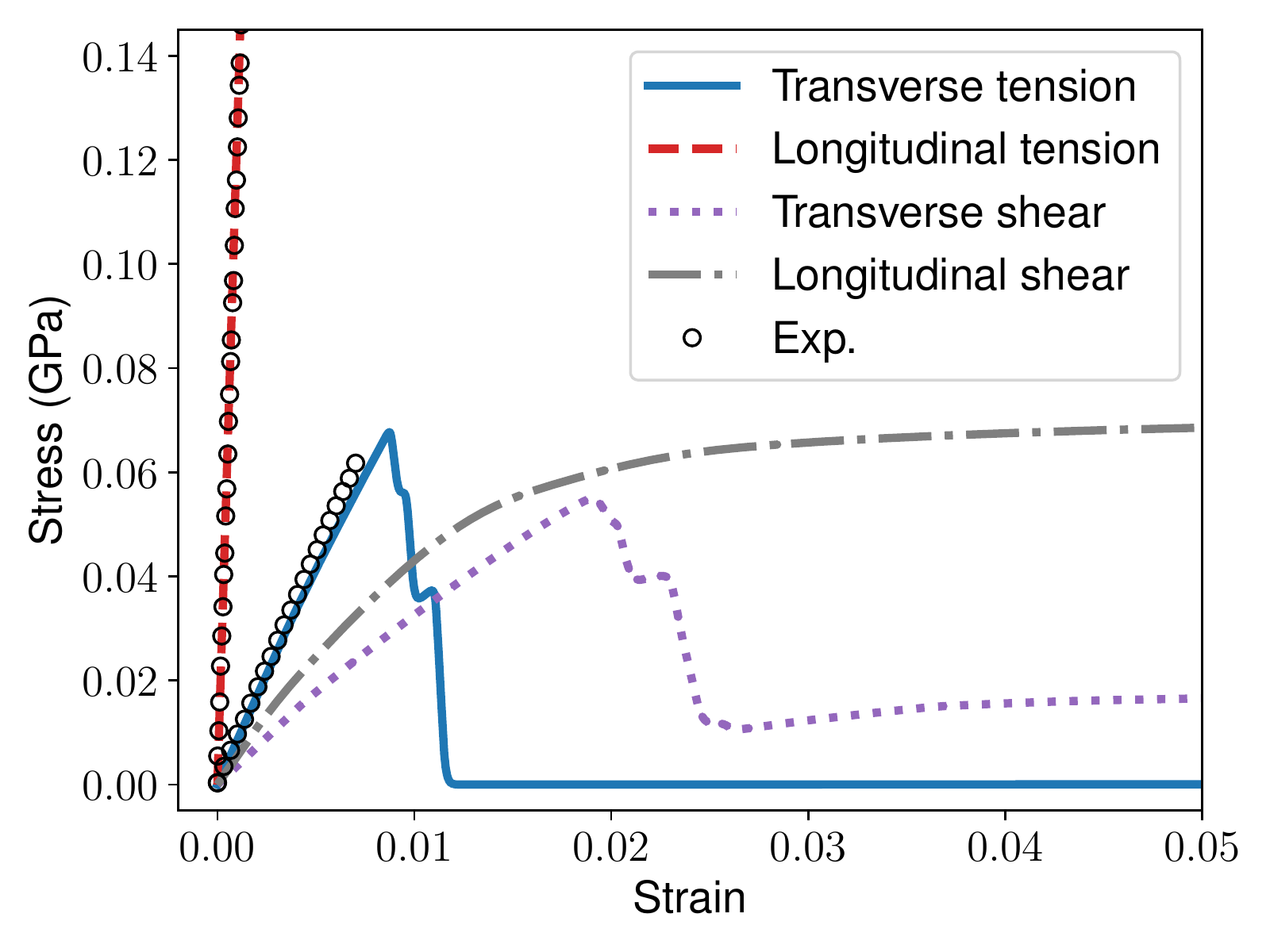}}
		\caption{Single material point tests on the 3-D unidirectional carbon fiber reinforced polymer composite. The macro scale tensor is $\textbf{A}^{macro} = 4/h^2\cdot \textbf{I}$. In (b), the macro length scale $h$ is 2.0 mm. Experimental results are plotted as circles ($\circ$).}
		\label{fig:ex3_calibration}
	\end{figure}

	We first evaluate the responses of a single material point. Similarly, its macro-scale tensor is assumed to be $\textbf{A}^{macro} = (4/h^2) \textbf{I}$, representing an isotropic sphere in space with diameter equal to $2h$. Figure \ref{fig:ex3_calibration} (a) shows the stress-strain curves under transverse (in-plane) tension for $h$ ranging from 0.4 mm to 10.0 mm. As expected, the magnitude of the softening stiffness increases with $h$, while the strength predictions are all at the same level, around 0.065 GPa. 
	
	The anisotropy of material responses induced by the microstructure is demonstrated in Figure \ref{fig:ex3_calibration} (b). Experimental data for longitudinal and transverse tension loadings are obtained from $0^\circ$ and $90^\circ$ coupon tests, respectively. Without any parameter fitting, the DMN predictions of the elastic stiffness match the experimental results very well. By adjusting the critical effective traction $t_c$, the UD composite's failure behaviors under transverse tension can also be properly captured. In Section \ref{sec:coupon}, the calibrated $t_c$ will be further validated by the $10^\circ$ coupon experiment. Another interesting finding is that the composite encounters  negligible failure under the longitudinal shear loading for the strain less than 0.05, because its deformation is shear-dominated and the matrix phase has less stress concentration comparing to other loading directions.
	
\begin{figure}[!t]
		\centering
		\graphicspath{{Figures/}}
		\includegraphics[clip=true,trim = 5cm 5cm 5cm 4cm,width=0.88\textwidth]{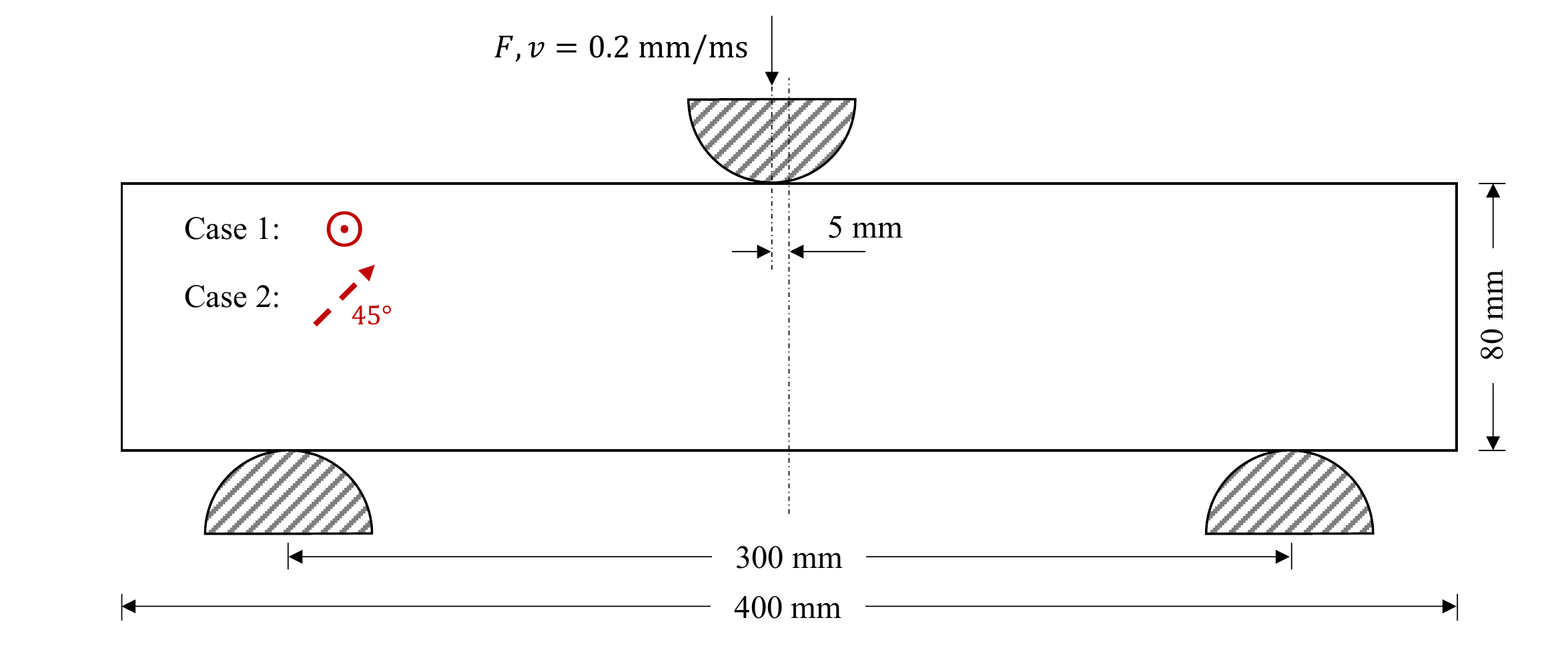}
		\caption{Illustration of the macroscale three-point bending model under 2-D plane-strain condition. Two cases with 1) out-of-plane and 2) in-plane $45^\circ$ fiber orientations are considered.}
		\label{fig:example3}
	\end{figure}

	We then apply the DMN to concurrent multiscale simulations for three-point bending tests of the UD composite. The geometry and boundary conditions are illustrated in Figure \ref{fig:example3}. No crack zone is predefined in the macroscale model so that damage initiations are handled inside the DMN. Although the microscale DMN model is still 3-dimensional, 2-D plane-strain conditions are considered in the macroscale model. The loading head moves downwards at a speed of $v=0.2$ mm/ms, and it is shifted left from the center by 5 mm to break the symmetry. Additionally, the head and the supports interact with the block through surface contacts, with the friction coefficient equal to 0.02.
	
	To demonstrate the effects of material anisotropy, we evaluate two cases with different fiber orientations in our study, as shown in Figure \ref{fig:example3}. Case 1 has fibers in the out-of-plane direction so that the microscale UD composite is dominated by transverse deformation. The fibers in Case 2 are aligned in the plane at a $45^\circ$ angle to the horizontal axis. Mesh refinement study is performed for each case. The coarse, medium and fine mesh sizes are 5, 2.5, and 1.25 mm, respectively. The corresponding numbers of 2-D plane-strain finite elements are 1280, 5120, and 20480, and each element has four integration points. No nonlocal regularization is applied here, so the macroscale softening zone due to matrix failure will localize in one layer of elements. Therefore, we set the in-plane macro length scale $h$ equal to the mesh size. Given the section thickness of 2 mm, the macro scale tensor of 3-D DMN can be written as
	\begin{equation}
	\textbf{A}^{macro} = \begin{bmatrix}4/h^2&&\\&4/h^2&\\&&1.0\end{bmatrix} \text{mm}^{-2}.
	\end{equation}
	\begin{remark}
		Under the 2-D plane strain condition, the results should be independent of the section thickness.  It requires that the crack surfaces activated in DMN are vertical to the 2-D plane in the global coordinate system, which has not been considered in this work. Under this extra constraint, the failure algorithms need to be modified since the crack surfaces are not necessarily activated in the direction with maximum effective traction of a 3-D stress state. 
	\end{remark}
	
	To avoid excessive element distortions, an element will be deleted if at least two of its integration points have the volumetric strain larger than 0.1,
	\begin{equation}
	\varepsilon_{11} + \varepsilon_{22} + \varepsilon_{33} > 0.1.
	\end{equation}
	According to our study, this is sufficient to guarantee the DMN with $h\geq 1.25$ mm to fail completely or have negligible residual stress.
	\begin{figure}[!t]
		\centering
		\graphicspath{{Figures/}}
		\subfigure[Case 1, out-of-plane fiber orientation.]{\includegraphics[clip=true,trim = 1.5cm 4.5cm 1cm 4.0cm,width=0.95\textwidth]{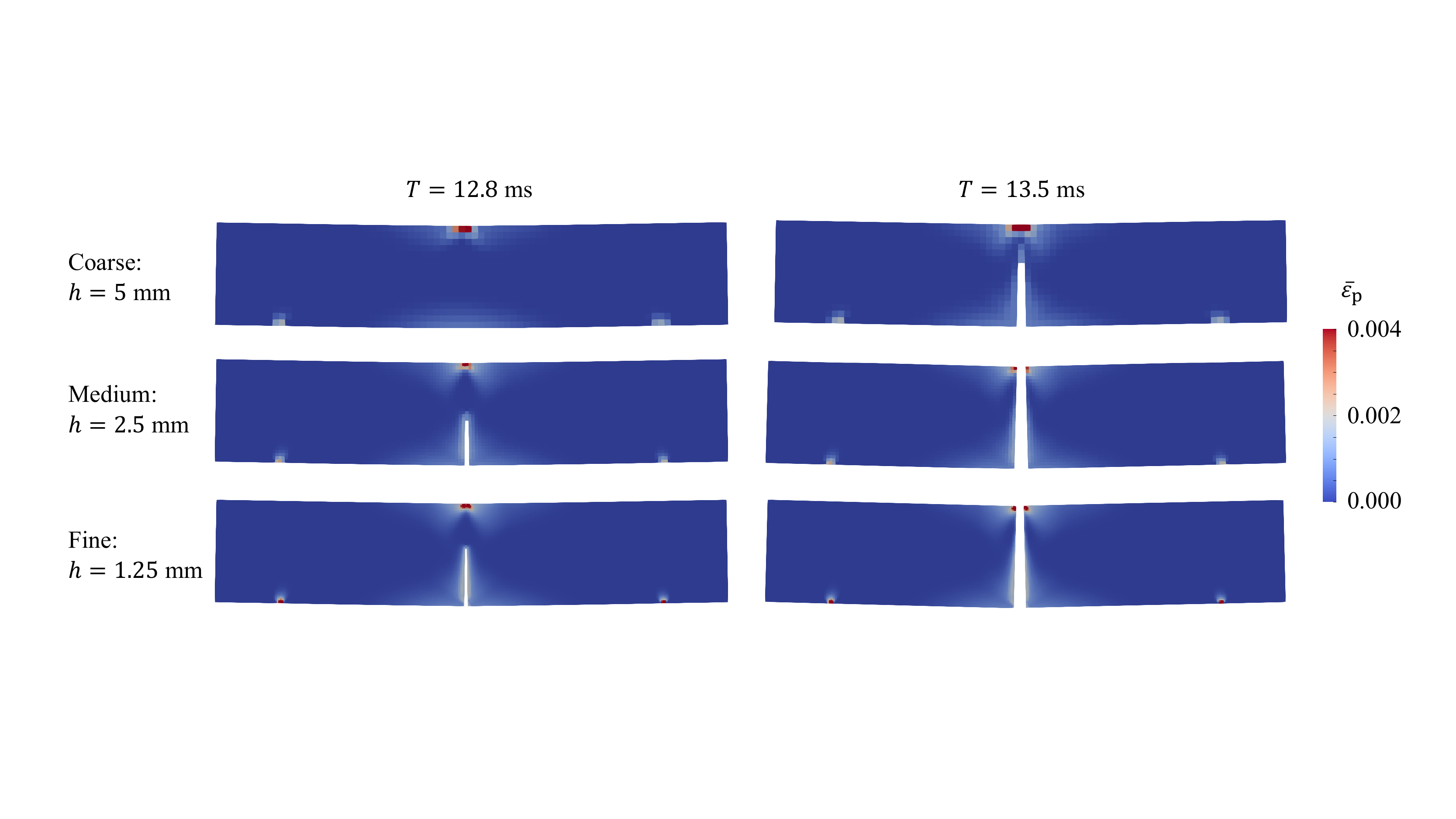}}
		\subfigure[Case 2, in-plane $45^\circ$ fiber orientation.]{\includegraphics[clip=true,trim = 1.5cm 4.5cm 1cm 4.0cm,width=0.95\textwidth]{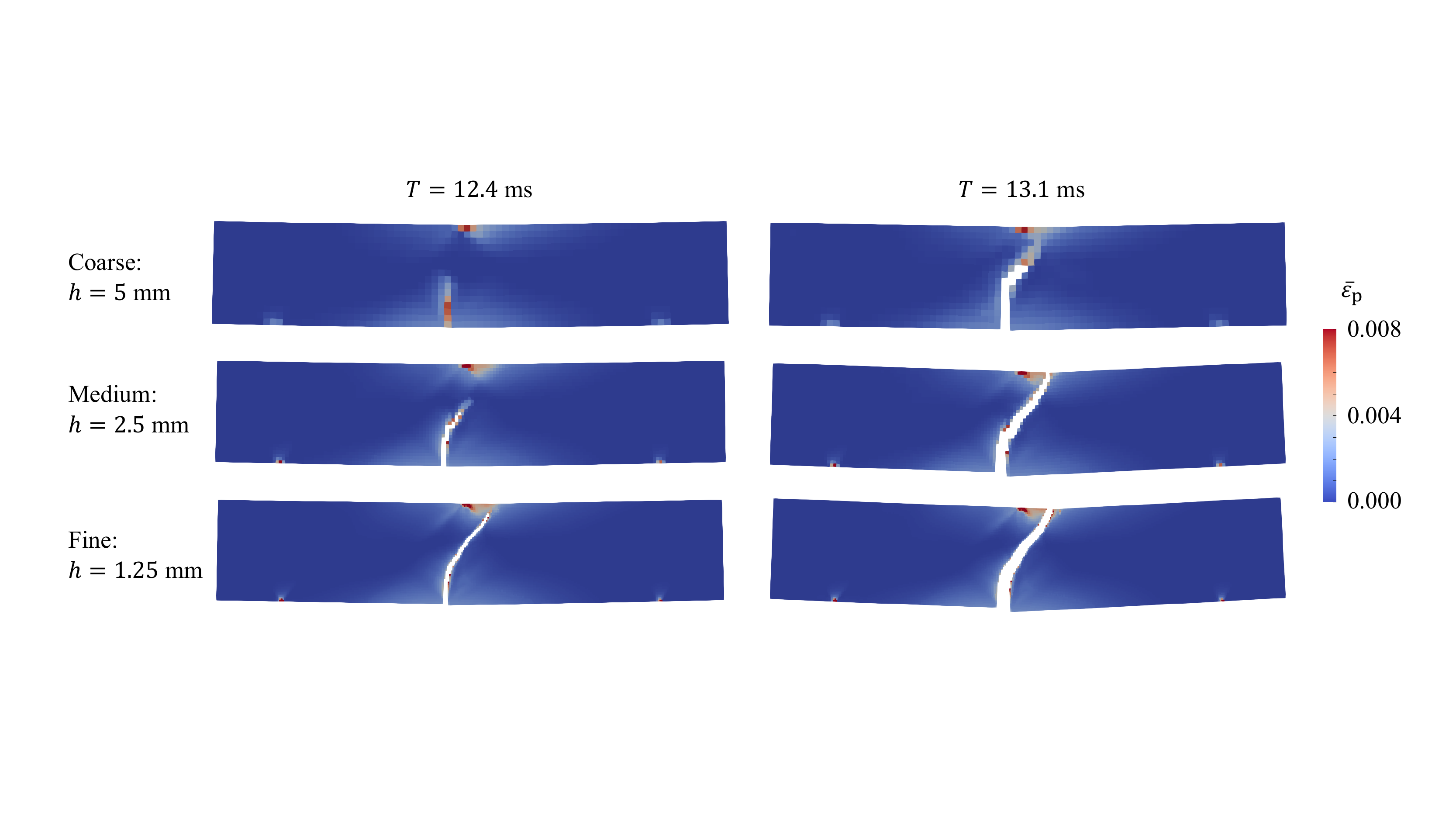}}
		\caption{Contour plots of the average effective plastic strain $\bar{\varepsilon}_p$ for different fiber orientations and mesh sizes. Element deletions are enabled in both cases.}
		\label{fig:ex3-case12}
	\end{figure}
	
	Figure \ref{fig:ex3-case12} summarizes the snapshots of the three-point bending tests for various fiber orientations and mesh sizes, rendered by the average effective plastic strain $\bar{\varepsilon}_p$ (see Eq. (\ref{eq:aplas})). For all the samples, the macroscopic crack initializes from the bottom side, despite that the average effective plastic strain is larger at the loading area on the top side. This is physically sound as the composite is stronger under compressive stress states, which is captured naturally by the multiscale DMN with cohesive layers. 
	
\begin{figure} [!t]
		\centering
		\graphicspath{{Figures/}}
		\subfigure[Case 1, out-of-plane fiber direction.]{\includegraphics[clip=true,trim = 0.0cm 0.0cm -0.5cm 0.0cm,width=0.44\textwidth]{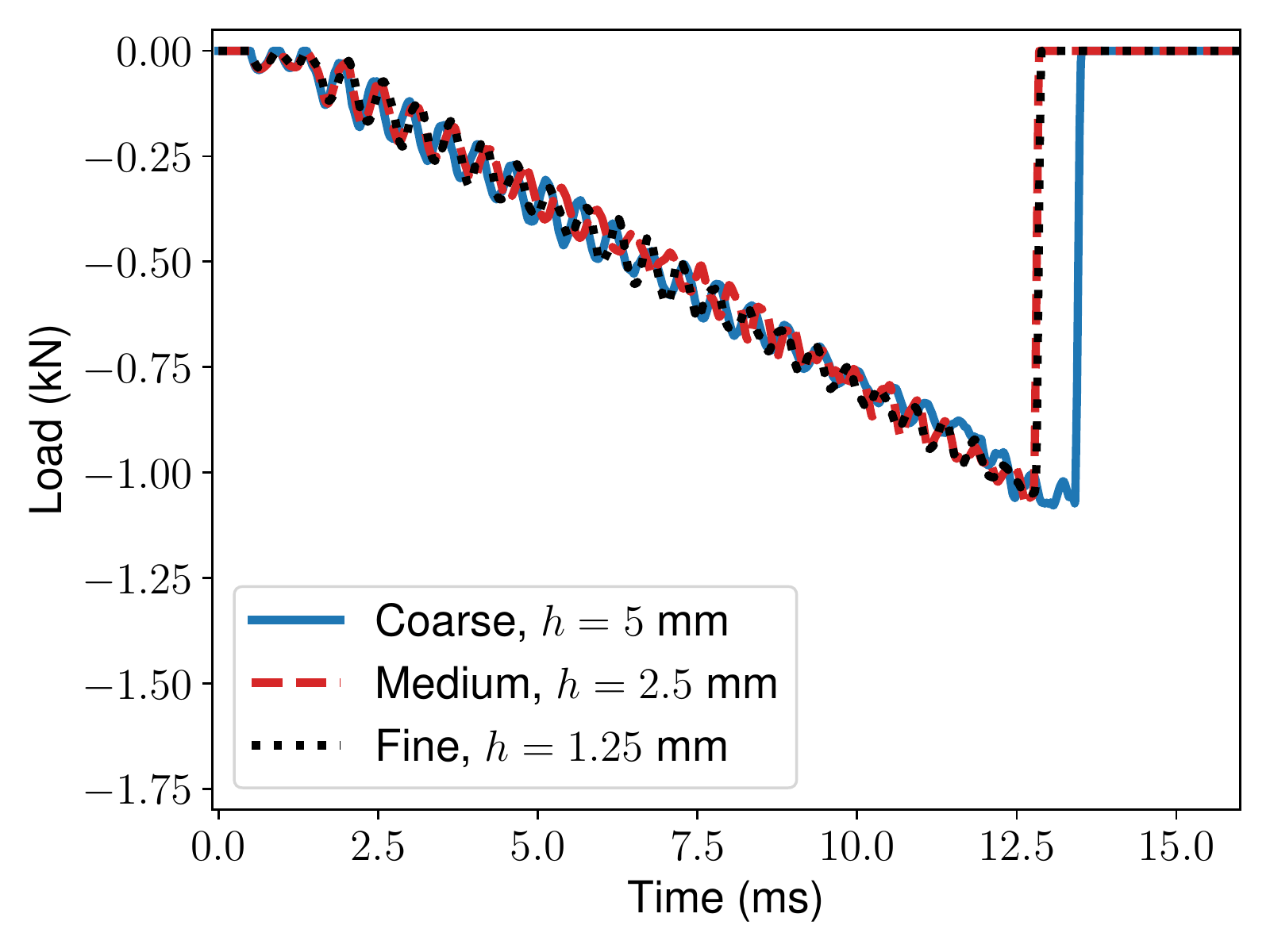}}
		\subfigure[Case 2, in-plane $45^\circ$ fiber direction.]{\includegraphics[clip=true,trim = 0.0cm 0.0cm -0.5cm 0.0cm,width=0.44\textwidth]{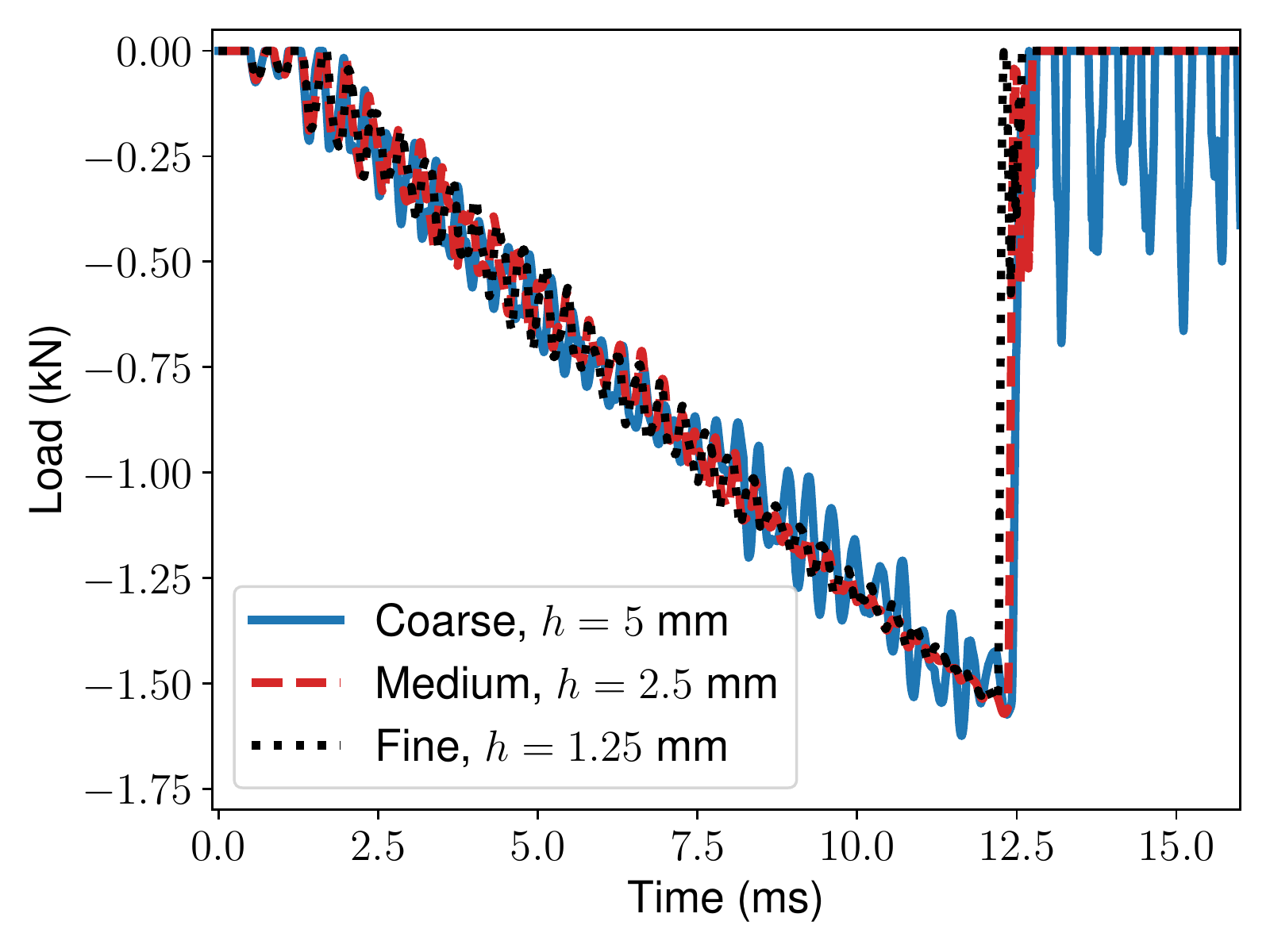}}
		\caption{Histories of the applied forces in the three-point bending tests for mesh sizes 5 mm (Coarse), 2.5 mm (Medium), and 1.25 mm (Fine). }
		\label{fig:ex3_load}
	\end{figure}

	In Case 1, the cracks propagate vertically for all the mesh sizes. At $T=13.5$ ms, the medium and fine meshes are fully separated, while the cracking process is more delayed in the coarse mesh with $h=5$ mm. In Case 2, the cracks first propagate in the vertical direction and then turn towards the fiber direction at 45$^\circ$. Different from Case 1, where the cracks are mainly triggered by transverse tension loading, Case 2 has a more complex stress state at the crack initiation point, with a combination of transverse tension and longitudinal shear loadings. One can also see from the deformed coarse mesh in Figure \ref{fig:ex3_load} (b) that shear deformations are prominent in the localization band. We did not consider the fiber failure and finite deformations in the DMNs, which could influence the Case 2 results. Overall, the predicted crack paths are consistent under mesh refinement.
	
	Figure \ref{fig:ex3_load} shows the histories of applied forces for different fiber orientations and mesh sizes. Due to the strengthening effect from the carbon fibers, the models in Case 2 are stiffer than the ones in Case 1. Slight plastic yielding can be observed in Case 2, indicating more shear deformations in the models. In both cases, the load curves achieve good convergence regarding the mesh size.
	
	\subsection{Composite coupon tests with experimental validation}\label{sec:coupon}
	\begin{figure}[!t]
		\centering
		\graphicspath{{Figures/}}
		\includegraphics[clip=true,trim = 3cm 6cm 2cm 6.8cm,width=0.98\textwidth]{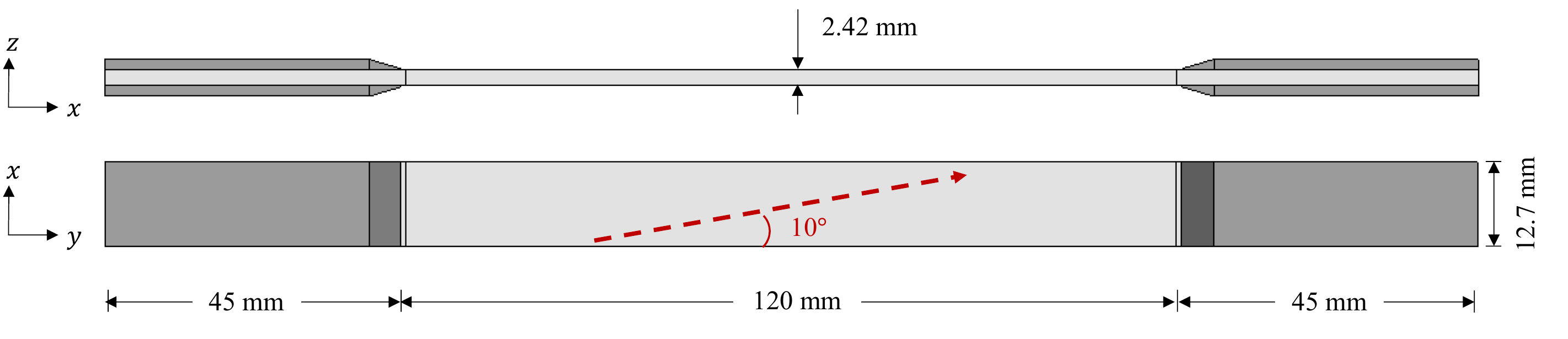}
		\caption{Illustration of $10^\circ$ off-axis tensile coupon. Axes are shown on the right of the plots. The red dashed arrow denotes the fiber direction of the composite. Only half of the model is simulated due to the symmetry of the middle $x-y$ plane.}
		\label{fig:example4}
	\end{figure}
	In the last example, we apply the DMN of the unidirectional CFRP composite to a $10^\circ$ off-axis tensile coupon test with experimental validations \cite{gao2020predictive}. All the material parameters remain the same as in Section \ref{sec:3point}. The geometry of the tensile coupon is shown in Figure \ref{fig:example4}. The length, width, and thickness of the coupon's middle measuring section are 120 mm, 12.7 mm, and 2.42 mm, respectively. In the experiment, tab sections at the two ends are tightly clamped so that no movements in the $x$ and $z$ directions are allowed. The tensile loading velocity in the $y$ direction is 0.0167 mm/s. However, to reduce the simulation time, we increase this velocity to 0.0167 mm/ms, considering that the material plasticity and crack activation models are not rate-dependent, and the velocity is still much less than the wave speed. Meantime, mass scaling is introduced so that the critical time step is no less than $1.8\times10^{-3}$ ms.
	
	Only the middle section is modeled by DMN, while the tab sections are described by an elastic material with the Young's modulus equal to 8 GPa. The middle section is meshed uniformly by linear 8-node solid elements with sizes $\Delta x=0.808$ mm, $\Delta y=0.808$ mm, and $\Delta z=0.606$ mm. The coupon is symmetric about the central $x-y$ plane, and the total number of elements in the symmetric model is 8320. Each solid element has one integration point, which is linked to a microscale DMN. Based on the mesh sizes, the macro scale tensor of DMN is defined as
	\begin{equation}
	\textbf{A}^{macro} = \begin{bmatrix}4/\Delta x^2&&\\&4/\Delta y^2&\\&&4/\Delta z^2\end{bmatrix} = \begin{bmatrix}6.127&&\\&6.127&\\&&10.892\end{bmatrix} \text{mm}^{-2}.
	\end{equation}
	Like the three-point bending tests, element deletion will be triggered if the volumetric strain at an integration point is larger than 0.1.
	
	\begin{figure} [!t]
		\centering
		\graphicspath{{Figures/}}
		\subfigure[Normal stress vs. normal strain.]{\includegraphics[clip=true,trim = 0.0cm 0.0cm -0.5cm 0.0cm,width=0.44\textwidth]{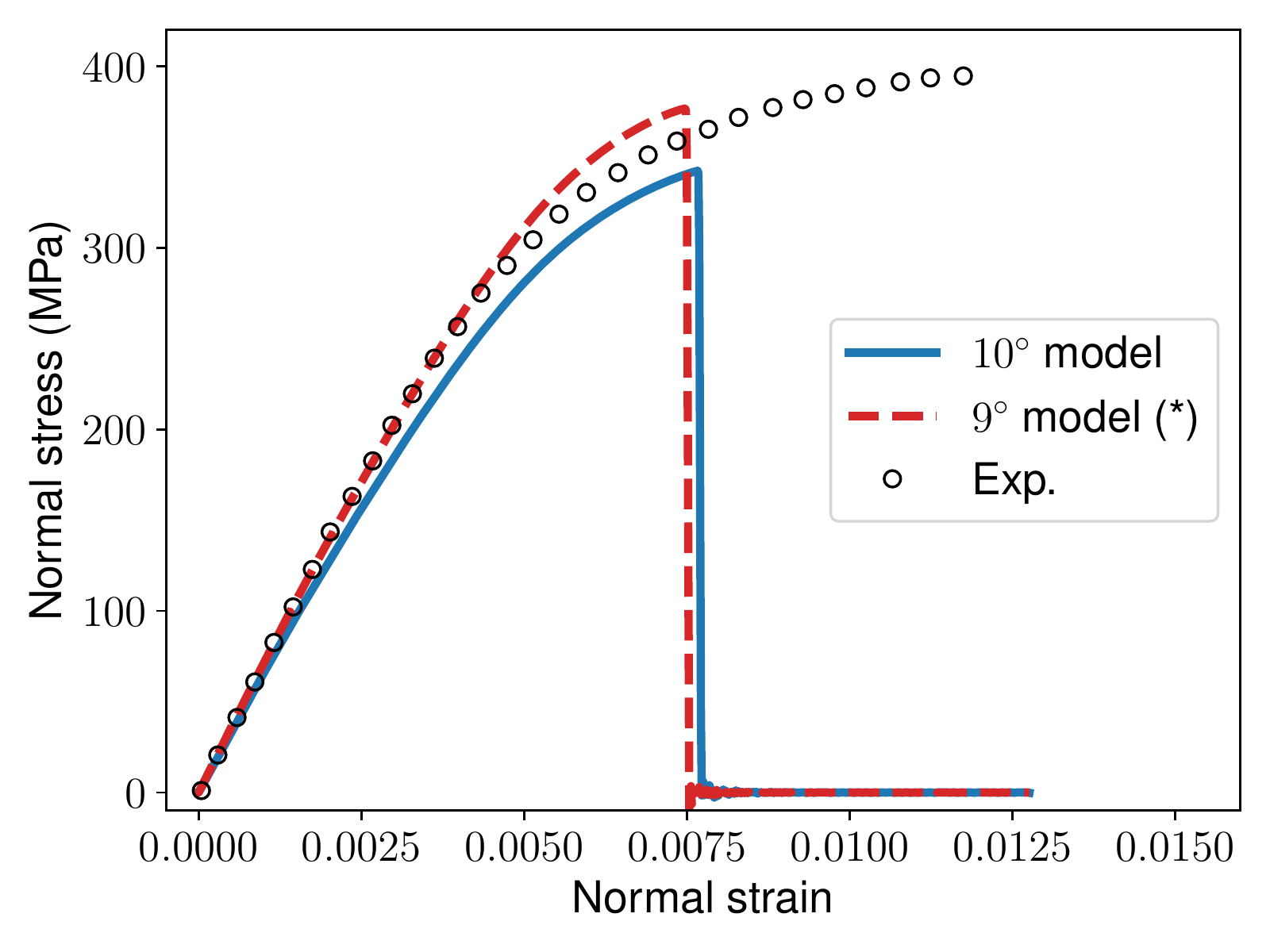}}
		\subfigure[Snapshots of crack propagation.]{\includegraphics[clip=true,trim = 6.0cm 1.2cm 8.0cm 3.0cm,width=0.44\textwidth]{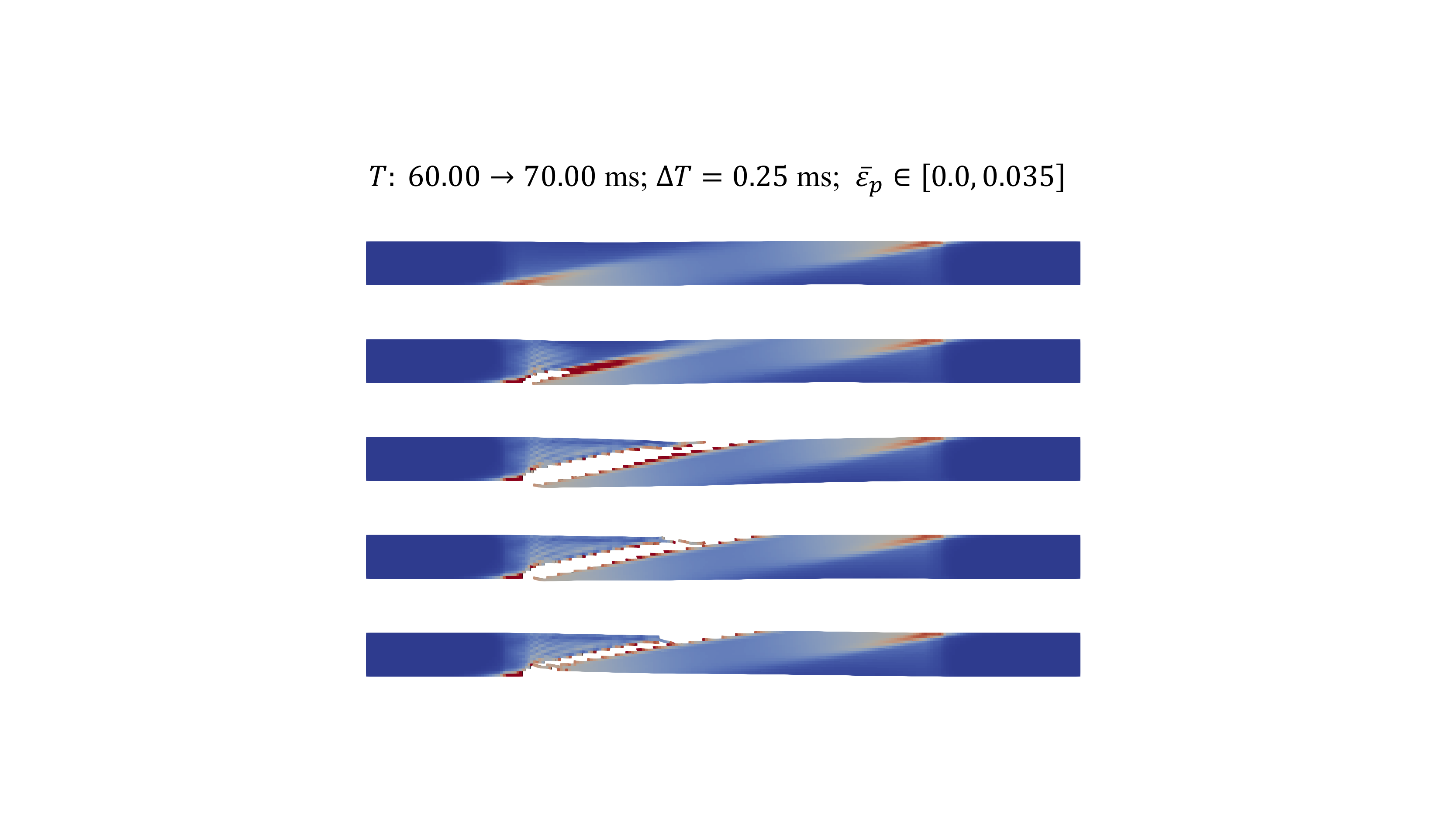}}
		\subfigure[Coupon crack formations in experiment and multiscale simulation. ]{\includegraphics[clip=true,trim = 3.0cm 6cm 3.0cm 6cm,width=0.95\textwidth]{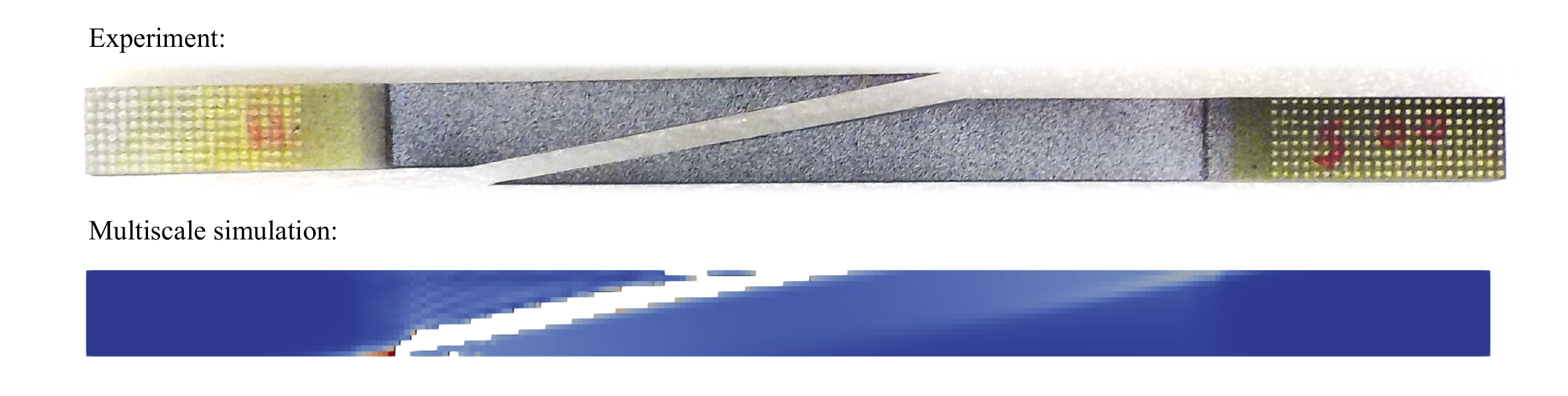}}
		\caption{Results of the $10^\circ$ off-axis tensile coupon. In (a), experimental data of the $10^\circ$ coupon is plotted as the circles. (*) Results of a $9^\circ$ off-axis model is also presented to show the sensitivity of stress-strain curve to fiber orientation. Snapshots in (b) are colored by the average effective plastic strain $\bar{\varepsilon}_p$.}
		\label{fig:ex4_evolve}
	\end{figure}
	
	The normal stress vs. normal strain curve predicted by the concurrent multiscale simulation is compared with the experimental results in Figure \ref{fig:ex4_evolve} (a). A good match of the curves before failure is observed. The multiscale simulation underestimates the maximum strain and stress, potentially due to higher stress concentrations at the clamping areas in the numerical model. To demonstrate the effects of fiber orientation, we also simulated a $9^\circ$ off-axis model. Although the angles are only differed by $1^\circ$, the stress-strain curves change notably due to the strong anisotropy of the UD composite.
	
	Moreover, Figure \ref{fig:ex4_evolve} (b) shows the evolution of crack formations in the macroscale model, and Figure \ref{fig:ex4_evolve} (c) compares the crack formations in the simulation and experiment. As we can see from the snapshots, the crack starts from the side close to the gripping area and propagates along the $10^\circ$ fiber direction, consistent with the experiment.  One discrepancy is that the crack initiates further away from the clamping area in the experiment, which may explain the difference of failure-strain predictions in Figure \ref{fig:ex4_evolve} (a). Nevertheless, with basic microscale material models (e.g., elasticity, isotropic plasticity, 1-D effective cohesive law) and minimal material calibrations, the DMN-enabled multiscale predictions agree well with the experimental validation data.
	
	\section{Advantages and limitations}\label{sec:ad}
	At this point, we have demonstrated DMN's capabilities of modeling multiscale systems with material failure. In terms of strain localization modeling, its advantages over traditional RVE-based methods are two-fold: 1) The macro length scales are naturally propagated to the microscale DOF so that the constraints on the RVE size and the need for extra energy regularization under different mesh sizes are eliminated; 2) As the homogenization within DMN is based on analytical functions, it overcomes the difficulties of applying boundary conditions on the RVE for arbitrary crack formations. The resulting concurrent multiscale models are robust and consistent under the mesh refinement.
	
	Another advantage of DMN, in general, is the efficiency gained from the model reduction. For example, the number of DOF in the DMN with 8 layers for the particle-reinforced composite is 28, while the DNS RVE model is meshed by 84,693 nodes and 59,628 10-node tetrahedron finite elements. For the cyclic loading test shown in Figure \ref{fig:plas} (a), the DNS model took around 7320 s on 10 CPUs, whereas the DMN with $N=8$ and $N_{dof}=28$, implemented in FORTRAN, only took 0.1 s for the same amount of loading steps on one CPU.  Thanks to the nature of the hierarchical binary-tree structure, the computational cost of DMN is proportional to the number of DOF in the network \cite{liu2019deep,liu2019exploring}.
	
	\begin{table}[t!]
		\captionabove{Wall times of various concurrent multiscale simulations on 10 CPUs. The simulated duration and the initial time step are provided for each example. Note that we have applied mass scaling in the $10^\circ$ off-axis tensile coupon test.} % title of Table
		\centering % used for centering table
		\label{table:time} % is used to refer this table in the text
		{\tabulinesep=1.1mm
			\begin{tabu}{>{\centering}p{0.08\textwidth} >{\centering}p{0.08\textwidth} >{\centering}p{0.08\textwidth}>{\centering}p{0.08\textwidth} >{\centering}p{0.08\textwidth} >{\centering}p{0.08\textwidth} >{\centering}p{0.08\textwidth}>{\centering}p{0.08\textwidth} >{\centering}p{0.08\textwidth}} % centered columns (4 columns)
				\hline
				\multirow{4}{*}{} &\multicolumn{2}{c}{\textbf{Crush tube (\ref{sec:crushtube})}} &\multicolumn{2}{c}{\textbf{Crush tube (\ref{sec:crushtube})}} & \multicolumn{2}{c}{\textbf{3-point bend (\ref{sec:3point})}} &\multicolumn{2}{c}{\textbf{$10^\circ$ coupon (\ref{sec:coupon})}}\\
				&\multicolumn{2}{c}{Damaged, 15 ms} &\multicolumn{2}{c}{No damage, 15 ms} & \multicolumn{2}{c}{Case 1, 16 ms} &\multicolumn{2}{c}{Mass scaling, 100 ms}\\
				\cline{2-9}
				&\parbox{0.08\textwidth}{\centering Timestep (ms)} & \parbox{0.08\textwidth}{\centering Cost (hour)} &\parbox{0.08\textwidth}{\centering Timestep (ms)} & \parbox{0.08\textwidth}{\centering Cost (hour)}&\parbox{0.08\textwidth}{\centering Timestep (ms)} & \parbox{0.08\textwidth}{\centering Cost (hour)}&\parbox{0.08\textwidth}{\centering Timestep (ms)} & \parbox{0.08\textwidth}{\centering Cost (hour)}\\
				\hline
				Coarse&$2.6e^{-4}$&1.1&$2.6e^{-4}$&0.9&$7.8e^{-4}$&1.0&\multirow{3}{*}{$1.8e^{-3}$}&\multirow{3}{*}{11.6}\\
				Medium&$1.3e^{-4}$&7.2&$1.3e^{-4}$&6.6&$3.9e^{-4}$&8.9&&\\
				Fine&$0.6e^{-4}$&\diagbox{ }{ }&$0.6e^{-4}$&45.5&$2.0e^{-4}$&69.9&&\\
				\hline
				%inserts single line
		\end{tabu}}
	\end{table}
	
	The wall times of various examples in Section \ref{sec:examples} are listed in Table \ref{table:time}, including the crush tubes with and without matrix failure, the three-point bending test with out-of-plane fibers, and the $10^\circ$ off-axis tensile coupon test. For the same shell geometry and material properties, the critical time step in the macroscale simulation is inversely proportional to the mesh size, while the number of elements is inversely proportional to the square of the mesh size. If the mesh size is refined by half (e.g., Coarse$\rightarrow$Medium), the computational time becomes approximately 8 times longer. By comparing the times of crash tubes with and without damage (see the snapshots in Figure \ref{fig:example2}), we can conclude that the enrichment of cohesive layers for modeling new crack surfaces increases the computational cost by around 10\% in our current implementation.
	
	The computational cost for the damaged crush tube with fine meshes is not provided in the table, as the simulation was terminated around 5 ms due to the convergence difficulties. After the tube was buckled under compression, some failed elements were highly distorted, and the DMN implicit solver did not converge after 10 time-step refinements. For all the other cases, the simulations were finished successfully. Note that convergence issues due to severe element distortions could be avoided by the element deletion introduced in the three-point bending tests and the $10^\circ$ off-axis tensile coupon test.
	
	Although not discussed in this paper, the network interpolation from transfer learning is also an appealing feature of DMN. In \cite{liu2019transfer,liu2020intelligent}, databases of multiple microstructures are unified to cover the design space of composites. The learned database can then be applied to concurrent multiscale simulations with local microstructure variations resulting from the manufacturing processes, such as the injection molding of short fiber-reinforced composites and the metallic additive manufacturing. 
	
	Several directions can be considered to further enhance the DMN framework in terms of both theories and implementations:
	\begin{enumerate}[itemsep=0mm]
		\item The finite-strain formulation of DMN should be considered for more general problems. In \ref{ap:a2}, we discuss several essential aspects of the finite-strain formulation, including its benefits for capturing local crack orientations, new algorithms for the crack initiation, and key steps of deriving the analytical functions in the cohesive building bock. 
		\item Nonlocal regularization methods can be used to reduce the mesh sensitivities in the macroscale model. In this work, the macro scale tensor of DMN is defined by the side lengths of a rectangular or cuboid element. However, for irregularly shaped elements, we need a more general method to relate the macro scale tensor to the mesh characteristics. Nonlocal regularization avoids this issue as the macro length scale is set by the nonlocal size. 
		\item Element deletions cause losses of mass and momentum in the system. In this regard, more advanced numerical techniques should be used for crack modeling in the macroscale model, such as \textcolor{black}{the extended finite element method \cite{belytschko2010coarse,bosco2015multi,wang2016diffuse}} \footnote{\textcolor{black}{Other than the overall stress-strain relationship of the base material in DMN, one needs to extract the overall traction-separation relationship from all localized cohesive layers for the equivalent macroscale discontinuity.}}, and meshfree methods with bond-breaking mechanisms \cite{wu2017three,ren20173d}. Other than the volumetric strain adopted in this work, it is also helpful to define a handy global damage indicator for triggering these crack separation schemes.
		\item To further speed up the concurrent multiscale simulation, one can distribute the macroscale material points to more CPUs. As the computational costs of a deforming DMN depend on the level of material nonlinearities and the number of enriching cohesive layers, the dynamic load balancing should be equipped for better parallelization efficiency. GPU computing is also a promising direction to investigate.
	\end{enumerate}
	
	\section{Conclusions and future work}\label{sec:conclusion}
	In this paper, we propose a new cell division scheme for consistent scale transitions within the DMN framework. In a multiscale failure analysis with strain localization effects, it overcomes the difficulties of choosing proper RVE sizes and applying boundary conditions. For a two-layer DMN building block, we derive the mathematical formulations of the cell division process, which enables the backward propagation of length scales from the macroscale material point to the microscale DOF. Algorithms for the activation of crack surfaces and the implicit failure analysis are proposed based on the cohesive-layer enrichment of DMN. The cohesive layers are modeled by an effective traction-separation law with viscous regularization on the damage parameter.
	
	We investigate two microstructures: the particle-reinforced composite and the unidirectional fiber composite. In particular, the network of the 3-D unidirectional-fiber composite is directly transferred from its 2-D cross-section model based on the physical interpretations of fitting parameters. Parametric studies on a single material point are performed to evaluate the effects of the macro length scale, the relaxation time, and the number of network layers under various loading conditions. We provide three examples of concurrent multiscale simulations with different element formulations in the macroscale, including 3-D thin shell, 2-D plane-strain, and 3-D solid elements. Simulation results of the $10^\circ$ off-axis CFRP tensile coupon test are further validated against the experimental data of the stress-strain responses and the crack formation.
	
	We believe the proposed cell division scheme for scale transition sets a strong basis of DMN for multiscale failure analysis and general problems with strain localization across scales. Together with its accuracy, efficiency, and extrapolation capabilities, the DMN framework provides a feasible way of pushing machine learning and data-driven multiscale materials modeling to computer-aided engineering at an industrial scale. It admits many straightforward extensions:
	\begin{enumerate}[itemsep=0mm]
		\item The DMN framework for multiscale strain localization modeling and concurrent simulations can be applied to other types of material systems, such as nano-particle reinforced rubber composites, polycrystalline materials, and various carbon fiber reinforced polymer composites \cite{liu2019exploring}.
		\item \textcolor{black}{More complex material behaviors with different failure modes can be considered. For instance, in addition to the matrix failure in the unidirectional fiber composite, one may model fiber failure and interfacial debonding \cite{liu2020deep} simultaneously.} More general cohesive laws could be adopted \cite{park2011cohesive}, while algorithms for the crack activation and tangent stiffness computation need to be modified accordingly. Multiphysics materials behaviors could also be explored under the framework.
		\item In the aspect of Integrated Computational Materials Engineering (ICME), it is important to derive the process-structure-property relationship for a multiscale material system. In \cite{liu2020intelligent}, we have demonstrated an integration of the manufacturing process simulation and microstructure-sensitive material models for short fiber reinforced composites based on DMN and transfer learning. With the new scale transition scheme, this approach can be applied to a broader class of materials.
		\item Efficiency and physical interpretability of DMN make it suitable for materials design, optimization, and uncertainty quantification. Trained by linear elastic data, DMNs are extrapolated to predict nonlinear material responses. Therefore, it reduces the sample-size requirement of offline data generation, which tends to be time-consuming for deep learning models without embedded physics.
	\end{enumerate}
	
	\section*{Acknowledgments}
	Z. Liu would like to acknowledge Dr. Haoyan Wei, Dr. C.T. Wu, and Dr. Yong Guo for the helpful discussions. Z. Liu would like to thank Dr. Cheng Yu for providing the adaptive time step refinement algorithm and Dr. Jiaying Gao for sharing the coupon model and the NIST experimental data. Finally, Z. Liu would like to thank Xinying Yu for stimulating creativity.
	
	\appendix
	\section{Machine Learning of DMN: Data generation and optimization}\label{ap:a1}
	Here, we summarize the key steps of training a DMN. For more detailed information, interested readers are referred to our previous papers for 2-D materials \cite{liu2019deep} and 3-D materials \cite{liu2019exploring}. First, for the training based on linear elastic data, the output function of a two-phase material in 3-D space can be written as
	\begin{equation}\label{eq:inout}
	\underbrace{\bar{\textbf{C}}^{rve}}_\text{Output}=\textbf{f}(\underbrace{\textbf{C}^{p1},\textbf{C}^{p2}}_\text{Inputs}, \overbrace{z^{j=1,2,...,2^{N-1}},\alpha_{i=1,...,N}^{k=1,2,...,2^{i-1}},\beta_{i=1,2,...,N}^{k=1,2,...,2^{i-1}},\gamma  _{i=1,2,...,N}^{k=1,2,...,2^{i-1}}}^\text{Fitting parameters}),
	\end{equation}
	where $\bar{\textbf{C}}^{rve}$ is the composite material's overall stiffness tensor predicted by the network, $\textbf{C}^{p1}$ and $\textbf{C}^{p2}$ are the elastic stiffness tensors of microscale phases. In the design of experiments (or sampling) for $\textbf{C}^{p1}$ and $\textbf{C}^{p2}$, we consider both phases to be orthotropically elastic.
	
	For a 3-D microstructure like the particle reinforced composite tested in this paper, we usually generate 500 samples from finite element analyses on the same RVE geometry but with different material inputs. The first 400 samples are selected as the training dataset, and the remaining 100 samples go into the test dataset. Thanks to the embedded physics in its building block, DMN requires substantially less training data than other machine learning problems relying on deep neural networks. According to our experiments, 400 training samples are sufficient to achieve comparable training and test errors.
	
	A cost function based on the mean square error (MSE) is formulated to quantify the distance between the DMN prediction and the training reference,
	\begin{equation}
	J =\dfrac{1}{2N_s} \sum_{s=1}^{N_s}\dfrac{||\bar{\textbf{C}}^{dns}_s-\bar{\textbf{C}}^{rve}||^2}{||\bar{\textbf{C}}^{dns}_s||^2},
	\end{equation}
	where $\bar{\textbf{C}}^{dns}_s$ is the overall stiffness matrix of $s$-th sample computed from DNS, and $N_s$ is the total number of training samples. The operator $||...||$ denotes the Frobenius matrix norm.
	
	Since the homogenization and rotation operations in the two-layer building block both have analytical functions, derivatives of the cost function with respect to all the fitting parameters in Eq. (\ref{eq:inout}) can be derived following the chain rule in a backward propagation process. To accelerate the training speed, we use the stochastic gradient descent (SGD) to optimize the cost function, and compression algorithms are also introduced to prune and merge the binary-tree network.
	
	For a given microstructure, the fitting parameters are initialized randomly following a uniform distribution. However, if one has pre-trained networks for a similar microstructure, transfer learning can be used to ease the training. As mentioned in our previous discussions, another application of transfer learning is to build a unified database of a set of microstructures within a material design space. As all the microstructures share the same topology structure through the initialization, DMNs for new intermediate microstructures can be created by directly interpolating the fitting parameters. 
	
	We implement the sampling, back-propagation, and optimization algorithms in Python, whereas the online module for concurrent multiscale simulations is realized in FORTRAN. The off-shelf automatic differentiation functions in TensorFlow or PyTorch could also be utilized for model training more conveniently.
	
	\section{Extension to finite-strain formulation}\label{ap:a2}
	\begin{figure}[!t]
		\centering
		\graphicspath{{Figures/}}
		\includegraphics[clip=true,trim = 2cm 8cm 2cm 7cm,width=0.98\textwidth]{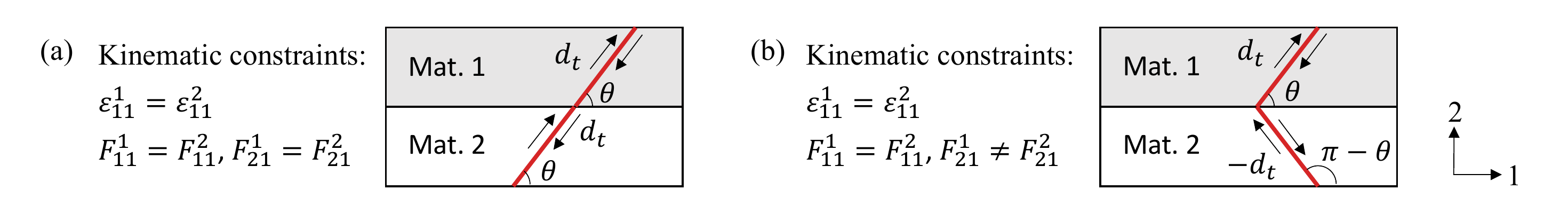}
		\caption{Small and finite formulations for two crack configurations of a 2-D building block.}
		\label{fig:small}
	\end{figure}
	
	Let us consider a 2-D building block with one crack surface in each child material, as shown in Figure \ref{fig:small}. Assume no normal displacement in the cohesive layers and no strain in the base material. Two cohesive layers share the same reciprocal length parameter $v_c$ and the same magnitude of sliding displacement $d_t>0$. The total strain in a child material under the small-strain formulation can be written as
	\begin{equation}\label{eq:small}
	\boldsymbol{\varepsilon} = {v}_c\textbf{R}_c {\textbf{d}} = {v} \begin{bmatrix}
	\cos^2\theta&\sin^2\theta&\sqrt{2}\sin\theta\cos\theta\\
	\sin^2\theta&\cos^2\theta&-\sqrt{2}\sin\theta\cos\theta\\
	-\sqrt{2}\sin\theta\cos\theta&\sqrt{2}\sin\theta\cos\theta&\cos^2\theta-\sin^2\theta\\
	\end{bmatrix}\begin{bmatrix}
	0\\0\\d_t/\sqrt{2}
	\end{bmatrix}
	=\dfrac{1}{2}{v}d_t\begin{bmatrix}
	\sin 2\theta\\-\sin 2\theta \\\sqrt{2}\cos 2\theta
	\end{bmatrix}
	\end{equation}
	with the notation
	\begin{equation*}
	\boldsymbol{\varepsilon} = \begin{bmatrix}
	\varepsilon_{11}&\varepsilon_{22} &\varepsilon_{12}
	\end{bmatrix}^T.
	\end{equation*}
	The kinematic constraint of the building block poses
	\begin{equation}
	\varepsilon^1_{11} = \varepsilon^2_{11}
	\end{equation}
	at the interface of the two child materials. As we can see Eq. (\ref{eq:small}), both configurations in Figure \ref{fig:small} will satisfy the kinematic constraint. Although the second case with two crack surfaces symmetric to the interface represents a non-physical crack geometry, it cannot be distinguished in the small-strain formulation.
	
	This issue can be resolved by describing the system in the finite-strain formulation. Similarly, the deformation gradient of a child material can be written as
	\begin{equation*}
	\textbf{F}= \textbf{I} + {v}_c\textbf{R}^f_c\tilde{\textbf{d}}^f =  \textbf{I} +{v}_c \begin{bmatrix}
	\cos^2\theta&\sin^2\theta&\sin\theta\cos\theta&\sin\theta\cos\theta\\
	\sin^2\theta&\cos^2\theta&-\sin\theta\cos\theta&-\sin\theta\cos\theta\\
	-\sin\theta\cos\theta&\sin\theta\cos\theta&\cos^2\theta&-\sin^2\theta\\
	-\sin\theta\cos\theta&\sin\theta\cos\theta&-\sin^2\theta&\cos^2\theta
	\end{bmatrix}\begin{bmatrix}
	0\\0\\d_t\\0
	\end{bmatrix},
	\end{equation*}
	\begin{equation}\label{eq:finite}
	\textbf{F}= \textbf{I}+\dfrac{1}{2}{v}_cd_t\begin{bmatrix}
	\sin 2\theta\\-\sin 2\theta \\\cos 2\theta + 1 \\ \cos 2\theta -1 
	\end{bmatrix}
	\end{equation}
	with
	\begin{equation*}
	\textbf{F} = \begin{bmatrix}
	F_{11}&F_{22} &F_{12}&F_{21}
	\end{bmatrix}^T \quad\text{and}\quad
	 \quad \textbf{I} = \begin{bmatrix}
	1&1 &0&0 
	\end{bmatrix}^T.
	\end{equation*}
	The kinematic constraints at the interface are
	\begin{equation}
	F^1_{11} = F^2_{11},\quad F^1_{21} = F^2_{21},
	\end{equation}
	which only prefer the first crack configuration in Figure \ref{fig:small} (a) for $\theta\in(0,\pi)$. Therefore, if one wants to determine the exact crack orientations more than the overall material responses, the finite-strain formulation should be used in general.
	
	In the remaining part of this section, we will briefly discuss the failure algorithms in DMN under the finite-strain formulation. Assume the deformation gradient and the first Piola-Kirchhoff (PK1) stress of the base material are $\textbf{F}$ and $\textbf{P}$, respectively. The unit normal of a surface in the deformed configuration $\textbf{n}$ is given by
	\begin{equation}
	{\textbf{n}} = \dfrac{{\textbf{F}}^{-T} {\textbf{N}}}{|{\textbf{F}}^{-T} {\textbf{N}}|},
	\end{equation}
	where $\textbf{N}$ is the surface's unit normal in the undeformed configuration. In contrast to Eq. (\ref{eq:activationlaw}), the plane with the maximum effective traction is obtained from
	\begin{equation}\label{eq:finite_nc}
	{\textbf{N}}_c = \argmax_{{\textbf{N}}'} \,t_m\left(\textbf{P}\cdot {\textbf{N}}', \dfrac{{\textbf{F}}^{-T} {\textbf{N}'}}{|{\textbf{F}}^{-T} {\textbf{N}'}|}\right).
	\end{equation}
	The definitions of the effective traction $t_m$ can be found in Eq. (\ref{eq:tm1}) and (\ref{eq:tm2}). Accordingly, the algorithms proposed in Section \ref{sec:activation} for finding the potential crack surfaces need to be modified.
	
	Once the crack surface is activated, the traction-separation law can be computed similar as before. For example, if $|\textbf{d} \cdot \textbf{n}_c|\geq 0$, the traction force per unit undeformed area can be written as
	\begin{equation}
	\textbf{t} = \dfrac{t_m}{d_m}\left[\beta^2 \textbf{d} + (1-\beta^2)(\textbf{d}\cdot \textbf{n}_c)\textbf{n}_c\right],
	\end{equation}
	where $\textbf{n}_c$ is the normal to the crack surface in the deformed configuration. Different from the small-strain formulation, $\textbf{t}$ is not only related to the displacement vector $\textbf{d}$, but also affected by the deformation gradient $\textbf{F}$ in the base material. Therefore, the incremental form of  the traction-separation law becomes
	\begin{equation}
	\Delta \textbf{t} = \tilde{\textbf{K}} | _{\textbf{F}} \Delta \textbf{d} + \tilde{\textbf{K}}_{geo} | _\textbf{d}\Delta \textbf{F} + \delta \textbf{t}
	\end{equation}
	with
	\begin{equation}
	\tilde{\textbf{K}} = \dfrac{\partial \textbf{t}}{\partial \textbf{d}}, \quad \tilde{\textbf{K}}_{geo} = \dfrac{\partial \textbf{t}}{\partial \textbf{n}_c}\dfrac{d \textbf{n}_c}{d \textbf{F}},
	\end{equation}
	where the extra geometric stiffness matrix $\tilde{\textbf{K}}_{geo}$ comes from the finite deformation of the base material. Note that if the deformation gradient is vectorized as $9\times1$ matrix, $\tilde{\textbf{K}}_{geo}$ has a shape of $3\times9$.
	
	The last thing to consider is the finite-strain interfacial condition in a cohesive building block with the cohesive layer in the $1-2$ plane (see \cite{liu2020deep}). For an arbitrary $\textbf{N}_c$ determined from Eq. (\ref{eq:finite_nc}),  a rotation of the coordinate system is needed so that it becomes $\textbf{N}_c = \begin{Bmatrix}
	0&0&1
	\end{Bmatrix}^T$. The equilibrium conditions at the interface are
	\begin{equation}
	\Delta t_1 = \Delta P_{13}, \quad \Delta t_2 = \Delta P_{23}, \quad \Delta t_3 = \Delta P_{33},
	\end{equation}
	and the cohesive layer does not impose any displacement or stress constraints on the base material in the other directions. Given the stress-strain relation in the base material, $\Delta \textbf{F} = \textbf{D} \Delta \textbf{P} + \delta \textbf{F}$, the contribution of the enriching cohesive layer to the total deformation gradient of the micro-cell can be derived as
	\begin{equation}
	\begin{Bmatrix}
	\Delta F^c_{13}\\\Delta F^c_{23}\\\Delta F^c_{33}
	\end{Bmatrix} = v_c\Delta \textbf{d} = v_c\left(\tilde{\textbf{K}}^{-1}\begin{Bmatrix}
	\Delta P_{13}\\\Delta P_{23}\\\Delta P_{33}
	\end{Bmatrix}- \tilde{\textbf{K}}^{-1}\tilde{\textbf{K}}_{geo}\textbf{D}\Delta \textbf{P} - \tilde{\textbf{K}}^{-1}\tilde{\textbf{K}}_{geo}\delta \textbf{F} - \tilde{\textbf{K}}^{-1}\delta \textbf{t}\right).
	\end{equation}
	
	\section{Adaptive time step refinement}\label{ap:ap3}
	For the implicit analysis of DMN with softening, the time step should be small enough so that the macroscale stiffness tensor at the top node is positive-definite. Meanwhile, the system with both nonlinear plasticity in the base material and softening of enriching cohesive layers may need a smaller increment for convergence at certain load steps. Since a concurrent multiscale simulation commonly has thousands of DMN instances with extensive loading paths, the adaptive load step refinement becomes necessary for better robustness and efficiency.
	
	\bigskip
	
	\noindent\fbox{\begin{minipage}{46em}\label{mp:2}
			\medskip
			\centering\textbf{Box C.1 Algorithm for adaptive time step refinement}
			\begin{enumerate}[itemsep=0mm]
				\item Initialize the macroscale strain increment $\Delta \boldsymbol{\varepsilon}^{macro}$ and time increment $\Delta t$, and set $N_{sub} = 1, i_{sub} = 0$
				\item Compute $\Delta \boldsymbol{\varepsilon}^{macro}_{sub}=\Delta \boldsymbol{\varepsilon}^{macro}/N_{sub}$, and $\Delta t_{sub} = \Delta t/N_{sub}$
				\item Use $\Delta \boldsymbol{\varepsilon}^{macro}_{sub}$ and $\Delta t_{sub}$ as the boundary conditions and time increments for the DMN failure analysis listed in Box \hyperref[mp:1]{3.3.1}
				\item If the Newton's method and the global crack configuration converge, $i_{sub} \leftarrow i_{sub} +1$; else, $N_{sub} \leftarrow 2N_{sub}$, $i_{sub}=2i_{sub}-1$
				\item If $i_{sub} = N_{sub} $, the load step is completed; else, go to 2.
			\end{enumerate}
			\par\smallskip
	\end{minipage}}
	\bigskip
	
	The initial/critical time step is determined by the dilatational wave speed and the shortest size of an element. For a general anisotropic material in either 2-D or 3-D spaces, the largest value in $C^{macro}_{11}$, $C^{macro}_{22}$, and $C^{macro}_{33}$ of the macroscale stiffness tensor can be used to estimate the wave speed safely. Since the UD composite is stiffer in the fiber direction, the critical time step of Case 2 is smaller than the one of Case 1 for the 2-D three-point bending tests in Section \ref{sec:3point}.
	
	%\section*{References}
	\bibliography{references_ROM}
\end{document}